\documentclass[prd,nofootinbib,preprintnumbers,preprint,superscriptaddress]{revtex4}
\usepackage{comment}
\usepackage{amsmath}
\usepackage{amsfonts}
\usepackage{amssymb}
\usepackage{dsfont}
\usepackage[hyperfootnotes=false]{hyperref}
\usepackage[dvipsnames]{xcolor}
\usepackage{blindtext}
\usepackage{graphicx}

\usepackage{latexsym}

\usepackage[font=small,labelfont=bf]{caption}

\usepackage{graphics}
\usepackage{epstopdf}
\usepackage{tabularx}
\usepackage{caption}
\usepackage{setspace}
\usepackage{float}
\usepackage{subcaption}
\usepackage{dcolumn}
\usepackage{bm}
\usepackage{hyperref}
\usepackage{tabularx}
\usepackage{tikz}
\usetikzlibrary{matrix}
\newsavebox{\tempbox}
\usepackage{textcomp}
\usepackage{gensymb}

\usepackage{setspace}
\usepackage{diagbox}
\usepackage{wrapfig}
\usepackage{lipsum}

\usepackage[section]{placeins}
\usepackage{comment}

\newcommand{\removesp{\vspace{-7.5cm}}}

\definecolor{ultramarine}{RGB}{0,32,96}

\newcommand{\debug}[1]{{\iffalse\color{ultramarine} \textbf{#1}\fi}}

\begin{document}

\title{Photon surfaces, shadows and accretion disks in gravity with minimally coupled scalar field}
\author{Igor Bogush}
\email{igbogush@gmail.com}
\affiliation{Faculty of Physics, Moscow State University, 119899, Moscow, Russia}

\author{Dmitri Gal'tsov}
\email{galtsov@phys.msu.ru}
\affiliation{Faculty of Physics, Moscow State University, 119899, Moscow, Russia}

\author{Galin Gyulchev}
\email{gyulchev@phys.uni-sofia.bg}
\affiliation{Faculty of Physics, Sofia University, 5 James Bourchier Boulevard, Sofia 1164, Bulgaria}

\author{Kirill Kobialko}
\email{kobyalkokv@yandex.ru}
\affiliation{Faculty of Physics, Moscow State University, 119899, Moscow, Russia}

\author{Petya Nedkova}
\email{pnedkova@phys.uni-sofia.bg}
\affiliation{Faculty of Physics, Sofia University, 5 James Bourchier Boulevard, Sofia 1164, Bulgaria}

\author{Tsvetan Vetsov}
\email{vetsov@phys.uni-sofia.bg}
\affiliation{Faculty of Physics, Sofia University, 5 James Bourchier Boulevard, Sofia 1164, Bulgaria}


\begin{abstract}
In this article, we conduct a sequential study of possible observable images of black hole simulators described by two recently obtained rotating geometries in Einstein gravity, minimally coupled to a scalar field. One of them, ``Kerr-like'' (KL), can be seen as a legitimate alternative to the rotating Fisher-Janis-Newman-Winicour (FJNW) solution, and the other (TSL) is a scalar generalization of the Tomimatsu-Sato solution.
Unlike the previous version of the rotating FJNW, these solutions do indeed satisfy the system's equations of motion. Our study includes both analytical and numerical calculations of equatorial circular orbits, photon regions, gravitational shadows, and radiation from thin accretion disks for various values of the object's angular momentum and scalar charge.
The  TSL solution was found to simulate Kerr for all valid parameter values with high accuracy. The maximum difference between the deviations of shadows from a circle for the Kerr and TSL cases does not exceed 1$\%$ and fits into the experimental observational data M87$^*$.  However, near-extreme objects show two times smaller peak values of the observed outflow luminosity of the accretion disk than for the Kerr black hole. The KL solution cannot be ruled out by the experimental data for small values of the scalar charge either. As the scalar charge increases, the optical properties change dramatically. The shadow can become multiply connected, strongly oblate, and the photon region does not hide the singularity, so it should be classified as a strong singularity.  
\end{abstract}


\maketitle

\setcounter{page}{2}

\setcounter{equation}{0}
\setcounter{subsection}{0}
\setcounter{section}{0}

\section{Introduction}
\label{sec:introduction}

The recent observation of the shadow of a supermassive compact object at the center of the M87 galaxy by the Event Horizon Telescope (EHT) collaboration \cite{EventHorizonTelescope:2019dse,EventHorizonTelescope:2021dqv}  has opened a new window for studying gravity in its strongest mode \cite{Cunha:2018acu, Dokuchaev:2019jqq,Bronzwaer:2021lzo,Perlick:2021aok}. Images of black holes are actually obtained in the microwave range, but since we are dealing with geometric optics, optical terminology is quite appropriate. 
The optical view of a remote ultra-compact object illuminated by external sources can be very different depending on the geometric position of the source relative to the line of observation, the angle of inclination to the angular velocity of the object's rotation, the lighting scheme (backlight, radiation from accretion disks, and so on) \cite{Dokuchaev:2020wqk} , the presence of plasma \cite{Perlick:2021aok},  dark matter \cite{Konoplya:2019sns,Saurabh:2020zqg,Lee:2021sws} and many other factors. 
A common feature is the presence of a dark spot, corresponding to the absorption of photons by the object and usually associated with the event horizon. 
In the case of uniform illumination from behind, the dark region limits the directions of null geodesics, which, being extended backwards, are gravitationally captured.

In principle, the presence of an event horizon is not absolutely necessary for formation of such images. Shadows can arise from naked singularities or wormholes, so an important task is to find differences between the shadows of real black holes and their  mimickers \cite{Liu:2018bfx,Jusufi:2018gnz,Abdikamalov:2019ztb}. 
As a matter of fact,
while the size of the shadow M87$^*$ can be consistently interpreted within the framework of the general theory of relativity with known estimates of the mass and distance from central black hole in the M87 galaxy, the error interval is still large enough to match the predictions of various theories involving scalar and vector fields with minimal and non-minimal interactions with gravity. 
Deviations from Einstein's gravity can be parameterized using phenomenological non-Kerr metrics  \cite{Psaltis:2018xkc}, which, when interpreted in terms of standard Einstein's equations, can violate fundamental principles such as energy conditions, cosmic censorship, and no-hair theorems.
The results of the EHT impose limitations on the parameters of such violations \cite{Rummel:2019ads}.  Using shadows, one can extract the information about the object's mass, rotation parameter, electric charge \cite{Tsukamoto:2014tja,Kumar:2018ple,Afrin:2021imp}, non-Kerr distortion parameters \cite{Atamurotov:2013sca,Grover:2018tbq},
the presence of scalar hair \cite{Cunha:2015yba,Khodadi:2020jij,Gan:2021pwu,Khodadi:2021gbc}, thermodynamics \cite{Cai:2021fpr}, dark matter \cite{Yuan:2021mzi}. Hypothetical ultra-compact objects (UCO) such as wormholes, regular black holes \cite{Dymnikova:2003vt},  gravastars or naked singularities can be distinguished from each other by their shadows \cite{Nedkova:2013msa,Gyulchev:2018fmd,Shaikh:2018kfv,Shaikh:2018lcc}, especially in the presence of luminous accretion disks \cite{Karimov:2019qfw,Karimov:2020fuj,Gyulchev:2020cvo,Khodadi:2020jij,Shaikh:2021yux, Afrin:2021imp,Boero:2021afh, Bisnovatyi-Kogan:2022ujt}. These perspectives have strongly stimulated the development of new methods for visualizing  the shadows of black holes, both analytical and numerical. 
 
From a mathematical point of view, general patterns of shadow formation can be understood not only based on the study of photon orbits,  i.e.,  one-dimensional submanifolds of spacetime, but also on the basis of the geometry of two-dimensional surfaces, an example of which is a photon sphere of radius $3M$ in Schwarzschild space \cite{Claudel:2000yi,Grenzebach,Grenzebach:2015oea,Yoshino1,Yoshino:2019dty,Galtsov:2019fzq,Galtsov:2019bty,Koga:2020akc}. In four-dimensional spacetime this is an umbilical hypersurface: for it the first and second quadratic forms are proportional. In the Kerr spacetime there are no photon surfaces, but there are spherical surfaces of constant radius in Boyer-Lindquist coordinates on which the spherical orbits of constant radius lie \cite{Teo:2020sey}, which do not fill these surfaces densely, but form a web at a certain angle of inclination of the thread, determined by the impact parameter. As three-dimensional hypersurfaces in spacetime, such spheres can be called partially umbilical \cite{Kobialko:2020vqf}: for them, the first and second quadratic forms proportional not for all tangent vectors, but for their part corresponding to a certain ratio of polar and azimuthal momenta, which can be related to the impact parameter. This understanding can be useful if the geodesic equations are not integrable \cite{Lukes}.
This leads to an alternative visualization of gravitational fields, allowing you to see properties independent of the choice of the observer and light sources when casting shadows. It is also possible to introduce non-closed umbilical and partially umbilical surfaces that visualize the properties of photons escape from the region of a strong gravitational field.
A detailed study of the geometry of such photon surfaces was given for the Plebanski-Demianski family of metrics  \cite{Kobyalko:2021zoc} containing both black holes and naked singularities, as well as asymptotically non-flat spaces. 

For decades, the alternative gravity of the first choice was the scalar-tensor theory, the simplest model being Einstein's general   relativity with a minimally coupled massless scalar field (MES). This raises attention to the cosmic censorship hypothesis \cite{Virbhadra:1995iy,Joshi:2011rlc}, which is often violated in the presence of a scalar field, remember the famous scalar no-hair theorems for black holes. The static spherically symmetric MES solution is a family parameterized by mass and scalar charge and is a naked singularity. Depending on the value of the scalar charge, this singularity can be invisible (weak) or visible (strong) from infinity, depending on whether the singularity is surrounded by a photon sphere or not  \cite{Virbhadra:2002ju}.
This solution is known as the Fisher-Janis-Newman-Winicour (FJNW) solution \cite{Fisher:1948yn,Bergmann:1957zza,Penney:1968zz,Janis:1968zz, Wyman:1981bd,Virbhadra:1997ie,Bhadra:2001fx}, sometimes also called the gamma-metric (for a more recent discussion, see \cite{Abdolrahimi:2009dc}). 
Recently, there has been renewed interest in this solution due to various dualities linking MES to non-minimal scalar-tensor theories such as Horndesky and DHOST and others \cite{Faraoni:2015paa,Galtsov:2018xuc,BenAchour:2020wiw,Domenech:2019syf,BenAchour:2020fgy}, where it can correspond to non-singular metrics or wormholes \cite{Galtsov:2021lmf}. 
Shadows and accretion disk profiles around a static FJNW were studied in \cite{Gyulchev:2019tvk,Gyulchev:2020cvo} (see also \cite{Abdikamalov:2019ztb}), where various differences from the case of a black hole were found. Other identifiers for naked singularities can be strong lens properties \cite{Gyulchev:2008ff,Jusufi:2018gnz} and negative precession of elliptical orbits \cite{Ota:2021mub,Solanki:2021mkt}. 
 
Especially interesting in scalar-tensor theories are Kerr-like naked singularities. Efforts to find rotating generalizations of the FJNW solution have a long history. The first rotating solution, proposed in \cite{Agnese:1985xj}, was obtained using the Janis-Newman (JN) algorithm, \cite{Newman:1965tw}, previously proposed as a formal way to obtain the Kerr metric from Schwarzschild. But, it is worth noting that the JN method was originally conceived as just a formal trick without a deep mathematical justification. Later it turned out that although the algorithm works in a number of supergravity models \cite{Erbin:2016lzq}, there are doubts about its applicability in general scalar-tensor theories, and an explicit check of the solution \cite{Agnese:1985xj} led to negative results \cite {Pirogov:2013wia,Hansen:2013owa,Bogush:2020lkp}). Still, this metric allows interpretation as a solution with additional matter sources, so its many applications in astrophysical modeling \cite{Gyulchev:2008ff,Kovacs:2010xm,Liu:2018bfx,Jusufi:2018gnz} can be considered from this point of view. 
 
Attempts to construct a rotating solution (and solutions with NUT) were also made in the framework of the Brans-Dicke (BD) theory \cite{Tiwari:1976vc,Kim:1998hc,Kirezli:2015wda}. Note that these solutions do not reproduce the FJNW metric in the Einstein frame. A rotating version of the BD analogue of the FJNW solution was proposed in \cite{Krori:1982} again using the JN trick, but the result  turned out to be in conflict with some of the equations of motion \cite{Bogush:2020lkp}.
Other stationary solutions to minimally coupled Einstein-scalar theory have recently been proposed in \cite{Chauvineau:2018zjy,Astorino:2014mda,Sultana:2015avz, Shahidi:2020bla,Bogush:2020lkp} based on known and new generation methods. The aim of this work is to study the photon structure, the shadows and optical images of accretion disks for two new solutions, constructed in \cite{Bogush:2020lkp}. A recent comparison of the observational predictions of one of them with the Kerr case are presented in \cite{Karimov:2022qdb}. 

The work plan is the following. In Sec. \ref{sec:review} we recall the basic ideas and tools of the analysis of the equatorial circular orbits, photon regions, gravitational shadows, and radiation from thin accretion disks. In Sec. \ref{sec:solution} we present two recently obtained rotating geometries in Einstein gravity minimally coupled to the scalar field  and briefly describe their basic features. In Sec. \ref{sec:circular_orbits} we give a fully analytical description of the equatorial circular orbits in order to define accretion disks and circular photon orbits further. In Sec. \ref{sec:photon_region} we develop a comprehensive study of the photon structure of these geometries, including the construction of photon surfaces and photon regions. In Sec. \ref{sec:shadows} we study the features of the gravitational shadows and the lensing patterns observed by a distant ZAMO observer. In Sec. \ref{sec:DiskImage} we numerically obtain the apparent images of thin accretion disk for both considered solutions and analyze them.
\section{Brief review}
\label{sec:review}
In this section, we recall methods of the analysis of the different observable optical properties of spacetimes, namely gravitational shadows, relativistic images, and radiation from thin accretion disks. Besides, we recall the ideas of the equatorial circular orbits and photon regions which are helpful for the description of the optical properties. For Kerr-like solutions, one assumes the general form of axially-symmetric metric with signature $(-+++)$ given by 
\begin{align} \label{eq:metric_g}
    &
    ds^2 = 
    g_{\mu\nu}dx^\mu dx^\nu = 
      g_{tt}dt^2
    + 2 g_{t\phi} dt d\phi
    + g_{rr}dr^2
    + g_{\theta\theta}d\theta^2
    + g_{\phi\phi}d\phi^2,
\end{align}
which is also equipped with the standard Killing vectors $\partial_t$ and $\partial_\phi$ along the time coordinate $t$ and the angular direction $\phi$ respectively. This leads to $(r,\theta)$ dependence of the metric coefficients. Since we also assume the $\mathbb{Z}_2$ symmetry with respect to the equatorial plane, in the approximation $(|\theta-\pi/2|\ll 1)$ the metric tensor $g_{\mu\nu}$ in Eq. (\ref{eq:metric_g}) is a function of $r$ up to $\mathcal{O}\left((\theta-\pi/2)^2\right)$.
\subsection{Equatorial circular orbits}
Consider equatorial circular orbits $\theta=\pi/2$ in a given metric (\ref{eq:metric_g}). The geodesic equations in the equatorial plane take the form:
\begin{equation}
    \frac{dt}{d\tau} = 
\frac{
          E g_{\phi\phi}
        + L g_{t\phi}}{g^2_{t\phi}
        - g_{tt} g_{\phi\phi}
    },
    \quad
    \frac{d\phi}{d\tau} =
     - \frac{
        E g_{t\phi} + L g_{tt}
    }{
        g^2_{t\phi} - g_{tt} g_{\phi\phi}
    },
    \quad
    g_{rr}\left(\frac{dr}{d\tau}\right)^2 = V(r),
\end{equation}
where $\tau$ is an affine parameter, $E$ and $L$ are the specific energy and specific angular momentum of the particles moving along the timelike or null geodesics $\gamma^\mu$
\begin{equation}
    E = - \dot{\gamma}^\mu g_{\mu t} , \quad
    L =   \dot{\gamma}^\mu g_{\mu \phi}, \quad \dot{\gamma}^\mu=d\gamma^\mu/ d\tau,
\end{equation}
and the radial potential $V (r)$ is defined from the condition $\dot{\gamma}_\mu\dot{\gamma}^\mu = \varepsilon$ by
\begin{equation}
   V(r) =
     \varepsilon
    + \frac{
          E^2 g_{\phi\phi}
        + 2 EL g_{t\phi}
        + L^2 g_{tt}
    }{ g_{t\phi}^2 - g_{tt} g_{\phi\phi}},
\end{equation}
with $\varepsilon = -1$ for timelike and $\varepsilon = 0$ for null geodesics. Quantities $E$ and $L$ are conserved since they are associated with Killing vectors $\partial_t$ and $\partial_\phi$ respectively. For circular orbits in the equatorial plane, the following conditions must hold: $V(r)=0$ and $\partial_r V(r)=0$. In the case of the timelike geodesic, these determine the specific energy $E$, the specific angular momentum $L$ and the angular velocity $\Omega$ of the particles moving on circular orbits:
\begin{subequations}
\label{eqAll}
\begin{eqnarray}\label{eqE}
    E &=& 
    - \frac{
        g_{tt}+g_{t\phi} \Omega
    }{
        \sqrt{-g_{tt}-2 g_{t\phi}\Omega-g_{\phi\phi}\Omega^2}
    },    \label{rotE}
    \\\label{eqL}
    L &=& 
      \frac{
        g_{t\phi}+g_{\phi\phi}\Omega
    }{
        \sqrt{-g_{tt}-2 g_{t\phi}\Omega-g_{\phi\phi}\Omega^2}
    },     \label{rotL}
    \\\label{eqOmega}
\Omega&=&\frac{d\phi}{dt}=\frac{-\partial_r g_{t\phi}+\sqrt{(\partial_r g_{t\phi})^2-\partial_r g_{tt}\partial_r g_{\phi\phi}}}{\partial_r g_{\phi\phi}}.
\end{eqnarray}
\end{subequations}
Timelike circular orbits in the equatorial plane exist only if the following inequality holds (See Ref. \cite{Harko:2009} for details):
\begin{equation} \label{eq:omega_condition}
    g_{tt}+2 g_{t\phi}\Omega+g_{\phi\phi}\Omega^2 < 0.
\end{equation}
For all stable circular orbits, the condition $\partial_r^2 V(r)< 0$ holds. The innermost timelike stable circular orbits $r_{ISCO}$ are determined by the limit of this inequality $\partial_r^2 V(r)|_{r=r_{ISCO}}=0$, i.e., 
\begin{equation}\label{eqIsco}
    E^2 \partial^2_r g_{\phi\phi}+2 E L\partial^2_r g_{t\phi}+L^2\partial^2_r g_{tt}-\partial^2_r(g^2_{t\phi}-g_{tt}g_{\phi\phi})=0.
\end{equation}
When inequality (\ref{eq:omega_condition}) saturates (i.e., the left-hand side is strictly equal to zero), one obtains the innermost unstable null circular orbits for massless particles $\varepsilon = 0$, called photon orbits with radius $r_{ph}$. More generally, for an arbitrary null circular orbit the conditions $V(r_{ph})=0$ and $\partial_r V(r_{ph})=0$ with $\varepsilon = 0$ leads to 
\begin{align} \label{eq:r_rho_system_2}
 &\rho^2 g_{tt}+2  \rho g_{t\phi}+g_{\phi\phi}=0,
    \quad \rho =\frac{1}{\partial_r g_{tt}}\left(- \partial_r g_{t\phi}\pm\sqrt{\partial_rg^2_{t\phi}-\partial_rg_{tt}\partial_rg_{\phi\phi}}\right),
\end{align}  
where we introduce an impact parameter $\rho=L/E$ instead $\Omega$ which is more suitable for describing null geodesics. Such orbits can be both stable and unstable. 
\subsection{Photon region} 
Consider now more general photon structures such as fundamental photon surfaces (FPS) and a photon region (see Refs. \cite{Kobialko:2020vqf,Grenzebach,Grenzebach:2015oea} for details). According to Ref. \cite{Kobialko:2020vqf}, null geodesics with a fixed value of the conserved impact parameter $\rho$ can propagate only in the regions defined by the inequality \vskip-1.5cm
\begin{align} \label{CR}
  \rho^2 g_{tt}+2  \rho g_{t\phi}+g_{\phi\phi} \geq0.
\end{align}
For example, a gravitational shadow for an asymptotic observer can be formed by null geodesic only with impact parameter $\rho$ such that the regions (\ref{CR}) are connected and contains both the horizon and spatial infinity. Thus, the determination of the range of the impact parameters, where the shadow is capable to be formed, can be considered as a preliminary estimation for further insights. As a rule, if the solution exhibits additional equatorial mirror symmetry $\mathbb{Z}_2$ (i.e., $\theta\rightarrow\pi-\theta$), the acceptable values $\rho$ lie in intervals $\rho_{i}\leq\rho\leq\rho_{j}$, where $\rho_{i}$ corresponds to different equatorial circular photon orbits $r^i_{ph}$ described by the equations (\ref{eq:r_rho_system_2}) \cite{Galtsov:2019fzq}\footnote{This follows from the fact that the circular equatorial orbits correspond to the local extremum on the left hand side of inequality (\ref{CR}).}. Solving the first equation (\ref{eq:r_rho_system_2}) with respect to $\rho$ we find an alternative expression 
\begin{align} \label{eq:r_rho_system_2_a}
    \rho_{i} =\frac{1}{g_{tt}}\left(- g_{t\phi}\pm\sqrt{g^2_{t\phi}-g_{tt}g_{\phi\phi}}\right)\Big|_{r=r^i_{ph},\theta = \pi/2}.
\end{align}  

After one has determined the acceptable range of the impact parameter $\rho$, we can proceed to the determination of the fundamental photon surfaces (FPS) \cite{Kobialko:2020vqf,Kobialko:2021aqg,Kobialko:2021uwy,Kobialko:2021qat} which play a key role in the process of shadow formation since they capture null geodesics with a fixed value of the impact parameter. Consider an arbitrary hypersurface $S$ of the form $r=f(\theta)$.
To determine the fundamental photon surfaces and regions, we will use the master equation \cite{Kobialko:2021uwy,Kobialko:2021qat} with the vector $\xi^\mu$ normal to $S$, i.e., $\xi^\mu\partial_\mu = N^{-1}\left(\partial_r - (g_{rr}/g_{\theta\theta})f'\partial_\theta\right)$ up to some factor $N^{-1}$. The master equation can be written as a non-linear differential equation of the second order (up to the multiplicative factors) 
\begin{subequations}
\begin{align}
&C(\rho) f'' + \frac{1}{2}\sum^{4}_{n=1}\sum^{3}_{m=1} P_{nm}(\rho-\omega)^{m-1}(f')^{n-1}=0, \label{eq:fps_a} \\
&C(\rho)=\rho^2 g_{tt}+2  \rho g_{t\phi}+g_{\phi\phi},\label{eq:fps_b}\\
&P_{nm}
    =\begin{pmatrix}
            g_{\theta\theta}\Tilde{g}_{\phi\phi} g^{rr}\cdot\partial_r \ln(g_{tt} g^{\theta\theta})
        &
            2 g_{tt} g_{\theta\theta} g^{rr}\cdot \partial_r\omega
        &
            g_{tt}g_{\theta\theta} g^{rr} \cdot\partial_r\ln(\Tilde{g}_{\phi\phi} g^{\theta\theta})
        \\
            -\Tilde{g}_{\phi\phi} \cdot\partial_\theta \ln(g_{tt}g_{\theta\theta}(g^{rr})^2)
        &
            -2 g_{tt} \cdot \partial_\theta \omega
        &
            -g_{tt} \cdot\partial_\theta \ln(g_{\theta\theta}\Tilde{g}_{\phi\phi} (g^{rr})^2)
        \\
            \Tilde{g}_{\phi\phi} \cdot\partial_r \ln(g_{tt}g_{rr}(g^{\theta\theta})^2)  
        &
            2 g_{tt} \cdot \partial_r \omega
        &
            g_{tt} \cdot\partial_r \ln(g_{rr}\Tilde{g}_{\phi\phi}(g^{\theta\theta})^2)
        \\
            - \Tilde{g}_{\phi\phi} g_{rr} g^{\theta\theta} \cdot\partial_\theta \ln(g_{tt}g^{rr})
        &
            -2 g_{tt} g_{rr} g^{\theta\theta} \cdot \partial_\theta \omega
        &
            - g_{tt}g_{rr} g^{\theta\theta} \cdot\partial_\theta\ln(\Tilde{g}_{\phi\phi}g^{rr})
    \end{pmatrix},\label{eq:fps_c}
\end{align}
\end{subequations}
where 
\vskip-1.5cm
\begin{align} 
  \Tilde{g}_{\phi\phi}=g_{\phi\phi}-g^2_{t\phi}/g_{tt},\quad \omega = - g_{t\phi}/g_{tt}.
\end{align}  
The metric tensor depends on the coordinate $\theta$ and function $f(\theta)$, and the problem is a second-order nonlinear ordinary differential equation. The inequality $C(\rho)\geq0$ coincides with definition (\ref{CR}). The additional equatorial mirror symmetry $\mathbb{Z}_2$ implies the same FPS symmetry. In this case, one of boundary conditions is formulated as $f'(\pi/2)=0$. The condition that FPS orthogonally intersects the boundary of the region defined by Eq. (\ref{CR}) \cite{Kobialko:2020vqf} ensures the compactness of FPS spatial section. The photon region is obtained as a union of the fundamental photon surfaces for all values of $\rho$ from the acceptable range. As in the case of the photon surface, geodesics winding around the FPS form the boundary of the gravitational shadow and set of the relativistic images. In many sophisticated spacetimes, the differential equation (\ref{eq:fps_a}) can be solved only numerically. For a chosen value $\rho$, the equation is solved with the finite difference method with condition $f'(\pi/2)=0$ and $f(\pi/2)=f_0$, where $f_0$ is a subject of the shooting method. The parameter $f_0$ is considered to be correct if FPS intersects the boundary of the region defined by Eq. (\ref{CR}) orthogonally up to a predefined precision.

\subsection{Local observer basis and conserved quantities}

Generally, the axially symmetric spacetimes do not permit to construct their shadows or accretion disk image analytically unless they posses the third constant of motion \cite{Grenzebach,Grenzebach:2015oea}. To overcome this issue, we will construct the gravitational shadows and thin accretion disk image numerically. In order to generate the image of the sky observed by a distant observer in the vicinity of the image of a gravitating object, we use the backward ray-tracing method \cite{Cunha:2018acu,Cunha:2015yba}. The observer with a camera is placed at some point $r_O$, $\theta_O$, $\phi_O$ (e.g., the red dot in Fig. \ref{OS}). Null geodesics are integrated in the backward direction starting from the camera towards the gravitating object (schematically, the black sphere in Fig. \ref{OS}). At some step, the geodesic can intersect a thin accretion disk or a distant sphere at $r_C$. Also, it can achieve a point that is very close to the horizon/singularity. The condition of the proximity to the horizon/singularity, where the geodesic is considered captured, is chosen as $r < r_{H,S} + \epsilon$, where $r_{H,S}$ is the coordinate of the horizon/singularity and $\epsilon$ is some small positive number. The tolerance $\epsilon$ is chosen in a such way that the final image is not changed appreciably for a further decrease of $\epsilon$. The distant sphere at $r_C$ or accretion disk is considered to be a source of light. 

The initial position of all geodesics is just the camera position. The initial momentum $p^\mu$ differs for each geodesic curve and depends on the angle at which it enters the camera. We express the momentum in terms of the local reference frame $\{\hat{e}_{(t)},\hat{e}_{(r)},\hat{e}_{(\theta)},\hat{e}_{(\phi)}\}$, where
\begin{itemize}
    \item $\hat{e}{}_{(t)}{}^\nu$ is the observer's four-velocity, which is a timelike vector;
    \item $\hat{e}{}_{(r)}{}^\nu$ is the direction opposite to the one the camera is pointing to. For example, the direction away from the gravitating object. This direction is opposite to the zenith of the observer's celestial sphere;
    \item $\hat{e}{}_{(\theta)}{}^\nu$ is the ``upward'' direction for the camera;
    \item $\hat{e}{}_{(\phi)}{}^\nu$ is the ``rightward'' direction for the camera.
\end{itemize}
Usually, one parameterizes the projection of geodesics onto the observer’s sky by a pair of celestial coordinates $\alpha\in[0,\pi]$ and $\beta\in[-\pi/2,\pi/2]$, which are related to the 4-momentum of the photon as \cite{Cunha}:
\begin{equation}
    p^\mu = 
        \hat{e}{}_{(t)}{}^\mu 
        + \sin \alpha \; \hat{e} {}_{(\theta)}{}^\mu
        + \sin \beta \, \cos \alpha \; \hat{e}{}_{(\phi)}{}^\mu
        + \cos \beta \, \cos \alpha \; \hat{e}{}_{(r)}{}^\mu 
    .
\end{equation}
It is also convenient to introduce the coordinates $(X, Y)$ of the stereographic projection of the celestial sphere onto the plane 
\begin{equation}
    X = \frac{
        2 \cos\alpha \, \sin \beta
    }{
          1
        + \cos \alpha \, \cos\beta
    },\qquad
    Y = \frac{
        2 \sin\alpha
    }{
          1
        + \cos \alpha \, \cos\beta
    },
\end{equation}
and 4-momentum of the photon reads as
\begin{equation}
    p^\mu = 
    \frac{1}{X^2+Y^2+4}\left(
          (X^2+Y^2+4) \hat{e}{}_{(t)}{}^\mu
        - (X^2+Y^2-4) \hat{e}{}_{(r)}{}^\mu
        + 4 Y \hat{e}{}_{(\theta)}{}^\mu
        + 4 X \hat{e}{}_{(\phi)}{}^\mu
    \right).
    \end{equation}

The general stationary observer's basis can be expanded in the coordinate basis $\{\partial_t,\partial_r,\partial_\theta,\partial_\phi\}$ as follows (see Ref. \cite{Novikov}):
\begin{equation}
\hat{e}_{(\theta)}=A^\theta\partial_\theta, \quad \hat{e}_{(r)}=A^r\partial_r, \quad \hat{e}_{(\phi)}=A^\phi\partial_\phi+ \xi \,\partial_t, \quad \hat{e}_{(t)}=\zeta\,\partial_t+\gamma\,\partial_\phi,
\end{equation}
imposing an orthonormal condition $\hat e_{(\mu)}{}^\lambda\hat e_{(\nu)}{}_\lambda=\eta_{\mu\nu}$.  The angular momentum of the observer is defined as 
\vskip-1cm
\begin{align}
L_O=\hat{e}_{(t)}{}^\mu g_{\mu \phi}=\zeta\,g_{t\phi}+\gamma\,g_{\phi\phi}.
\end{align}
The locally measured linear momentum of any particular photon is given by $p^{(t)}=-\hat{e}_{(t)}{}^\mu \,p_\mu $ and  $p^{(i)}=\hat{e}_{(i)}{}^\mu \,p_\mu $ (where $i=1,2,3$), hence:
\begin{equation}\label{locally measured r momentum}
    p^{(t)} = \zeta E -\gamma L,\quad
    p^{(\theta)} = A^\theta p_{\theta},\quad
    p^{(\phi)} = A^\phi L -\xi E,\quad
    p^{(r)} = A^r p_{r}.
\end{equation}
The conserved quantities $E=-p_{t}$ and $L=p_{\phi}$, defined in Eqs. (\ref{eqAll}), describe respectively the energy and the angular momentum of the photon measured by an observer at spatial infinity \cite{Bardeen}.

In this article, we will consider the Zero Angular Momentum Observers (ZAMO) reference frame $L_O=0$ (see Ref. \cite{Novikov}). In this particular case 
\begin{align}
    A^\theta = \frac{1}{\sqrt{g_{\theta\theta}}},\quad
    A^r = \frac{1}{\sqrt{g_{rr}}},\quad
    A^\phi = \frac{1}{\sqrt{g_{\phi\phi}}},\quad
    \zeta = \sqrt{\frac{g_{\phi\phi}}{g_{t\phi}^2-g_{tt}g_{\phi\phi}}},\quad
    \gamma = -\frac{g_{t\phi}}{g_{\phi\phi}}\zeta, \quad \xi=0,
\end{align} %
and 
\begin{align}
&p_{(\theta)}=\sqrt{g_{\theta\theta}}\sin\alpha,\qquad\qquad\,\,\,L=\sqrt{g_{\phi\phi}}\sin\beta\,\cos\alpha,\\\nonumber
&p_{(r)}=\sqrt{g_{rr}}\cos\beta\,\cos\alpha,\qquad E=\frac{1+\gamma\sqrt{g_{\phi\phi}}\sin\beta\,\cos\alpha}{\zeta}.
\end{align}
\subsection{Shadow}
\begin{figure}[tb!]
\centering
 \subfloat[][]{
  \includegraphics[width=0.45\textwidth]{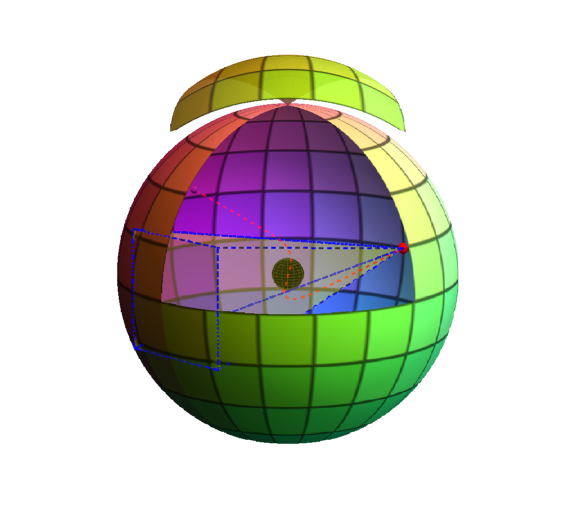} \label{OS}
 }
  \subfloat[][]{
  \includegraphics[width=0.35\textwidth]{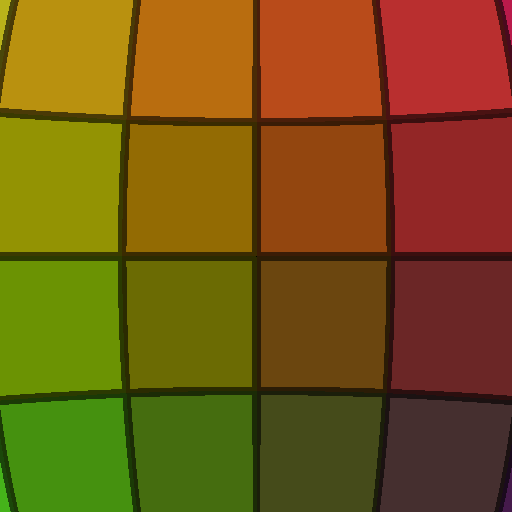} \label{FS}
 }
\caption{(\ref{OS}) the sphere coloring and a schematic example of the ray tracing, the red dot is an observer, the black sphere represent schematically a horizon or singularity (or another light-absorbing surface), the outer sphere is a distant surface emitting the light with the corresponding color, depending on its position $\theta$, $\phi$; (\ref{FS}) an image of the flat space obtained by an observer $\theta_O=\pi/2$.}
\label{RT}
\end{figure}

This allows us to build a mapping from the celestial sphere of the camera to the distant sphere decoded by some color $S^2_O\rightarrow S^2_C$, i.e., to assign a certain color to each point of the observer's celestial sphere $S^2_O$. For example, if the observer captures an image of an empty spacetime (i.e., Minkowski spacetime parameterized by the spherical coordiantes), he or she will get the frame depicted in Fig. \ref{FS}. In the flat spacetime, we see the undistorted image of the celestial sphere. However, the situation changes significantly in the presence of a strong gravitational field, where the light path significantly deviates (red geodesic in Fig. \ref{OS}), creating pronounced effects of gravitational lensing (shadow, relativistic images and Einstein rings) \cite{Cunha:2018acu}. 

If the shadow (set of geodesics cought by the horizon, singularity or another absorbing surface) forms some simply connected domain $D_{SH}$ in plane ($X,Y$), we will characterize it by the following parameters: the vertical $\Delta Y$ and horizontal $\Delta X$ sizes of the shadow, the average radius $R_c$, and the deviation from a circle $\Delta C$ \cite{Cunha:2015yba}. Let us introduce polar coordinates $(R, \varphi)$ in the camera plane as follows
\begin{equation}
    X = R\cos\varphi,\qquad
    Y = R\sin\varphi.
\end{equation}
Then, the simply connected boundary of the shadow can be parameterized by the function $R=F_{SH}(\varphi)$ (which is convenient for the numerical interpolation). Additionally, one can introduce functions $X_{SH}(\varphi) = F_{SH}(\varphi) \cos\varphi$ and $Y_{SH}(\varphi) = F_{SH}(\varphi) \sin\varphi$. The horizontal and vertical sizes can be found from the difference
\begin{align}
    \Delta X = X_{SH}^\text{max} - X_{SH}^\text{min},
    \qquad
    \Delta Y = Y_{SH}^\text{max} - Y_{SH}^\text{min},
\end{align}
where max/min denote global maximum/minimum of the corresponding quantities. The center of mass $(X_c, Y_c)$ of the shadow can be estimated as
\begin{equation}
    X_c = \frac{ \int_{D_{SH}} X dA }{\int_{D_{SH}} dA} =
    \frac{
        2\oint X_{SH} F_{SH}^2 d\varphi
    }{
        3\oint F_{SH}^2 d\varphi
    },\qquad
    Y_c = \frac{ \int_{D_{SH}} Y dA }{\int_{D_{SH}} dA} =
    \frac{
        2\oint Y_{SH} F_{SH}^2 d\varphi
    }{
        3\oint F_{SH}^2 d\varphi
    },
\end{equation}
where $dA$ is the area element of the shadow domain $D_{SH}$, and similar formula is used for $Y_c$. One can construct a circle with a center in $(X_c, Y_c)$ and radius $R_c$ defined from the following expression
\begin{align}
R^2_c=\frac{1}{2\pi}\oint d\varphi\{ (X_{SH}(\varphi)-X_c)^2+(Y_{SH}(\varphi)-Y_c)^2\}.
\end{align}
The dimensionless deviation from the circle is given by (see Ref. \cite{Cunha:2015yba})
\begin{align} \label{deviation_sphericity}
\Delta C^2=\frac{1}{2\pi R^2_c}\oint d\varphi\left(\sqrt{(X_{SH}(\varphi)-X_c)^2+(Y_{SH}(\varphi)-Y_c)^2}-R_c\right)^2.
\end{align}
The latter characteristic depends on the distance from the observer to the gravitating object weakly and deserves attention as reliable. Similarly, since absolute characteristics $\Delta Y$, $\Delta X$ and $R_c$ depend on the position of the observer strongly, it is useful to introduce more suitable ratios
\begin{align}
\mu_{Y/R}=\Delta Y/(2R_c), \quad \mu_{X/R}=\Delta X/(2R_c), \quad \mu_{Y/X}=\Delta Y/\Delta X.
\label{eq:ratios}
\end{align}

Note that for an equatorial ($\theta_O=\pi/2$) ZAMO observer the equatorial size $\Delta X$ of the $Z_2$-symmetric shadow can be calculated analytically if the equatorial photon orbits are known. Indeed, for a $Z_2$-symmetric shadow, we can expect that the shadow will have a maximum size at the middle of the shadow $Y=0$ \cite{Galtsov:2019fzq}. Therefore comparing the impact parameter for the circular null orbit and for the equatorial geodesic in the ZAMO observer's point, we can obtain the explicit expression for the horizontal shadow size ($Y=0$) since circular null orbits correspond to the boundary of the shadow.
\begin{align} \label{eq:sh_sise}
 \Delta X = |X_i-X_j|, \quad X_i = \frac{
        2 \sin \beta_i
    }{
          1
        + \cos\beta_i
    }, \quad \sin\beta_i=\frac{\rho_i}{g_{\phi\phi}+g_{t\phi}\rho_i }\sqrt{g^2_{t\phi}-g_{tt}g_{\phi\phi}}, 
\end{align}
where $\rho_i$ and $\rho_j$ correspond to forward and backward equatorial circular null orbits \cite{Grenzebach,Grenzebach:2015oea} obtained from the equations (\ref{eq:r_rho_system_2}). For an asymptotic observer in an asymptotically flat spacetime, the horizontal size can be estimated as
\begin{align}
\Delta X=|\rho_i-\rho_j|/r_O+\mathcal{O}(r^{-2}_O). 
\label{eq:sh_sise_b}
\end{align}
Therefore, it is convenient to define a dimensional characteristics such as $\Delta x = \Delta X (r_O - r_H)$ (and similarly for other quantities from the stereographic projection), where the shift is introduced to get rid of the divergence near the horizon or singularity at $r=r_{H}$. Such quantity is finite for an infinitely distant observer and the corrections for a finite $r_O$ are of the order $\mathcal{O}(r^{-1}_O)$.

\subsection{Radiation from the accretion disk}

The method which generates an image of the accretion disk is similar to the method for shadows and relativistic images, but the source of light is an accretion disk instead of a distant sphere. The accretion disk is considered to consist of stable circular timelike orbits in the equatorial plane $\theta=\pi/2$ and in the range $r_{ISCO}\leq r\leq 30M$. The set of celestial angles $\alpha$ and $\beta$, which correspond to the null geodesics that goes through the accretion disk forms the image of the accretion disk on the observer’s sky within an initially defined precision. The numerical ray-tracing algorithm reproduces the complete image of the disk composed of all relativistic images of different orders. 

We consider thin accretion disk formed by a neutral anisotropic fluid moving in the equatorial plane. The disk height is considered negligible compared to its extension in the horizontal direction. The model was studied by Novikov  and Thorne in \cite{Novikov2,Page:1974}, where they derived an expression for the radiation flux from the surface of the disk as a function of the radial coordinate $r$:
\begin{equation}\label{F_r}
 F(r)=-\frac{\dot{M}_{0}}{4\pi \sqrt{-g^{(3)}}}\frac{\partial_r\Omega
_{}}{(E-\Omega
L)^{2}}\int_{r_{ISCO}}^{r}(E-\Omega
L)\partial_r L dr.
\end{equation}
Here, $g^{(3)}$ is the determinant of the induced metric of the equatorial plane and $\dot M_0$ is a constant representing the mass accretion rate. The specific energy $E$, specific angular momentum $L$ and angular velocity $\Omega$ of the particles on a given circular orbit are given in the general form for a stationary axisymmetric spacetime in Eqs. (\ref{eqAll}). The observable radiation flux for a distant observer is modified by the gravitational redshift. The apparent intensity $F_{O}$ in each point of the observer's sky is given by
\begin{equation}\label{F_obs}
F_{O} = \frac{F}{(1+z)^4},
\end{equation}
where $z$ is the gravitational redshift. For a general stationary and axially symmetric metric, $z$ can be expressed in the form \cite{Luminet:1979}:
\begin{equation}
1+z=\frac{1+\Omega \rho}{\sqrt{ -g_{tt}-2 g_{t\phi}\Omega-g_{\phi\phi}\Omega^2 }},
\end{equation}
where $\Omega$ is the angular velocity on the circular orbit, and the impact parameter $\rho$ is related to the celestial coordinates $\alpha$ and $\beta$ for ZAMO reference frame by
\begin{equation}
    \rho=\frac{L}{E}=\frac{ \sqrt{g_{\phi\phi}}\,\zeta \sin\beta\,\cos\alpha}{1+\sqrt{g_{\phi\phi}}\, \gamma\sin\beta\,\cos\alpha}.
\end{equation}

\subsection{Analysis of solutions}

In what follows, we analyze two rotating solutions of the scalar-tensor gravity from Ref. \cite{Bogush:2020lkp} performing the following steps.

\textbf{Step one:} Analyze the equatorial circular orbits. Both considered solutions have non-integrable geodesic dynamical systems (as  Zipoy-Voorhees metric \cite{Lukes}), but they still admit analytical description of the equatorial circular orbits, which provide important information about various physical phenomena. In particular, null circular orbits usually correspond to the maximum and minimum values of the impact parameter at which geodesics from the vicinity of the horizon/singularity can reach spatial infinity, i.e. in principle be observable. For an equatorial observer, such orbits correspond to the equatorial dimensions of the shadows (\ref{eq:sh_sise}). The results are described in Sec. \ref{sec:circular_orbits}.

\textbf{Step two:} Analyze the fundamental photon surfaces and regions. As is well known, knowledge of the properties of circular orbits is not enough to understand all the features of the optical behavior of geometries. Hence, we need to analyze the structure of photon surfaces, regions and their generalizations \cite{Kobialko:2020vqf} which are closely related to determining the shape of the gravitational shadow. To visualize the photon regions, we use the standard diagram method \cite{Grenzebach,Grenzebach:2015oea}. The results are described in Sec. \ref{sec:photon_region}.

\textbf{Step three:} Construct the relativistic images and shadows for these gravitating objects for different values of the solution parameters. At this stage, we carry out a complete numerical simulation of the shadows and patterns of gravitational lensing arising under external illumination of the gravitating object \cite{Cunha:2018acu,Cunha:2015yba}. The results are described in Sec. \ref{sec:shadows}.

\textbf{Step four:} Study the observable radiation emitted by a geometrically thin and optically thick accretion disk. Astrophysical  ultra-compact objects are supposed to possess accretion disks, e.g., the accretion disk was directly observed in the vicinity of M87* \cite{EventHorizonTelescope:2019dse}.  
Understanding the features of accretion disks near compact objects that go beyond the Kerr paradigm can help find new physics. The results are described in Sec. \ref{sec:DiskImage}.

\section{Solutions}
\label{sec:solution}
Solutions that go beyond the Kerr paradigm are of interest for searching for new physics and obtaining constraints on a wide variety of extended theories of gravity. Einstein  gravity, minimally coupled to a scalar field, often appears as a consistent truncation of many larger theories. Assuming primary scalar hair, a rotating generalization of FJNW static solutions becomes non-trivial both for their analytic expression and for their interpretation. We consider here two such generalizations recently proposed in Ref. \cite{Bogush:2020lkp}. One of them has a metric tensor in the form of a Kerr solution with an additional conformal factor in the sector $(r,\theta)$. The other exists only for a phantom scalar field and looks like the Tomimatsu-Sato solution with an additional factor in the same sector $(r, \theta)$. The first will be called "Kerr-Like" (KL) and the second "Tomimatsu-Sato-Like" (TSL). The first one is mostly similar to the Kerr solution, while the second exhibits new interesting  physical features (see Ref. \cite{Bogush:2020lkp} for details).

\subsection{Kerr-like geometry}

One of the new rotating solutions for gravity with minimally coupled scalar field given in Ref. \cite{Bogush:2020lkp} reads: 
\begin{align}\label{eq:solution_I}
ds^2=-\frac{\Delta -a^2\sin^2\theta}{r^2+a^2\cos^2\theta}\left(dt-\omega d\phi\right)^2+H( dr^2+ \Delta d\theta^2) + \frac{
        r^2+a^2\cos^2\theta
    }{
        \Delta -a^2\sin^2\theta
    } \Delta\sin^2\theta d\phi^2,
\end{align}
where 
\vskip-1.5cm
\begin{align}
   & \omega = -\frac{2aMr\sin^2\theta}{\Delta-a^2\sin^2\theta}, \quad
    H = \frac{r^2+a^2\cos^2\theta}{\Delta}
    \left(1+\frac{b^2}{\Delta}\sin^2\theta\right)^{-\Sigma^2/b^2}, \\\nonumber
    &\Delta = (r-M)^2-b^2, \quad
    b = \sqrt{M^2-a^2}.
\end{align}
The solution is a rotating generalization of the Fisher solution with the Zipoy-Voorhees oblateness parameter $\delta$  constrained by the condition $M\delta = \sqrt{\Sigma^2 + M^2}$. The outer horizon/singularity is at the point $r_H=M+b$. A scalar field can be either normal or phantom (for imaginary $\Sigma$).  
The solution has the form of a Kerr black hole with an additional common factor for $dr^2$ and $d\theta^2$. The Ricci scalar is
\begin{equation}
    R = \frac{2\Sigma^2}{\Delta(r^2 + a^2\cos^2\theta)}
    \left(1+\frac{b^2}{\Delta}\sin^2\theta\right)^{\Sigma^2/b^2}.
\end{equation}
Near the horizon $\Delta \sim 0$, Ricci scalar is $R\sim \Delta^{- 1 - \Sigma^2/b^2}$.
If $\Sigma^2/b^2 > -1$ and $\Sigma \neq 0$, the solution is a naked singularity. If $\Sigma = 0$, the solution is the vacuum Kerr solution. Since $\sqrt{g_{rr}}dr \sim \Delta^{-1+\Sigma^2/b^2}$ is not an integrable function for $\Sigma^2/b^2 < 0$ near $r_H$, in this case the point $r_H$ is infinitely far, and it is not a genuine ``horizon''. However, we will call this surface ``horizon'' for $\Sigma^2/b^2>0$ since this solution generates shadow similar to the Kerr's ones.

\subsection{Tomimatsu-Sato-like geometry}
A method for obtaining rotating generalizations of FJNW using the Tomimatsu-Sato metric was presented in Ref. \cite{Bogush:2020lkp}.
The oblateness parameter $\delta=2$  of the Tomimatsu-Sato metric fixes the value of the phantom scalar charge. As an example, the following rotating FJNW solution with the constrained phantom scalar charge ${\Sigma^2 = -3(M^2-a^2)}$ was generated
\begin{align} \label{eq:metric}
ds^2=-\frac{A}{B}\left(dt-\omega d\phi\right)^2+\frac{B}{p^4W_{-}^8} dr^2+\frac{B k^2}{p^4W_{-}^6} d\theta^2 + \frac{B}{A}k^2 W_{-}^2\sin^2\theta  d\phi^2,
\end{align}
where
\begin{subequations}
\begin{align}
    \label{eq:solution_II}
    A = & p^4 W_{-}^8 + q^4\sin^8\theta
    - 2p^2q^2W_{-}^2\sin^2\theta\left[
        2W_{-}^4 + 2\sin^4\theta + 3W_{-}^2\sin^2\theta
    \right],
    \\
    B = & \left[
        p^2W_{+}^2W_{-}^2-q^2(\cos^2\theta+1)\sin^2\theta+2pWW_{-}^2
    \right]^2 +
    \\\nonumber &
    + 4q^2\cos^2\theta \left[
        pWW_{-}^2 + (pW+1)\sin^2\theta
    \right]^2,
    \\
    C = & -p^3WW_{-}^2\left[
        2W_{+}^2W_{-}^2 + (W^2+3)\sin^2\theta
    \right] -
    \\\nonumber &
    -p^2W_{-}^2\left[
        4W^2W_{-}^2 + (3W^2+1)\sin^2\theta
    \right]
    +q^2(pW+1)\sin^6\theta,
    \\\nonumber
    \omega = & \frac{2 q M \sin^2\theta C}{A}, \quad  
    W=\left(r - M\right)/k,
    \quad
    W^2_\pm=W^2\pm1,
    \\
    k = & M p/2,\quad
    q=a/M,\quad
    p=\sqrt{1 - a^2/M^2},
\end{align}
\end{subequations}
with $M$ and $a$ being mass and rotation parameter respectively. The constants $p, q$ satisfy a constraint $p^2+q^2=1$. The regular ``horizon'' is placed at $r_H = M + \frac{1}{2}\sqrt{M^2-a^2}$. As we will see further, this surface is infinitely distant from any point with finite $r$, so this is not a genuine horizon. However, we will call this surface ``horizon'' for the same reason as for the KL solution. The scalar field is purely imaginary (phantom) and has a singularity at the horizon. The curvature scalar is
\begin{equation} \label{eq:r_ts}
    R = -24 \; \frac{M^2-a^2}{M^4} \; \frac{ W_{-}^4 }{ B }.
\end{equation}
The function $B$ has zero at the equatorial plane (the ring singularity), where the scalar field is regualar. The ring singularity is always outside the horizon for $0<a/M<1$ and is defined by the equation
\begin{equation}
      (2 r_S - 3M) (2 r_S - M)^3 
    + M a^2 (4 r_S - 3M)
    = 0.
\end{equation}
The largest difference between the singularity radius and horizon radius is $\displaystyle\max_{a/M} \frac{r_S - r_H}{M} \approx 0.054$ at $a/M \approx 0.95$.
In the extremal limit, this solution has a trivial scalar field \cite{Bogush:2020lkp}, which gives the extremal vacuum TS $\delta=2$ solution, which is known to coincide with the extremal Kerr one.

The near-horizon geometry is not typical for standard black holes
\begin{equation}
    ds^2=
        - f_1\left(dt-f_5 d\phi\right)^2
        + f_2(\delta r)^{-4} d(\delta r)^2
        + f_3(\delta r)^{-3} d\theta^2 
        + f_4(\delta r)  d\phi^2,
\end{equation}
where $f_i$ are some functions of $\theta$, $\delta r$ is a small deviation from the horizon $r-r_H=\delta r$. The line element along the radial direction is $ds = \sqrt{g_{rr}} dr \sim (\delta r)^{-2} dr$, which is not integrable, and the ``horizon'' is infinitely far. In the static limit, the near-horizon geometry has another asymptotic behavior
\begin{equation}
    ds^2 = 
    - (\delta r)^2 dt^2
    + (\delta r)^{-2} d(\delta r)^2
    + (\delta r)^{-1} \left(
          d\theta^2
        + \sin^2 \theta d\phi^2
    \right).
\end{equation}

The chronology boundary defined by $g_{\phi\phi} = 0$ is compact and covers the horizon. The boundary at the equator is defined as
\begin{subequations}
\begin{align}
    &
    16 (r-M)^3 r - Ma^2 (4r-3M) = 0,
    \\ &
    16 (r-M) r^3 + Ma^2 (4r-M) = 0,
    \\ &
    64 r^6 + 3 Ma^4 (4 r-7M) + 4 a^2 r (r-M) (4 r^2 + 20 r M - 25M^2) = 0,
\end{align}
\end{subequations}
where the second equation defines the root of multiplicity two (but it splits into two roots of multiplicity one out of the equator). All of these equations have one root out of the horizon. Nevertheless, $g_{\phi\phi} \to 0$ for $\cos\theta\to\pm1$, so there are no Misner strings.

\section{Circular orbit}
\label{sec:circular_orbits}
\subsection{KL geometry}
\begin{figure*}[t]
    \centering
    \begin{subfigure}[t]{0.47\textwidth}
        \includegraphics[width=\textwidth]{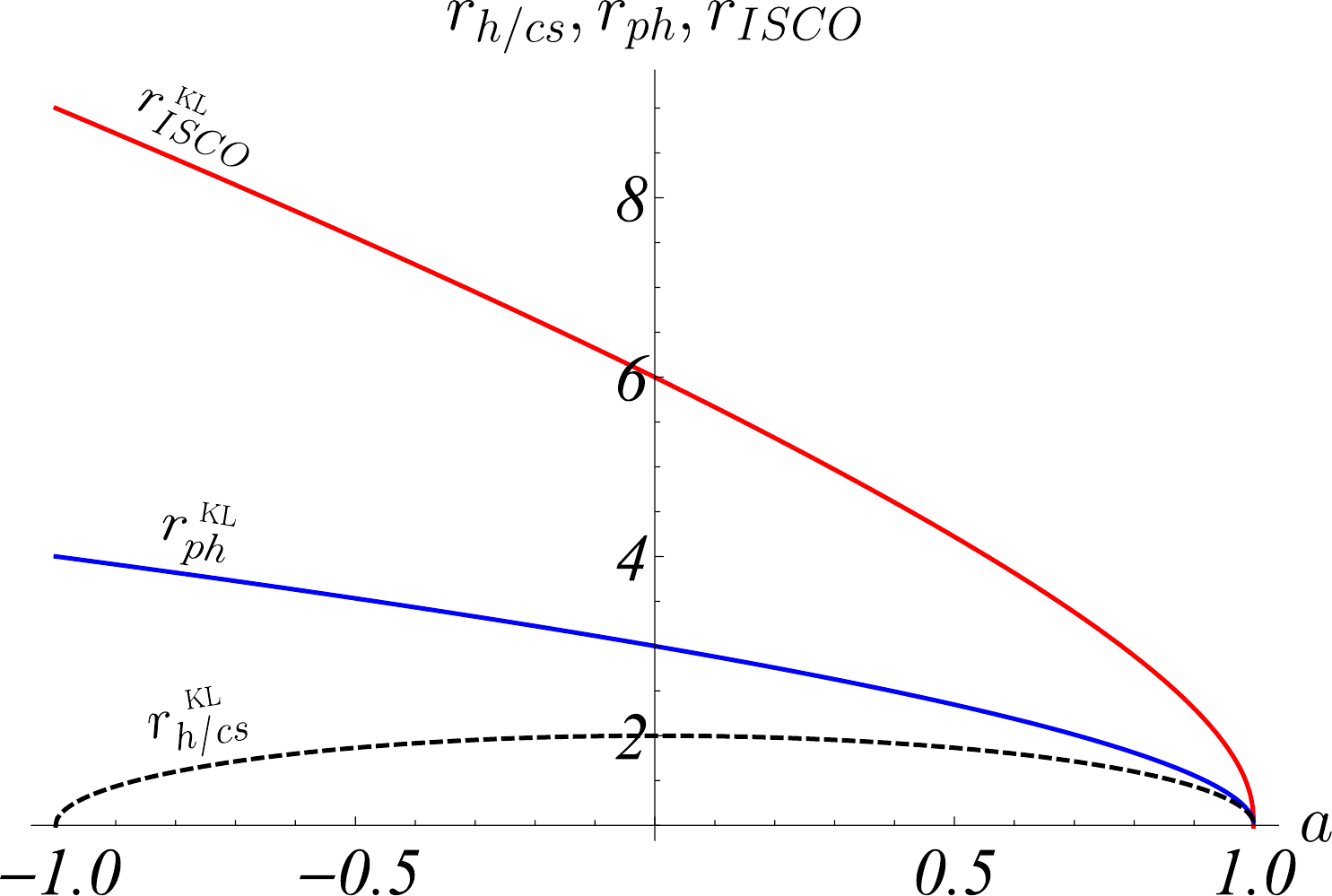}
           \end{subfigure}
           \caption{\label{fig:ISCO}\small Location of the event horizon/curvature singularity $r_{h/cs}$ (black dashed curve) of the KL geometry, the equatorial circular photon orbit  $r_{ph}$ (blue curve) and the inner marginally stable orbits $r_{ISCO}$ (red curve) as a function of the specific angular momentum $a$. In the extremal limit $a=M$ the positions of the curvature singularity, the equatorial circular photon orbit and the inner marginally stable orbit approach each other similar to the Kerr black hole. The positive values of the specific angular momentum $a$ correspond to prograde orbits, while the negative values of $a$ correspond to retrograde orbits of the particles. Mass $M$ is considered to be equal to 1.}
\end{figure*}

In the KL geometry (\ref{eq:solution_I}) equations (\ref{eqAll}) become:
\begin{subequations}
\begin{align}
   & E(r) = \frac{{1 - \frac{{2M}}{r} + \frac{a}{M}{{\left( {\frac{M}{r}} \right)}^{3/2}}}}{{\sqrt {1 - \frac{{3M}}{r} + \frac{{2a}}{M}{{\left( {\frac{M}{r}} \right)}^{3/2}}} }},
   \\
    &L(r) = \frac{{1 - \frac{{2a}}{M}{{\left( {\frac{M}{r}} \right)}^{3/2}} + \frac{{{a^2}}}{{{r^2}}}}}{{\sqrt {1 - \frac{{3M}}{r} + \frac{{2a}}{M}{{\left( {\frac{M}{r}} \right)}^{3/2}}} }}\sqrt {Mr},
    \\
    &\Omega(r)=\frac{\sqrt{M}}{r^{3/2}+a \sqrt{M}}.
\end{align}
\end{subequations}
One notes that $E$, $L$ and $\Omega$ do not depend on the scalar charge $\Sigma$, i.e., they coincide with those from the traditional Kerr black hole solution. Equation (\ref{eqIsco}) for the ISCO takes the form:
\begin{equation}
   \left(E^2(r)-1\right) r^3 +2 a^2 M E^2(r)-4 a M E(r) L(r)+2 M L^2(r)=0,
\end{equation}
which again coincides with the one in the Kerr black hole case. The numerical values of the location of the event horizon/curvature singularity $r_{h/cs}$, the equatorial circular photon orbit $r_{ph}$, and the inner marginally stable orbits $r_{ISCO}$ are shown in Fig. \ref{fig:ISCO} as a function of the specific angular momentum of the solution $a$. The figure includes both prograde and retrograde motion of the massive particles from the disk. In the extremal case $a=M$, these three functions approach each other. Because of that, we consider only orbits that are in the vicinity of the curvature singularity. In this case, our numerical method produces physically meaningful results starting from a radial coordinate at least $\sim 10^{-16}\,M$ away from $r_{h/cs}$. Throughout the paper, we will assume that $a>0$ for prograde\footnote{Prograde motion of particles from the thin accretion disk requires the same signs of the particle's angular momentum $L$ and the specific angular momentum $a$ of the compact object, i.e., $L a\geq 0$. On the other hand,  for retrograde orbits one has $L a\leq 0$. Traditionally, we consider $L > 0$, thus the left-hand side of Fig. \ref{fig:ISCO} $a < 0$ represent retrograde orbits and the right-hand side $a > 0$ represent prograde orbits.} motion of particles of the thin accretion disk. Moreover, the full image of the disk, constructed by the  ray-tracing solver, includes the path of both prograde and retrograde photons emitted by the surface of the disk.

We note that the scalar charge $\Sigma$ has no impact on the equatorial circular orbits for KL solution, thus Fig. \ref{fig:ISCO} depicts also the Kerr case. However, the scalar charge $\Sigma$, as a part of the oblateness parameter $\delta = \sqrt{1+\Sigma^2/M^2}$ entering explicitly $g_{rr}$ and $g_{\theta\theta}$, highly influences the overall disk image creation as the ray-tracing procedure shows further. In the static case $a=0$ the ISCO of the massive particles coincides with that for the Schwarzschild black hole at $r_{ISCO}=6M$.

\subsection{TSL geometry}

\begin{figure*}[t]
    \centering
    \begin{subfigure}[t]{0.47\textwidth}
        \includegraphics[width=\textwidth]{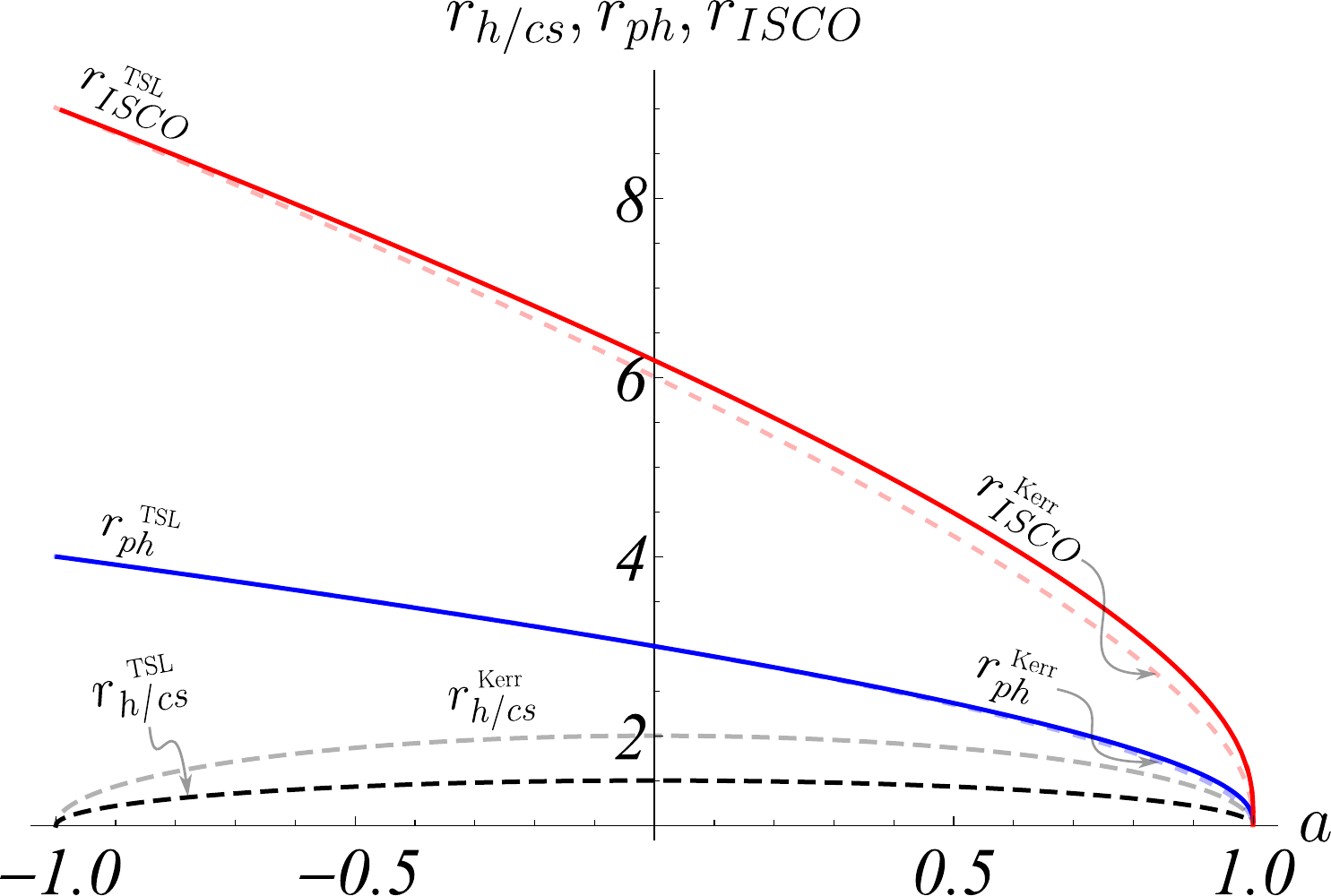}
           \end{subfigure}
           \caption{\label{fig:ISCO_KL_II}\small Location of the event horizon/curvature singularity $r_{h/cs}$ (black dashed curve) of the TSL geometry, the equatorial circular photon orbit $r_{ph}$ (blue curve) and the inner marginally stable orbit $r_{ISCO}$ (red curve) as a function of the specific angular momentum $a$. In the extremal limit $a=M$ the positions of the curvature singularity, the equatorial circular photon orbit and the inner marginally stable orbit approach each other. The mass $M$ is considered to be equal to 1.}
\end{figure*}

In the TSL geometry (\ref{eq:metric}) the ISCO equation (\ref{eqIsco}) becomes very complicated, hence we solve it numerically as shown in Fig. \ref{fig:ISCO_KL_II}. The structure of the diagram is qualitatively similar to the previous case. For comparison we have also presented the location of the horizon, the equatorial photon orbit and the ISCO for the Kerr black hole observing only minimal deviations in these quantities for the two solutions. As in the previous case positive values of the specific angular momentum $a>0$ correspond to prograde orbits, approaching the compact object more closely. On the other hand, negative values of specific angular momentum $a<0$ denote retrograde orbits of the particles, which appear further away from the central object.

\section{Photon region}
\label{sec:photon_region}

\subsection{KL geometry}

The inequality (\ref{CR}) depends on the components of the tensor metric along the Killing vectors $\partial_{t}$, $\partial_{\varphi}$. As far as they are the same for solution (\ref{eq:solution_I}) and for the Kerr metric, the inequality does not depend on the scalar charge $\Sigma$ and coincides for both of them. First, we find values of $\rho$ when the horizon/singularity is not accessible from the spatial infinity. Null geodesics with such $\rho$ cannot get into the horizon/singularity, so they will not correspond to the shadow. From the Eq. (\ref{eq:r_rho_system_2}) we find
\begin{align}
    \rho_{\pm} =
     -a - 6M \sin\left(
          \frac{1}{3} \arccos(a/M)
        - \frac{\pi (1 \pm 4)}{6}
    \right), \quad
    r_{\pm} = \sqrt{\rho^2_{\pm}-a^2}/\sqrt{3}, 
\end{align}
where $r_{\pm}$ correspond to forward and backward equatorial circular orbits. 

In the subextreme case $|a|\leq M$ null geodesics arriving from the infinity can reach the horizon/singularity only for the following interval of the impact parameters
\begin{align} \label{rho:range}
\rho_{-}<\rho<\rho_{+}.
\end{align}
In particular, since the position of these orbits does not depend on $\Sigma$, the horizontal size of the shadow remains the same for different values of the scalar charge at least for an observer with $\theta_O=\pi/2$. However, the shadow changes as the structure of the surfaces of the photon region changes.

In the trivial case $a=0$ and $\Sigma=0$, the circular orbits $r_{\pm}$ coincide and correspond to the standard photon sphere of the Schwarzschild metric $r_{\pm}=r_{PS}=3M$. At nonzero $\Sigma$ even in static regime ($a=0$), the photon sphere immediately decays into the photon region (similarly to the Zipoy-Voorhees solution \cite{Galtsov:2019fzq}).

\begin{figure}[tb!]
\centering
\subfloat[][]{
  \includegraphics[scale=0.45]{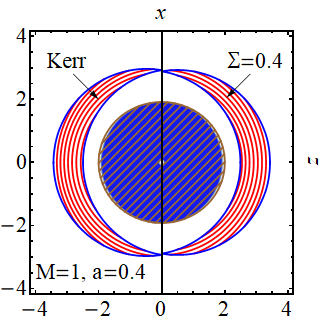} \label{PR1a}
 }
\subfloat[][]{
	\includegraphics[scale=0.45]{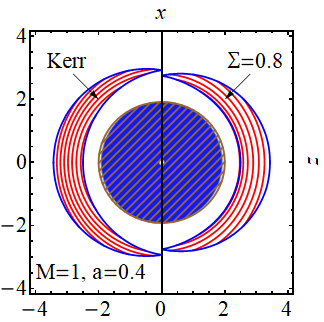} \label{PR1b}	
 }
\subfloat[][]{
	\includegraphics[scale=0.45]{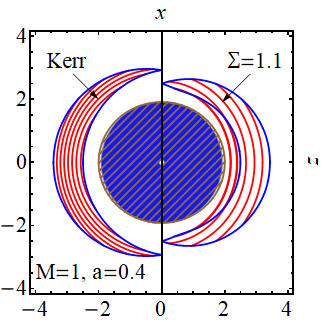} \label{PR1c}	
 }
 \\
 \subfloat[][]{
  \includegraphics[scale=0.45]{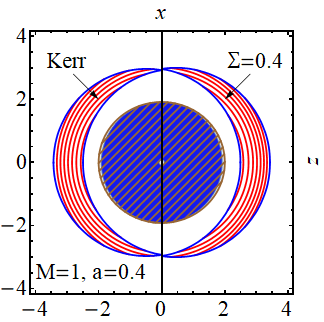} \label{PR3a}
 }
\subfloat[][]{
	\includegraphics[scale=0.45]{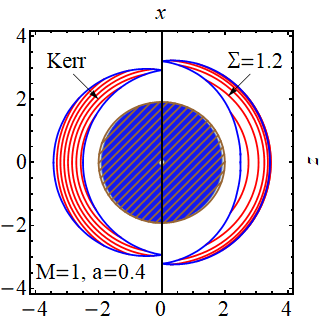} \label{PR3b}	
 }
\subfloat[][]{
	\includegraphics[scale=0.45]{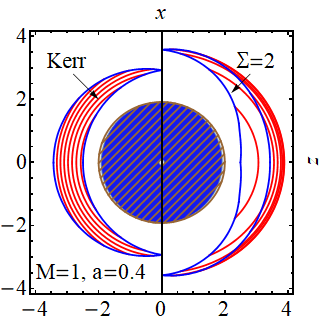} \label{PR3c}	
 }
\caption{Photon regions of the KL solution (\ref{eq:solution_I}) in comparison with Kerr metric with the same $M=1$, $a=0.4$: (\ref{PR1a}-\ref{PR1c}) for the scalar field, (\ref{PR3a}-\ref{PR3c}) for the phantom scalar field ($\Sigma \to i\Sigma$). Red lines denote the fundamental photon surface, blue line is formed by the boundaries of the fundamental photon surfaces, the brown line denotes the ergo-region, the blue region denotes the interior region of the UCO.}
\label{PR_Scalar}
\end{figure}

In order to construct the photon region, first we numerically construct the continuous family of fundamental photon surfaces (\ref{eq:fps_a}) for all $\rho$ from the range (\ref{rho:range}) as described before. The result of the numerical study is shown in Figs. \ref{PR_Scalar}. Here we use the standard method for photon region visualizations proposed in Refs. \cite{Grenzebach,Grenzebach:2015oea}. The left part of each panel in Figs. \ref{PR_Scalar} is an illustration of a photon region for the Kerr metric, while the right one illustrates the photon region of the solution (\ref{eq:solution_I}) with various nonzero $\Sigma$ and the same rotation parameter $a=0.4$.

For the real scalar field (Figs. \ref{PR1a}-\ref{PR1c}), the photon region shrinks vertically for higher values of the scalar charge $\Sigma$. Phantom field behaves differently -- the photon region enlarges for higher $\Sigma$ (Figs. \ref{PR3a}-\ref{PR3c}). As a result, the shadow created by this gravitational source is deformed in comparison with the Kerr metric. Since the equatorial circular orbits are not changed, the horizontal size $\Delta X$ of the shadow remains the same and the entire shadow is flatter/elongated. At the same time, the presence of a well-defined Kerr-like photon region indicates that singularity is a weak one (in the sense of Ref. \cite{Virbhadra:2002ju}) and the corresponding shadow has a more or less smooth boundary and a set of relativistic images near the shadow \cite{Virbhadra:1999nm}. Nevertheless, as we will see further, some additional optical structure may appear inside the shadow.

\begin{figure}[tb!]
\centering
 \subfloat[][Scalar $\Sigma=1.1$] {
  \includegraphics[scale=0.45]{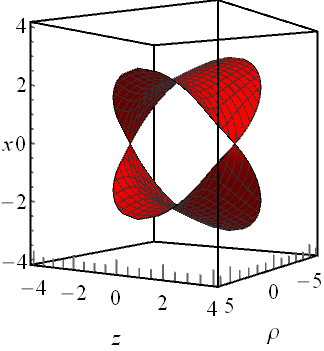} \label{PR2a}
 }
  \subfloat[][Kerr $\Sigma=0$] {
	\includegraphics[scale=0.45]{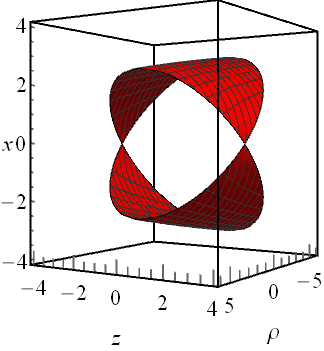} \label{PR2b}	
 }
\subfloat[][Phantom $\Sigma=1.1 i$] {
	\includegraphics[scale=0.45]{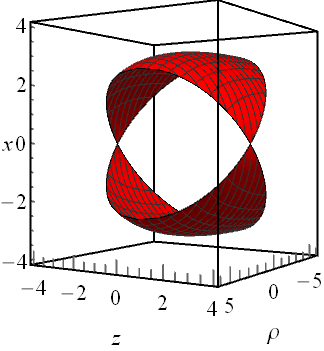} \label{PR2c}	
 }
\caption{Photon function for the KL solution (\ref{eq:solution_I}): (\ref{PR2a}) real scalar field, (\ref{PR2b}) Kerr and (\ref{PR2c}) phantom scalar field with $M=1$, $a=0.4$. }
\label{PR3_Scalar}
\end{figure}

At large values of the parameter $\Sigma$, the fundamental photon surfaces in the photon region turn out to be more and more deformed and displaced in comparison with Kerr. For example, for the KL solution with $a=0.4M$, $\Sigma=1.1M$, the maximal deviation from $r=\text{const}$ surfaces is of the order of $10^{-1}M$. For the real scalar field, the surfaces are curved toward the singularity, and the photon region is not covered in a one-to-one manner by fundamental photon surfaces since there are such points in the photon region that correspond to FPS surfaces with different $\rho$ (since the photon region goes beyond the boundary points of individual FPS surfaces depicted with blue lines in Fig. \ref{PR1c}), as it was in the case of the Zipoy-Voorhees or Tomimatsu-Sato metrics. Particularly, in the static case, the photon region will be two-sheeted since each photon surface in the photon region corresponds to two impact parameters $\rho$ and $-\rho$. 

Such many-valuedness of the photon region foliation by FPS is apparently an important feature of non-integrable dynamical systems. In particular, in all integrable dynamical systems of the Kerr type, the photon region is single-valued.

As shown in Ref. \cite{Kobialko:2020vqf}, in the case of two-sheeted photon region, the {\it fundamental photon function} is more instructive, representing a mapping from the impact parameter to the corresponding fundamental photon surface (see visualization of the function in Fig. \ref{PR3_Scalar}). A slice $\rho = \text{const}$ represent an individual FPS. From Fig. \ref{PR2a} follows that the minimal distance between singularity and photon region corresponds not to a circular orbit with $\rho_\pm$, but to an FPS with some intermediate value of the impact parameter. For such intermediate values of $\rho$, the corresponding null geodesics are not equatorial. This leads to the fact that the shadow created by the solution (\ref{eq:solution_I}) with the real scalar field for an equatorial observer has a minimal deviation from the center of the object not at the equator since the non-equatorial geodesics forming the shadow pass closest to the gravitational source. For a phantom solution, the situation is opposite (Fig. \ref{PR2c}). At a sufficiently large $\Sigma$, the photon region and FPS differ significantly from the case of KL.  

In the case of the real scalar field, if the scalar charge is large enough $\Sigma > \Sigma_\text{cr}(M, a)$, FPS can partially sink into the singularity. As a result, a part of the outer shadow contour ceases to create relativistic images and even disappears since there are no longer photon surfaces on which geodesic can wind. In this case, the singularity becomes stronger \cite{Virbhadra:2002ju} as it is not longer covered by photon surfaces for all valid $\rho$. For the higher rotation parameter $a$, the critical value $\Sigma$ is smaller and tends to zero in the extreme limit.

\subsection{TSL geometry}

\begin{figure}[tb!]
\centering
 \subfloat[][]{
  \includegraphics[width=0.4\textwidth]{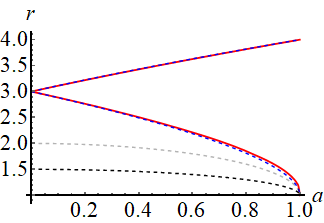} \label{CO1}
 }
  \subfloat[][]{
  \includegraphics[width=0.4\textwidth]{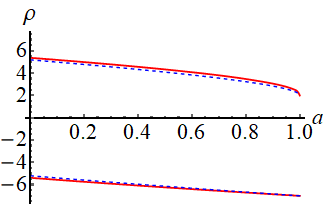} \label{CO2}
 }
\caption{Radius $r$ (a) and parameter $\rho$ (b) for the circular orbits as a function of the rotating parameter $a$ for Kerr (blue) and rotating TSL (red), $M=1$. Dashed black lines denote the horizons for Kerr and TSL.}
\label{CO}
\end{figure}

In the static case, two null circular orbits coincide and correspond to the photon surface
\begin{align} 
r_{\pm}=r_{PS}=3M, \quad \rho_\pm=\pm \frac{25}{6}\sqrt{\frac{5}{3}}M.
\end{align}
Expressions for the null circular orbits in the extreme limit can be evaluated explicitly as well
\begin{align} 
r_{-}=4M, \quad r_{+}=M,\qquad
\rho_{-}=-7M,  \quad \rho_{+}=2M.
\end{align}
In the general rotating case, values $r_\pm$ and $\rho_\pm$ are different and do not have a simple analytical expression and are determined from the equations (\ref{eq:r_rho_system_2}). Their numerical values as a function of parameter $a$ are shown in Fig. \ref{CO1} (red curves). For comparison, the circular orbits in the Kerr metric are depicted with blue curves, while the black dashed lines correspond to the horizons. The plots of the impact parameters $\rho_\pm$ are very close to each other as well and coincide in the extreme limit (Fig. \ref{CO2}). Note that at a value of $a/M\approx 0.686714$, the circular photon orbit $r_-$ coincides with $r_{ISCO}$ and subsequently for greater $a/M$ exceeds it $r_->r_{ISCO}$. As a consequence, the accretion disks will fall into the photon region which can lead to a change in the shape of the images generated by such a light source. 

\begin{figure}[tb!]
\centering
\subfloat[][]{
  \includegraphics[scale=0.45]{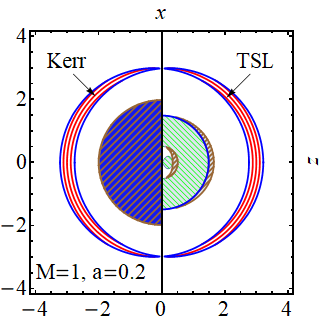} \label{PR4a}
 }
\subfloat[][]{
	\includegraphics[scale=0.45]{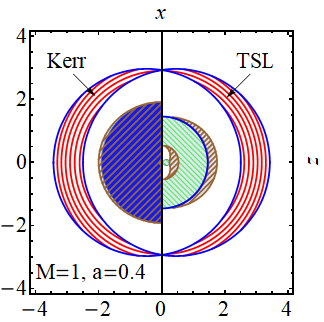} \label{PR4b}	
 }
 \subfloat[][]{
	\includegraphics[scale=0.45]{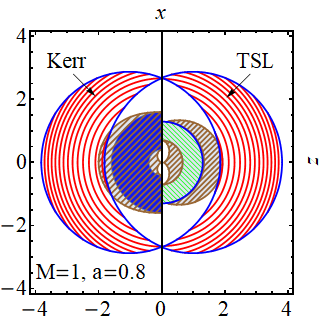} \label{PR4c}	
 }
\caption{Photon regions of the TSL solution (\ref{eq:metric}) in comparison with Kerr metric with the same parameters $M,\,a$ for different values of $a$ and $M=1$. Red lines denote the fundamental photon surface, blue line is formed by the boundaries of the fundamental photon surfaces, gray denotes the ergo-region, the blue region denotes the interior region of the UCO (under horizon or singularity), green encodes the region of causality violation.}
\label{PR_Scalar_1}
\end{figure}

The continuous family of FPS for all $\rho$ from the range found numerically is depicted in Figs. \ref{PR_Scalar_1}. The photon region almost exactly resembles the case of the Kerr metric. In particular, the photon function is almost one-sheeted, and each FPS can be approximated with the surfaces $r=\text{const}$ with good precision. For example, for $a=0.9M$ the deviation from the $r=\text{const}$ surfaces has the order of $10^{-3}M$ which is two orders smaller than the deformation of surfaces in KL geometry.

\begin{figure}[tb!]
\centering
  \subfloat[][Kerr]{
	\includegraphics[scale=0.45]{PR3b}
 }
\subfloat[][Solution (\ref{eq:metric})]{
  \includegraphics[scale=0.45]{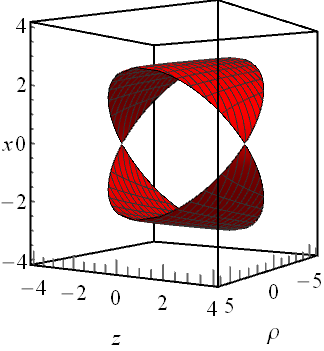} \label{PR5a}
 }
\caption{Photon function for the TSL solution (\ref{eq:metric}) and Kerr with $M=1$, $a=0.4$.}
\label{PR3_Scalar_2}
\end{figure}

This can also be observed from the structure of the photon function in Figs. \ref{PR3_Scalar_2}. Thus, this solution belongs to the class of Kerr mimickers, and one can expect that the shadow of this solution is difficult to distinguish from the Kerr one.

\section{Shadows}
\label{sec:shadows}

\subsection{KL geometry}

\begin{figure}[tb!]
\centering
 \subfloat[][$\Delta X$]{
  \includegraphics[width=0.4\textwidth]{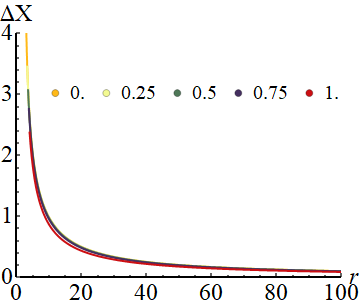} \label{HSKa}
 }
 \subfloat[][$\Delta x$]{
  \includegraphics[width=0.4\textwidth]{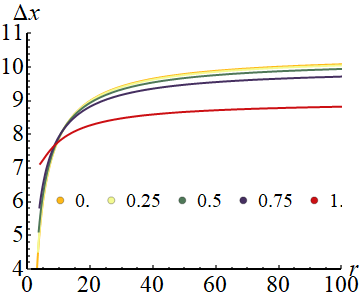} \label{HSKb}
 }
\caption{The dependence of (\ref{HSKa}) the horizontal size $\Delta X$ and (\ref{HSKb}) the normalized horizontal size $\Delta x=\Delta X(r_O-r_H)$ as a function of the rotation parameter $a=0,1/4,...$ and the position of the observer $r_O$ for the Kerr or KL solution with $M=1$.}
\label{HSK}
\end{figure}

Let us estimate the equatorial size of the shadow for an equatorial observer using the analytical result (\ref{eq:sh_sise}). The horizontal size does not depend on the scalar charge $\Sigma$ and is given by the following analytical formula \cite{Grenzebach,Grenzebach:2015oea} 
\begin{align}
\Delta X=|X_+-X_-|, \quad \quad X_\pm = \frac{
        2 \sin \beta_\pm
    }{
          1
        + \cos\beta_\pm
    }, \quad \sin\beta_\pm= \frac{r_O \sqrt{\Delta(r_O)}\rho_{\pm}}{r_O(r^2_O+a^2)+2 a M(a-\rho_{\pm})}.
\end{align}
The dependence of quantities $\Delta X$ and $\Delta x$ as a function of the rotation parameter $a$ and the position of the observer $r_O$ is shown in Fig. \ref{HSK}. For an asymptotic observer $r_O \gg M$ function $\Delta x$ tends to some finite limit and we can obtain the following asymptotic series 
\begin{align}
\Delta X= 6\sqrt{3} \cos\left(\frac{1}{3}\arccos(a/M)-\frac{\pi }{6}\right)\left(\frac{M}{r_O}-\frac{M^2}{r^2_O}\right)+O\left(\frac{M^3}{r^3_O}\right).
\end{align}
For static and extreme limits, there is a simple precise analytical expression for the horizontal size of the shadow
\begin{align}
    \Delta X_{\text{static}} = 6\sqrt{3}M/r_O, \quad
    \Delta X_{\text{extreme}} = 9M/r_O.
\end{align}

First, consider the strong gravitational lensing image for the static limit. The zero scalar charge $\Sigma = 0$ corresponds to the Schwarzschild solution whose shadow is shown in Fig. \ref{as1}, where one can find the main features of strong gravitational lensing -- shadow formation (black area) and relativistic images.

For the case with a small non-zero scalar charge $\Sigma\leq b=M$, the corresponding image of the strong gavitational lensing is shown in Fig. \ref{as2}. As expected, the shadow and the image of gravitational lensing slightly differ from the ones for Schwarzschild solution. The main difference is that the shadow is flattened in comparison with the Schwarzschild case. In particular, deviation from the circle $\Delta C$ increases monotonically for the higher $\Sigma$. The evolution of the image with different $\Sigma$ is similar to one for Zipoy-Voorhees metric with another $\delta=\sqrt{1+\Sigma^2/M^2}>1$ \cite{Galtsov:2019fzq,Abdikamalov:2019ztb}. Thus, the solution in this range of parameters mimics the Schwarzschild solution and has the type of weakly naked singularities. 

However, further increase of the scalar charge $\Sigma$ leads to more appreciable effects associated with the presence of a scalar field. The shadow does not represent a (more or less deformed) solid disk, and one can find narrow regions with scattered (not absorbed) geodesics inside the convex envelope of the shadow (Fig. \ref{as3}). There are geodesics scattered inside the shadow (e.g., narrow colored regions inside the shadow disk) since these photons enter the photon region, fly through the region sufficiently close to the singularity (i.e., without FPS), and then they can leave the trapped region. This case fundamentally differs from the Schwarzschild or Kerr cases \cite{Galtsov:2019bty}. 

For some large value $\Sigma$, the outer shadow contour merges with the inner contours (Figs. \ref{as4}-\ref{as9}). In these cases, the shadow drastically differs from the Kerr one due to the fact that the FPS surfaces are very deformed and very close to the singularity as explained in the previous section.


For a rotating solution, the result is similar. At small values of $\Sigma$, the shadow and the picture of gravitational lensing slightly differ from Kerr (see Figs. \ref{as10}-\ref{as11}), but shadow becomes more circular than for the Kerr solution, i.e., the deviation from the circle $\Delta C$ decreases and reaches a certain minimum value and then monotonically increases (we use a convex hull of the shadow to calculate circularity since the shadow boundary has a very non-circular shape).

At large values $\Sigma$, the solution is strongly deformed and its boundary is not well defined in numerical solutions (Figs. \ref{as12}-\ref{as15}). The transition value $\Sigma$ at which the shadow differs significantly from the Kerr's shadow (i.e., the shadow is not ellipse-like and it is not simply connected) occurs at lower values than in the non-rotating limit. Particularly, in the extreme limit, the shadow differs from the extreme Kerr shadow significantly even for small values of $\Sigma$ (Figs. \ref{as16}-\ref{as18}). This optical behavior confirms our previous analysis of fundamental photon surfaces and functions.

As expected from the previous analysis, solutions with phantom scalar field demonstrate shadows with an opposite behavior -- shadows elongate in the vertical direction (see Fig. \ref{RT4} and Fig. \ref{RT5}). For all values of $\Sigma$, the shadow does not have additional internal structure, but deforms stronger for the larger value of $\Sigma$. Thus, this solution is a good Kerr mimicker if $\Sigma$ is not very large.


\subsection{TSL geometry}

\begin{figure}[tb!]
\centering
\subfloat[][]{
  \includegraphics[width=0.4\textwidth]{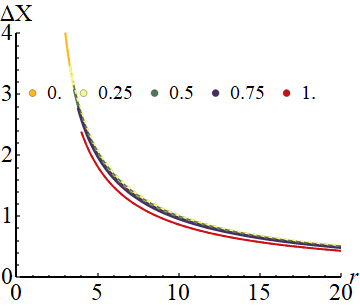}
 }
 \subfloat[][]{
  \includegraphics[width=0.4\textwidth]{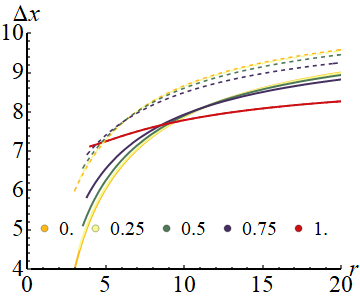}
 }
\caption{The plot of the dependence of quantities (a) $\Delta X$ and (b) $\Delta x=\Delta X(r_O-r_H)$ on the parameters of rotation $a$ and the position of the observer $r_O$ for Kerr and TSL solution (dashed).}
\label{HSKc}
\end{figure}

Let us consider a shadow from the solution (\ref{eq:metric}) observed by an equatorial observer. Using Eq. (\ref{eq:sh_sise}), we can determine the equatorial size of the shadow. Since there is no a simple analytical expression for $\rho$, the function of the equatorial size of the shadow $\Delta x$ is constructed numerically in Fig. \ref{HSKc}. An asymptotic observer of a rotating solution far from the extreme limit beholds a larger shadow for the solution (\ref{eq:metric}) than for the Kerr solution.

For static and extreme limits, the precise analytical expressions for the equatorial size of the shadow are
\begin{align}
    \Delta X_\text{static}=\frac{25}{3}\sqrt{\frac{5}{3}}M/r_O, \quad
    \Delta X_\text{extreme}=9M/r_O.
\end{align}
The sizes $\Delta X$ of the shadow of the extreme solution (\ref{eq:metric}) and Kerr solution completely coincide as expected. For a static solution, the equatorial size differs even for an asymptotic observer, namely, it is 3.5\% larger for the solution (\ref{eq:metric}) than for the Kerr solution. Furthermore, we analyze size ratios (e.g. $\mu_{X/R}$) as they are less dependent on the observer's position (which is inevitably finite in numerical calculations).

The result of the numerical modeling of the shadow is shown in Fig. \ref{TS1}. For the sake for comparison, we also present an image of a gravitational lensing in the Kerr metric with the same values of the rotation parameter $a$. As expected from the foresaid analysis of the photon region, the image of gravitational lensing is almost identical to the Kerr metric. However, there are a number of differences, whose investigation requires a more detailed analysis. 

\begin{figure}[]
\centering
  \subfloat[][Kerr]{
  \includegraphics[scale=0.5]{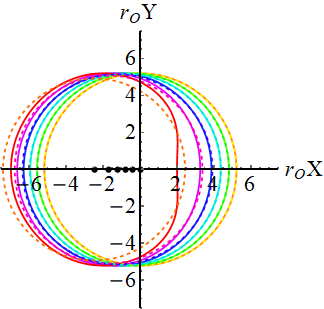} \label{1a}
 }
  \subfloat[][TSL]{
  \includegraphics[scale=0.5]{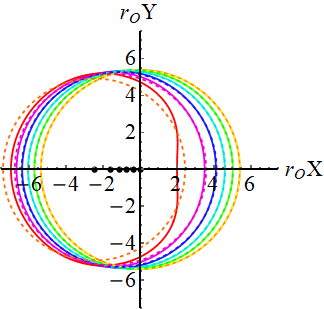} \label{1b}
 }
\caption{
Shadow boundaries for (a) Kerr and (b) the TSL solution (\ref{eq:metric}) with $M=1$ and $a=0,\,0.2,\,0.4,\,0.6,\,0.8,\,1$ for an equatorial observer $r_O/M=10\,000$, $\theta_O=\pi/2$.}
\end{figure}

\begin{figure}[]
    \subfloat[][]{
  \includegraphics[scale=0.45]{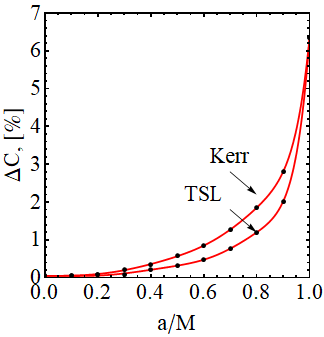} \label{2ts}
 }
 \subfloat[][]{
  \includegraphics[scale=0.475]{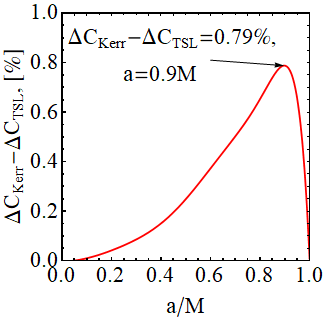} \label{4ts}
 } \\
  \subfloat[][]{
  \includegraphics[scale=0.5]{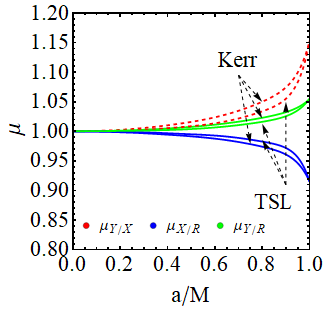} \label{3ts}
 }
 \subfloat[][]{
  \includegraphics[scale=0.47]{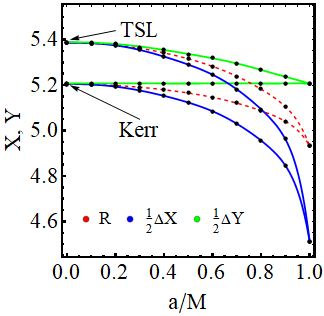} \label{DXYR}
 }
\caption{
Comparison of the shadows between Kerr and the TSL solution (\ref{eq:metric}) for an equatorial observer $r_O/M=10\,000$, $\theta_O=\pi/2$:
(c) deviation from the circle, (d) difference between deviations, (e) ratios $\mu_{Y/X}$ (red dashed), $\mu_{X/R}$ (blue) and $\mu_{Y/R}$ (green), and (f) horizontal and vertical shadow size and average radius. Black dots denote values calculated numerically and other curves are obtained with interpolation.}
\end{figure}

In order to clarify these differences, we construct a family of shadow boundaries for different values of the rotation parameter (Figs. \ref{1a} and \ref{1b}) as well as the corresponding fitted circles. These families look quite similar, but in fact they differs in the following way. The ratios $\mu_{Y/X}$ (red dashed), $\mu_{X/R}$ (blue) and $\mu_{Y/R}$ (green) from Eq. (\ref{eq:ratios}) are shown in Fig. \ref{3ts}, as well as the deviation from a circle $\Delta C$ (\ref{deviation_sphericity}) in percents as a function of the rotation parameter in Fig. \ref{2ts}. The difference between the family of Kerr and TSL metrics becomes more obvious. The shadow boundary of the solution (\ref{eq:metric}) is more circular than the corresponding shadow of the Kerr metric for all rotation values. In fact, the difference between deviations has a maximum at $a\approx0.9M$ equal to $\approx0.79\%$ as shown in Fig. \ref{4ts}. 
Note that, the deviation from circle does not exceed $10\%$, which is consistent with the observations of M87 \cite{EventHorizonTelescope:2019dse}. Thus, at the given level of observation precision, this solution is optically indistinguishable from the Kerr solution. Fig. \ref{DXYR} illustrates more observer dependent characteristics such as the vertical $\Delta Y$ and $\Delta X$ horizontal size of the shadow and its average radius $R_c$. The shadow of the TSL solution is generally larger, and the vertical size of the shadow decreases with increasing $a$, in contrast to the Kerr family. In the extreme limit, the shadows coincides as expected, since both solutions coincide for $a=M$.

\section{Image of thin accretion disk in the geometry of KL and TSL solutions}\label{sec:DiskImage}

In this section we will numerically obtain the apparent images of thin accretion disk for the KL and TSL solutions. The accretion process is modeled by the collective movement of neutral massive particles (dust) on stable circular orbits in the equatorial plane. It is assumed also that the radiation from the disk occurs isotropically \cite{Novikov2}. Our study is going to consider restricted values for the scalar charge to mass ratio in the range $-9 \leq \Sigma^2/M^2 \leq 50$ and values of the specific angular momentum $a/M=\{0, 0.5, 0.9, 0.998\}$. The latter correspond to the static, the rotating and the near-extremal cases. In order to capture the strongest relativistic effects around a compact object the physical extension of the disk is  taken up to the radial distance $r/M=30$. The inner edge of the disk is defined by a particular ISCO as depicted on Fig. \ref{fig:ISCO}. The observer is assumed to be positioned at $r_{O}/M=10\,000$, which represents an effective asymptotic infinity in the spacetime. 

\subsection{KL geometry}

\begin{figure}[!hbt]
    \centering
    \begin{subfigure}[t]{0.45\textwidth}
        \includegraphics[width=\textwidth]{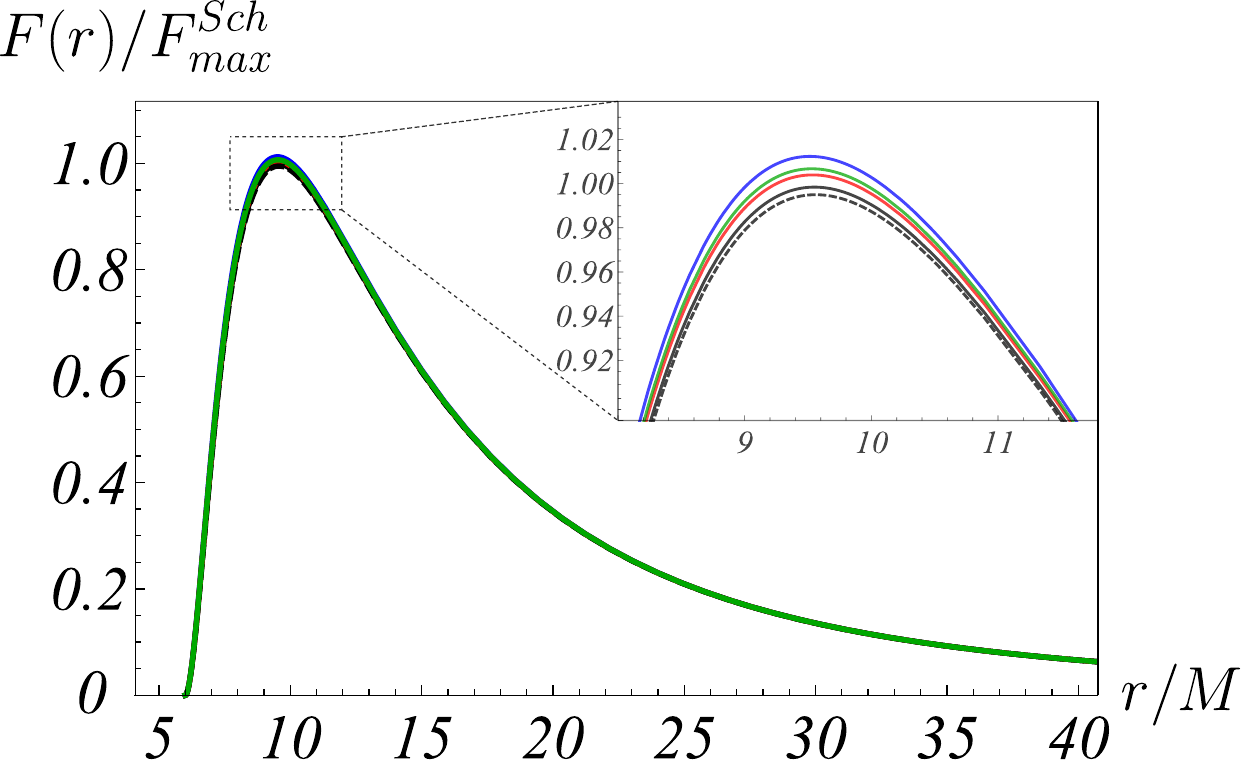}
        \caption{Static case $a/M=0$}\label{fig:KL_I_Fluxs_a}
           \end{subfigure}
         \begin{subfigure}[t]{0.45\textwidth}
        \includegraphics[width=\textwidth]{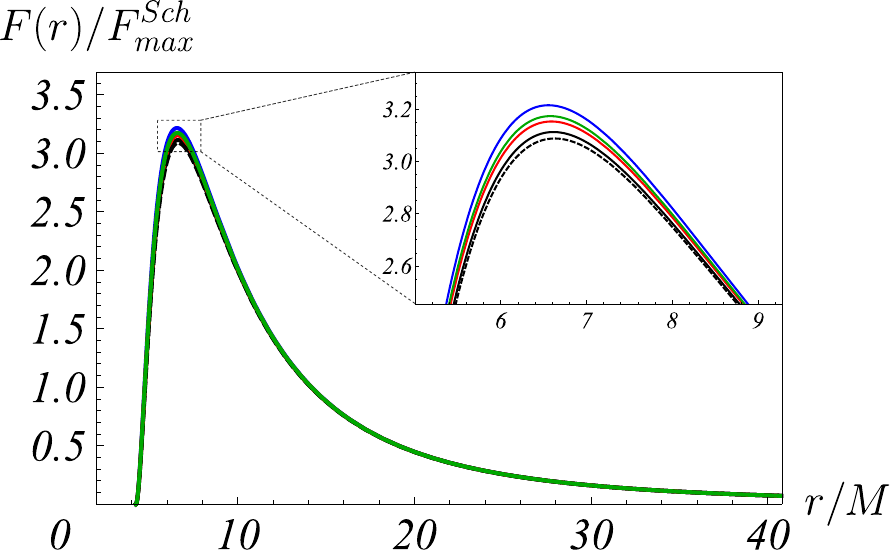}
        \caption{Rotating case $a/M=0.5$}\label{fig:KL_I_Fluxs_b}
           \end{subfigure}
    \begin{subfigure}[t]{0.45\textwidth}
        \includegraphics[width=\textwidth]{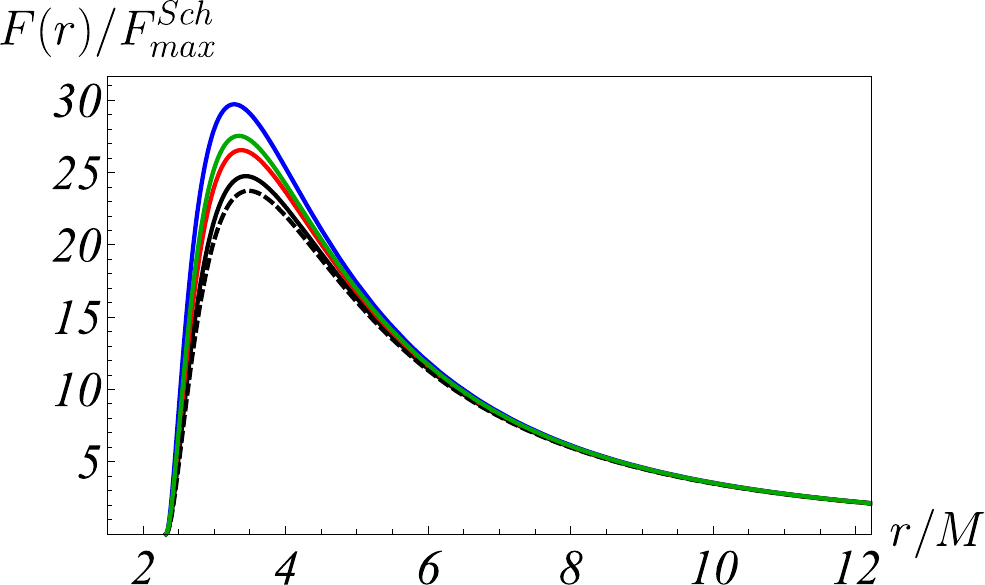}
        \caption{Rotating case $a/M=0.9$}\label{fig:KL_I_Fluxs_c}
           \end{subfigure}
         \begin{subfigure}[t]{0.46\textwidth}
        \includegraphics[width=\textwidth]{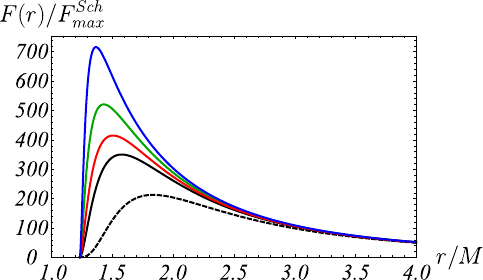}
        \caption{Near-extremal case $a/M=0.998$}\label{fig:KL_I_Fluxs_d}
           \end{subfigure}
         \caption{\label{fig:KL_I_Fluxs}\small Dependence of the radiation energy flux over the disk
on the radial distance in the KL geometry. a) Static case $a=0$ and $\Sigma^2/M^2$ in the range $\{-0.5, 0, 0.8, 1.2, 2.0\}$ (dashed black, black, red, green, blue). In this case, the reference is the black solid curve, corresponding to the Schwarzschild black hole with the maximum of the flux function $F(r)$ given by $F^{Sch}_{max}=1.37\times 10^{-5} M \dot M$. b) Rotating case for $a/M=0.5$ and $\Sigma^2/M^2$ in the range $\{-0.5, 0, 0.8, 1.2, 2.0\}$ (dashed black, black, red, green, blue). c) Rotating case for $a/M=0.9$ and $\Sigma^2/M^2$ in the range $\{-0.5, 0, 0.8, 1.2, 2.0\}$ (dashed black, black, red, green, blue).  d) Near-extremal case for $a/M=0.998$ and $\Sigma^2/M^2$ in the range $\{-0.5, 0, 0.1, 0.2, 0.3\}$ (dashed black, black, red, green, blue).} 
\end{figure}

The radial distribution of the radiation energy flux over the surface of the disk is depicted in Fig. \ref{fig:KL_I_Fluxs}. In the static case $a=0$ (Fig. \ref{fig:KL_I_Fluxs_a}) the flux $F(r)$ depends on the values of the scalar charge $\Sigma$. Compared to the Schwarzschild case ($\Sigma^2/M^2=0$) the maximum value of $F(r)$ increases slowly for larger positive values of $\Sigma^2/M^2$ and also slowly decreases for $\Sigma^2/M^2<0$ (the phantom case). The flux values are normalized to the maximum energy flux $F^{Sch}_{max}=1.37\times 10^{-5} M \dot M$  of the Schwarzschild black hole. The radial location of the peak values of the flux function tend to get closer to the ISCO ($r_{ISCO}=6M$) for an increasing positive $\Sigma^2/M^2$, while in the phantom case the peak is shifted away from ISCO. This behavior is shown in Table \ref{table:1}.

Similar situation occurs in the rotating case. For example, at $a/M=0.5$ (Fig. \ref{fig:KL_I_Fluxs_b}) the peaks of the flux are at least 3 times larger than the Schwarzschild peak $F^{Sch}_{max}$. However, the peaks occur closer to the compact object between $r/M=5.42$, corresponding to  $\Sigma^2/M^2=50$, and $r/M= 6.88$, corresponding to $\Sigma^2/M^2=-9$ (Table \ref{table:1}). For comparison, the ISCO for the  Kerr black hole  ($\Sigma^2/M^2=0$, $a/M=0.5$) is located at $r/M=4.23$, while its flux peak is at $r/M=6.62$, (Table \ref{table:1}).

Increasing the specific angular momentum of the compact object to $a/M=0.9$ the peaks of the flux function increase at least 10 times that of the Schwarzschild case (Fig. \ref{fig:KL_I_Fluxs_c}). The radial location of the peaks occurs in $2.41\leq r/M\leq 4.15$ for $-9\leq \Sigma^2/M\leq 50$. The ISCO for the  Kerr black hole  ($\Sigma^2/M^2=0$, $a/M=0.5$) is located at $r/M=2.32$, while its flux peak is at $r/M=3.44$. 

Although this tendency also holds in the near-extremal case $a/M=0.998$, the maximal values of the flux function increase very fast when increasing the scalar charge. The location of these peaks is almost at the ISCO ($r/M=1$), which in this case approaches the event horizon/curvature singularity.

\setlength{\tabcolsep}{0.5em} 
\renewcommand{\arraystretch}{1.2}
\begin{table}[t]
		\centering
	\begin{tabular}{||c|cc|cc|cc|cc||}
		\hline
	\diagbox{$\Sigma^2/M^2$}{$a/M$} 	& \multicolumn{2}{c|}{0.0}    & \multicolumn{2}{c|}{0.5}    & \multicolumn{2}{c|}{0.9}    & \multicolumn{2}{c||}{0.998}    \\ \hline
		& \multicolumn{1}{c|}{$r/M$} & {$F/F^{Sch}_{max}$} & \multicolumn{1}{c|}{$r/M$} & {$F/F^{Sch}_{max}$}  & \multicolumn{1}{c|}{$r/M$} & {$F/F^{Sch}_{max}$} & \multicolumn{1}{c|}{$r/M$} & {$F/F^{Sch}_{max}$}  \\ \hline
	$-9.0$	& \multicolumn{1}{c|}{9.72} & 0.94 & \multicolumn{1}{c|}{6.88} & 2.71 & \multicolumn{1}{c|}{4.15} & 13.7 & \multicolumn{1}{c|}{3.28} & 35.2 \\ \hline
	$-0.5$	& \multicolumn{1}{c|}{9.56} & 0.99 & \multicolumn{1}{c|}{6.63} & 3.09 & \multicolumn{1}{c|}{3.48} & 23.7 & \multicolumn{1}{c|}{1.84} &  212 \\ \hline
	\hspace{35pt} $0$ \hspace{2pt} \text{(Kerr)}	& \multicolumn{1}{c|}{9.55} & 1.00 & \multicolumn{1}{c|}{6.62} & 3.11 & \multicolumn{1}{c|}{3.44} & 24.7 & \multicolumn{1}{c|}{1.58} & 350 \\ \hline
	0.8	& \multicolumn{1}{c|}{9.54} & 1.00 & \multicolumn{1}{c|}{6.59} & 3.15 & \multicolumn{1}{c|}{3.38} & 26.6 & \multicolumn{1}{c|}{1.27} & 11807 \\ \hline
	1.2	& \multicolumn{1}{c|}{9.53} & 1.01 & \multicolumn{1}{c|}{6.58} & 3.17 & \multicolumn{1}{c|}{3.34} & 27.5 & \multicolumn{1}{c|}{{1.26}} & 214670  \\ \hline
	2.0	& \multicolumn{1}{c|}{9.52} & 1.01 & \multicolumn{1}{c|}{6.56} & 3.22 & \multicolumn{1}{c|}{3.28} & 29.7 & \multicolumn{1}{c|}{{1.25}} & $1.27 \times 10^8$ \\ \hline
	50	& \multicolumn{1}{c|}{8.70} & 1.47 & \multicolumn{1}{c|}{5.42} & 8.88 & \multicolumn{1}{c|}{2.41} & $ 970036$ & \multicolumn{1}{c|}{{1.24}} &  $7.24\times 10^{197}$\\ \hline
	\end{tabular}
		\caption{The location of the energy flux peak of the disk in the KL geometry shown for different values of the specific angular momentum $a/M$ and the scalar charge $\Sigma/M$.}
\label{table:1}
\end{table}

\setlength{\tabcolsep}{0.5em} 
\renewcommand{\arraystretch}{1.2}
\begin{table}[t]
		\centering
	\begin{tabular}{||c|c|c|c|c||}
		\hline
	\diagbox{$\Sigma^2/M^2$}{$a/M$} 	& \multicolumn{1}{c|}{0.0}    & \multicolumn{1}{c|}{0.5}    & \multicolumn{1}{c|}{0.9}    & \multicolumn{1}{c||}{0.998}    \\ \hline
		& \multicolumn{1}{c|}{$\delta F, [\%]$}  & {$\delta F, [\%]$}  &  {$\delta F, [\%]$}  & {$\delta F, [\%]$}  \\ \hline
	$-9.0$	& \multicolumn{1}{c|}{6.0 $\downarrow$} & \multicolumn{1}{c|}{13 $\downarrow$} & \multicolumn{1}{c|}{45 $\downarrow$} & \multicolumn{1}{c||}{90 $\downarrow$}  \\ \hline
	$-0.5$	& \multicolumn{1}{c|}{1.0 $\downarrow$} & \multicolumn{1}{c|}{0.6 $\downarrow$} & \multicolumn{1}{c|}{4.0 $\downarrow$} & \multicolumn{1}{c||}{39 $\downarrow$} \\ \hline
	 \hspace{35pt} $0$ \hspace{2pt}\text{(Kerr)}& \multicolumn{1}{c|}{0} & \multicolumn{1}{c|}{0} & \multicolumn{1}{c|}{0} & \multicolumn{1}{c||}{0} \\ \hline
	0.8	& \multicolumn{1}{c|}{0} & \multicolumn{1}{c|}{1.3 $\uparrow$} & \multicolumn{1}{c|}{7.7$\uparrow$} & \multicolumn{1}{c||}{$3.3\times10^3$ $\uparrow$} \\ \hline
	1.2	& \multicolumn{1}{c|}{1.0 $\uparrow$} & \multicolumn{1}{c|}{1.9 $\uparrow$} & \multicolumn{1}{c|}{11 $\uparrow$} & \multicolumn{1}{c||}{{$6.1\times10^4$ $\uparrow$}} \\ \hline
	2.0	& \multicolumn{1}{c|}{1.0 $\uparrow$} & \multicolumn{1}{c|}{3.5 $\uparrow$} & \multicolumn{1}{c|}{20 $\uparrow$} & \multicolumn{1}{c||}{{$3.6\times10^7$ $\uparrow$}} \\ \hline
	50	& \multicolumn{1}{c|}{47 $\uparrow$} & \multicolumn{1}{c|}{186 $\uparrow$} & \multicolumn{1}{c|}{$3.9\times10^6$ $\uparrow$} & \multicolumn{1}{c||}{{$2.1\times10^{197}$ $\uparrow$}} \\ \hline
	\end{tabular}
		\caption{The relative difference $\delta F=|F_{(KL)}^{\, max}-F_{(Kerr)}^{\, max}|/F_{(Kerr)}^{\, max}$ of the radial flux function peaks between Kerr and KL geometry. The arrow $\uparrow$ indicates $F^{\,max}_{(KL)}>F^{\,max}_{(Kerr)}$, while $\downarrow$ indicates  $F^{\,max}_{(KL)}<F^{\,max}_{(Kerr)}$. The values of the $F_{(Kerr)}^{\,max}$ are given in Table \ref{table:1}.}
\label{table:fluxPercentageKLI}
\end{table}

Looking at the apparent images of the disks in the static KL case, one notes that the observable energy flux $F_O$ is more widely  distributed over the surface of the disk around its corresponding peaks. Unlike the static case, in the rotating case  the most significant part of $F_O$ tends to be distributed over smaller areas around the peaks, thus making the apparent image of the rest of the disk to look darker. 

For the rotating compact objects the maximal values of the observable flux $F_O$ are significantly larger than in the static case. However, most of the radiation is concentrated around the vicinity  of the peak and there exists a larger discrepancy between the value of the flux maximum and the remaining part of the flux distribution. Since the observable flux in our images is normalized by its maximum value, it causes the disks to appear dimmer than in the static case. We should note however, that this is a relative effect characterizing the flux distribution in each particular case. The apparent  brightness of the disk images should not be compared for different values of the solution parameters due to different normalization of the presented observable flux.  In order to be able to draw conclusions about  the disk luminosity when we vary  the spin or the oblateness of the compact objects, we have tabulated the absolute value of the maximum observable flux in Tables \ref{table:3ab} and  \ref{table:3a}.

The study of the observable images of thin accretion disks, as seen by a distant observer, is presented in the following subsections.

\subsubsection{Static case \texorpdfstring{$a/M=0$}{aM0}}
\label{sec:kl_static}

Here we consider the apparent images of a thin accretion disk for $-9\leq\Sigma^2/M^2\leq 3.9$ in the static KL geometry (Figs. \ref{fig:CD_KL_I_a=0_90deg}-\ref{fig:CD_KL_I_Fantom_a=0_90deg}). The asymptotic observer is located at $r/M=10\,000$ and the inclination angle is $\theta_{O}=\pi/2$. In all cases the disk extends from ISCO at $r/M=6$ to $r/M=30$. This interval places the matter of the disk in the strong part of the gravitational field, where the most interesting relativistic effects can be captured.

Disk images in the static case, shown in Fig. \ref{fig:KL_I_IsoR_Disk}, have the same qualitative behavior as the obtained images of the shadows from Fig. 10. For a small scalar charge, there is a visible difference between KL and Kerr solution only in the central part of the disk image, 
where we observe multiple bright rings. Looking at Fig. \ref{fig:CD_KL_I_a=0_90deg} one notes that for small charges in the interval $0\leq\Sigma^2/M^2\leq 1$ the apparent image of the disk in the KL case mimics the image of the Schwarzschild accretion disk. The only difference is that the disk becomes more oblate with the subsequent increasing of the charge. For $\Sigma^2/M^2>1$, more higher-order images appear in the central part of the overall image, and their structure tends to become more complicated. The additional increase of the scalar charge $\Sigma$ has no further significant effect on the direct (zeroth order) image of the disk and its energy flux distribution. Still, it leads to a noticeable change in the shape of the higher-order images, as shown in Fig. \ref{fig:KL_I_IsoR_Disk}. In the presence of a real scalar field, the relativistic images of the accretion disk become more and more oblate due to the deformation of the photon region surface toward the singularity. The formed images appear symmetrically above and below with respect to the plane of the accretion disk, and for relatively small values of the scalar charge, they are simply connected. At a scalar charge near $\Sigma^2/M^2 = 1.6$, the images of the third and higher orders become non-simply connected. For a scalar charge $\Sigma^2/M^2 > 2$, the direct images begin to split, and at higher scalar charge values above $\Sigma^2/M^2 > 3.8$, separate sets of symmetrically distributed multi-connected images begin to form. 

The presence of a phantom scalar field leads to distortion of the photon region surface in the direction opposite to the singularity, leading to the formation of vertically elongated images of all orders (Fig. \ref{fig:CD_KL_I_Fantom_a=0_90deg}). The distribution of the observed energy flux at the disk surface does not change significantly and follows the structural deformations of the images. Unlike the cases with a real scalar field, in the presence of the phantom field, the increase of the scalar charge does not lead to the formation of relativistic images with non-trivial morphology. The indirect image, together with the higher-order images, remain stacked around the shadow's visible boundary, similar to the Schwarzschild case, with the difference that the images are highly elongated.
    
In all cases, the flux distribution function is normalized by the maximal value of the observable flux $F_O^{\,max}$, which is different for different values of the scalar charge $\Sigma^2/M^2$ as shown in Table \ref{table:3ab}. This means that every disk image shown in Fig. \ref{fig:CD_KL_I_a=0_90deg} is normalized by the maximum value of its own observable flux. In this way,  due to the fact that some of disks have several times larger observable flux from other disks, we can avoid some images to appear darker and risk loosing information about their flux distribution. The observable flux of the disk does not change significantly when varying the scalar charge in the range $-9\leq \Sigma^2/M^2\leq 2$. 

In Tables \ref{table:3ab} and \ref{table:3abc} we have compared the observable flux of the Kerr accretion disk with respect to those in the KL solution for $a/M=0$ and $a/M=0.9$. In the static case $a/M=0$, the disk in the KL solution is almost indistinguishable (at most 6.6\% difference) from the Kerr disk. In the rotating case $a/M=0.9$, the accretion disk for the KL solution gradually increases its maximum of the observable flux from that of the Kerr disk for positive scalar charge and decreases its maximum peak for negative $\Sigma^2/M^2$. Nevertheless, we observe that the phantom scalar field tends to suppress the maximum of the observable flux, while the presence of a real scalar field amplifies it. These effects become more noticeable for larger values of the specific angular momentum.


\setlength{\tabcolsep}{0.5em} 
\renewcommand{\arraystretch}{1.2}

\begin{table}[t!]
    \centering
      \begin{tabular}{||c|c|c|c|c|c|c||}
       \hline
         \multicolumn{7}{||c||}{$a/M=0$}          \\ \hline
          $\Sigma^2/M^2$     & -9 & 0 & 0.9 & 1.35 & 1.6 & 2.0  \\  \hline
          $F_{O}^{\,max}\dot{M}_{0} \times 10^{-5}$, $\theta_{O}=\pi/2$ & 2.223 & 2.232 & 2.230 & 2.230 & 2.229 & 2.228  \\   \hline
          $F_{O}^{\,max}\dot{M}_{0} \times 10^{-5}$, $\theta_{O}=4\pi/9$ & 3.585 & 3.841 & 3.869 &3.882 & 3.889 & 3.902  \\   \hline
          \hline
         \multicolumn{7}{||c||}{$a/M=0.9$}          \\ \hline
          $\Sigma^2/M^2$     & -9 & 0 & 0.2 & 0.4 & 0.8 & 2.0  \\  \hline
          $F_{O}^{\,max}\dot{M}_{0} \times 10^{-5}$, $\theta_{O}=\pi/2$ & 60.90 & 96.96 & 97.37 & 97.66 & 97.84 & 93.33  \\   \hline
          $F_{O}^{\,max}\dot{M}_{0} \times 10^{-5}$, $\theta_{O}=4\pi/9$ & 70.13 & 137.4 & 140.2 & 142.9 & 148.8 & 169.2  \\   \hline
       \end{tabular}
      \caption{Numerical estimation of the maximal value of the observable energy flux shown for different values of the scalar charge $\Sigma^2/M^2$ and specific angular momentum $a/M$ for the KL geometry. The grid step size of the ray tracing solver for $\alpha$ and $\beta$ is $10^{-2} M$.}
    \label{table:3ab}
\end{table}
\setlength{\tabcolsep}{0.5em} 
\renewcommand{\arraystretch}{1.2}

\begin{table}[t!]
    \centering
      \begin{tabular}{||c|c|c|c|c|c|c||}
          \hline
         \multicolumn{7}{||c||}{Relative flux difference between Kerr and KL solutions for $a/M=0$}          \\ \hline
          $\Sigma^2/M^2$     & -9 & 0 & 0.9 & 1.35 & 1.6 & 2.0  \\  \hline
          $\delta F_{O}, [\%]$, $\theta_{O}=\pi/2$ & 0.40 $\downarrow$ & 0  & 0.32 $\uparrow$ & 0.32 $\uparrow$ & 0.27 $\uparrow$ & 0.23 $\uparrow$  \\   \hline
          $\delta F_{O}, [\%]$, $\theta_{O}=4\pi/9$ & 6.66 $\downarrow$ & 0 & 0.76 $\uparrow$ & 1.10 $\uparrow$ & 1.30 $\uparrow$ & 1.60 $\uparrow$ \\   \hline
          \hline
         \multicolumn{7}{||c||}{Relative flux difference between Kerr and KL solutions for $a/M=0.9$}          \\ \hline
          $\Sigma^2/M^2$     & -9 & 0 & 0.2 & 0.4 & 0.8 & 2.0  \\  \hline
          $\delta F_{O}, [\%]$, $\theta_{O}=\pi/2$ & 37.2 $\downarrow$ & 0 & 0.42 $\uparrow$ & 0.72 $\uparrow$ & 0.91 $\uparrow$ & 3.74 $\uparrow$ \\   \hline
          $\delta F_{O}, [\%]$, $\theta_{O}=4\pi/9$ & 49.0 $\downarrow$ & 0 & 2.0 $\uparrow$ & 4.00 $\uparrow$ & 8.30 $\uparrow$ & 23.1 $\uparrow$ \\   \hline
       \end{tabular}
      \caption{Relative flux difference $\delta F_{O}=|F_{O \, (Kerr)}^{\,max}-F_{O \, (KL)}^{\,max}|/F_{O \, (Kerr)}^{\, max}$ between Kerr and KL solutions. Here $F_{O \, (Kerr)}^{\,max}$ for $\theta_{O}=\pi/2$ and $\theta_{O}=4\pi/9$ is given in Table \ref{table:3ab} for $\Sigma^2/M^2=0$. The $\uparrow$ indicates $F^{\,max}_{O \, (KL)}>F^{\,max}_{O \, (Kerr)}$, while $\downarrow$ indicates  $F^{\,max}_{O \, (KL)}<F^{\,max}_{O\, (Kerr)}$.}
    \label{table:3abc}
\end{table}

The apparent images of the thin accretion disk in the KL geometry for small enough scalar charge $\Sigma^2/M^2 <30$, as seen by an asymptotic observer with an inclination angle $\theta_{O}=4\pi/9$, resemble the structure  of the observable disk, presented  in the previous subsection. This situation is depicted in (Fig. \ref{fig:KL_I_a=0_80deg}a -\ref{fig:KL_I_a=0_80deg}d). However, for large charges $\Sigma^2/M^2>30$ (Fig. \ref{fig:KL_I_a=0_80deg}e-\ref{fig:KL_I_a=0_80deg}f.) the disk image undergo a drastic change even in the zeroth order. The direct images become extremely oblate and disjoint in the upper part, while the higher-order images are almost or entirely missing from the picture.

In the presence of a phantom field, larger negative charge $\Sigma^2/M^2<0$ leads to more prolate look of the apparent disk image (Fig. \ref{fig:KL_I_Stat_Phantom_80deg_a}).

\FloatBarrier

\subsubsection{Rotating case}

In the rotating case $a/M=0.9$, for an asymptotic observer at an inclination angle $\theta_{O}=\pi/2$, the apparent images of the disk become dislocated to the left in the $x$-direction due to the frame dragging generated by the rotation of the compact object. The oblateness of the disk image does not change significantly for different values of the scalar charge. The observable flux $F_O$ peaks closer to the ISCO of the disk and at higher values with respect to the same quantity in the static case (Fig. \ref{fig:CD_KL_I_a=0.9_90deg}).

In the near-extremal case $a/M=0.998$ (Fig. \ref{fig:CD_KL_I_а=0.9_90}) the disk image becomes significantly darker from all other cases. Here, the peak of the observable flux $F_O$ is concentrated very close to the ISCO. The most intensive observable radiation is distributed in a very small area around the peak, thus only a small portion of the flux can be depicted in a high contrast. This is true for all values of the scalar charge.

Finally, the phantom scalar field continuous to stretch the disk image in the $y$-direction, when the scalar charge tends to larger negative values, thus making the image looks more prolate (Fig. \ref{fig:CD_KL_I_Phantom_a=0.9_90deg}), as in the previous cases.

The situation for $\theta_O = 4\pi/9$ is similar to the one presented in the static case. The difference now is that due to the rotation of the compact object the entire image of the disk is shifted to the left in the $x$-axis (Fig. \ref{fig:CD_KL_I_a=0.9_80deg}). Furthermore, the disk image appears dimmer, because the maximal values of the observable flux, although much higher (see Table \ref{table:3ab}), are distributed in a smaller area as compared to the disks in the static case.

In the near-extremal case $a/M=0.998$ (Fig. \ref{fig:KL_I_Extremal_a=0_80deg}) the peak of the observable flux $F_O$ is concentrated very close to the ISCO and the brightest part appears only in a small area around the peak. For this reason the rest of the disk looks predominately darker. 

As in the previous cases the phantom scalar field influences the image of disk to assume more prolate form in the $y$-direction for larger negative values of the scalar charge (Fig. \ref{fig:KL_I_Rot_Phantom_80deg}).

\FloatBarrier

\subsection{TSL geometry}
In this section we present the apparent images of the thin accretion disk in the TSL geometry. The situation is similar to the previous considerations, where we show the disk at inclination angles $\theta_O=\pi/2$ (Fig. \ref{fig:ColorDisk_KL_II_90deg}) and $\theta_O=4\pi/9$ (Fig. \ref{fig:ColorDisk_KL_II_80deg}). The values of the specific angular momentum are given by $a/M=\{0, 0.5, 0.998\}$. The images of the disks are compared to the corresponding ones in the Schwarzschild and Kerr cases. Following Figs. \ref{fig:ColorDisk_KL_II_90deg} and \ref{fig:ColorDisk_KL_II_80deg}, the Kerr and TSL accretion disk images are very similar for small angular momenta.  For fast rotation some visible differences arise, for example in the near extreme mode for $a=0.998$. The main difference is the appearance of additional disk images inside the usual shadow of the Kerr solution. This effect can be explained by the fact that the lower edge of the accretion disk is located near the ring singularity inside the photon region. Note that this additional structure did not appear in the previous section \ref{sec:shadows} because the shadows were created by an external light source outside of the photon region.

\begin{figure}[t!]
    \centering
    \begin{subfigure}[t]{0.55\textwidth}
        \includegraphics[width=\textwidth]{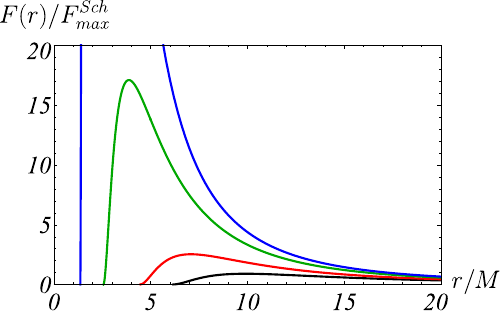}
    \end{subfigure}
    \caption{\label{fig:KL_II_Fluxs}\small Dependence of the radiation energy flux over the disk on the radial distance in the TSL geometry: static case $a/M=0$ (black curve); rotating case $a/M=0.5$ (red curve); rotating case $a/M=0.9$ (green curve); near-extremal case $a/M=0.998$ (blue curve). In all cases, the flux function $F(r)$ is normalized to the maximal flux $F^{Sch}_{max}=1.37\times 10^{-5} M \dot M$ of the Schwarzschild black hole.} 
\end{figure}

A more significant difference is observed from the analysis of the radial flux. Indeed, the peaks of the radial flux function for the Kerr black hole are larger than the corresponding values of the flux function in the TSL case for all values of their angular momentum (Table \ref{table:2}). The deviation increases when the spin parameter grows reaching up to 36\% for fast rotation. The radial distribution of the flux function $F(r)$ is shown in Fig. \ref{fig:KL_II_Fluxs} where the peaks for the TSL solution occur at similar radial distances as for teh Kerr black hole.

Both solutions continue to differ notably when comparing the observable fluxes of their corresponding disks. From the relative observable flux difference between Kerr and TSL solutions, shown in Table \ref{table:3a}, one can make several observations. Firstly, one notes that the observable flux of TSL is always smaller than the corresponding flux in the Kerr case for all values of the specific angular momentum. Secondly, the difference is about 12 \% for the static case (for $\theta_O=\pi/2$ and $4\pi/9$) and then gradually increases up to 31 \% for $\theta_O=\pi/2$ and 80 \% for $\theta_O=4\pi/9$ when approaching the near-extremal situation. This suggests that for fast rotation astrophysical observations can strongly discern between Kerr and TSL solutions both by means of the substantial deviation in the disk luminosity and by the appearance of additional optical structure at the central part of the image.

\setlength{\tabcolsep}{0.5em} 
\renewcommand{\arraystretch}{1.2}
\begin{table}[t]
		\centering
	\begin{tabular}{||c|cc|cc|cc|cc||}
		\hline
	\, $a/M$ \, 	& \multicolumn{2}{c|}{0.0}    & \multicolumn{2}{c|}{0.5}    & \multicolumn{2}{c|}{0.9}    & \multicolumn{2}{c||}{0.998}    \\ \hline 
		& \multicolumn{1}{c|}{$r/M$} & {$F/F^{Sch}_{max}$} & \multicolumn{1}{c|}{$r/M$} & {$F/F^{Sch}_{max}$}  & \multicolumn{1}{c|}{$r/M$} & {$F/F^{Sch}_{max}$} & \multicolumn{1}{c|}{$r/M$} & {$F/F^{Sch}_{max}$}  \\ \hline
	\text{TSL}:	& \multicolumn{1}{c|}{9.40} & 0.89 & \multicolumn{1}{c|}{6.58} & 2.52 & \multicolumn{1}{c|}{3.38} & 17.1 & \multicolumn{1}{c|}{1.81} & $ 224 $ \\ \hline	
		\text{Kerr}:	& \multicolumn{1}{c|}{9.55} & 1 & \multicolumn{1}{c|}{6.62} & 3.11 & \multicolumn{1}{c|}{3.44} & 24.7 & \multicolumn{1}{c|}{1.58} & $ 350 $ \\ \hline	
	$\delta F, [\%]$	& \multicolumn{1}{c|}{--} & 11 $\downarrow$ & \multicolumn{1}{c|}{--} & 19 $\downarrow$ & \multicolumn{1}{c|}{--} & 31 $\downarrow$ & \multicolumn{1}{c|}{--} & 36 $\downarrow$ \\ \hline
	\end{tabular}
		\caption{The location of the disk's energy flux peak in the TSL geometry, shown for different values of the specific angular momentum $a/M$; $\delta F=|F_{(TSL)}^{\, max}-F_{(Kerr)}^{\, max}|/F_{(Kerr)}^{\, max}$. The $\downarrow$ indicates  $F^{\, max}_{(TSL)}<F^{\, max}_{(Kerr)}$.}
\label{table:2}
\end{table}
\setlength{\tabcolsep}{0.5em} 
\renewcommand{\arraystretch}{1.2}

\begin{table}[t!]
    \centering
      \begin{tabular}{||c|c|c|c|c|c|c|c||}
       \hline
         \multicolumn{8}{||c||}{Kerr}      \\  \hline
          $a/M$     & 0 & 0.5 & 0.7 & 0.8 & 0.9 & 0.95 & 0.998    \\  \hline
          $F_{O}^{\,max}\dot{M}_{0} \times 10^{-5}$, $\theta_{O}=\pi/2$ & 2.232 & 7.991 & 18.91 & 35.77 & 96.96 & 235.2 & 3302
          \\  \hline
          $F_{O}^{\,max}\dot{M}_{0} \times 10^{-5}$, $\theta_{O}=4\pi/9$ & 3.840 & 13.77 & 31.63 & 57.19 & 137.4 & 280.4 & 1445
          \\  \hline
          \hline
         \multicolumn{8}{||c||}{TSL}      \\ \hline
         $a/M$     & 0 & 0.5 & 0.7 & 0.8 & 0.9 & 0.95 & 0.998     \\ \hline
         $F_{O}^{\,max}\dot{M}_{0} \times 10^{-5}$, $\theta_{O}=\pi/2$ & 1.976 & 6.283 & 13.66 & 24.22 & 59.38 & 132.2 &  1838  \\   \hline
         $F_{O}^{\,max}\dot{M}_{0} \times 10^{-5}$, $\theta_{O}=4\pi/9$ & 3.375 & 10.81 & 22.81 & 39.96 & 90.85 & 180.9 & 1106  \\   \hline
          \hline
         \multicolumn{8}{||c||}{Relative flux difference between Kerr and TSL solutions}      \\ \hline
         $a/M$     & 0 & 0.5 & 0.7 & 0.8 & 0.9 & 0.95 & 0.998     \\ \hline
         $\delta F_{O}, [\%]$, $\theta_{O}=\pi/2$ & 11.5 $\downarrow$ & 21.4 $\downarrow$ & 27.8 $\downarrow$ & 32.3 $\downarrow$ & 38.7 $\downarrow$ & 43.8 $\downarrow$ & 80.0 $\downarrow$  \\   \hline
         $\delta F_{O}, [\%]$, $\theta_{O}=4\pi/9$ & 12.1 $\downarrow$ & 21.5 $\downarrow$ & 27.9 $\downarrow$ & 30.1 $\downarrow$ & 33.9 $\downarrow$ & 35.5 $\downarrow$ & 31.0 $\downarrow$  \\   \hline
       \end{tabular}
      \caption{Numerical estimation of the maximal value of the observable energy flux shown for different values of the specific angular momentum $a/M$. Here $\delta F_{O}=|F_{O \, (Kerr)}^{\,max}-F_{O \, (TSL)}^{\,max}|/F_{O \, (Kerr)}^{\,max}$. The grid step size of the ray tracing solver for $\alpha$ and $\beta$ is $10^{-2} M$. The $\uparrow$ indicates $F^{max}_{O \, (TSL)}>F^{max}_{O \, (Kerr)}$, while $\downarrow$ indicates  $F^{max}_{O \, (TSL)}<F^{max}_{O \, (Kerr)}$.}
    \label{table:3a}
\end{table}

\section{Conclusions}
\label{sec:conslusions}
The aim of this article was to give detailed optical pictures for shadows and accretion disks around two black hole mimickers described by  solutions of Einstein gravity minimally coupled to scalar field \cite{Bogush:2020lkp}.   Both of them can be considered as rotating generalizations of FJNW static naked singularities, which, in contrast to the previously known ones, are legitimate solutions that satisfy the full set of equations of motion of the theory.
 
For both considered solutions, the geodesic equations are non-integrable, but for the TSL solution, numerical integration shows only small deviations from the results of the integrable Kerr case. The corresponding fundamental photon  surfaces approach spherical ones, and the optical properties of space-time resemble those of the Kerr solution. Thus, in this case, we get ultra-compact objects that can be considered as good simulators of ordinary vacuum black holes in general relativity. The edge of the shadow is even closer to the circle than in the Kerr metric for all values of the rotation parameter. In fact, its deviation from the circle does not exceed $10\,\%$, which agrees with the observations of M87 \cite{EventHorizonTelescope:2019dse} and differs by less than $1\,\%$ from the Kerr case. Thus, at the level of accuracy of the EHT observations, TSL solutions are practically indistinguishable from Kerr. Note that the maximum difference is observed in the near-extreme regime with $a\approx  0.9M$ and can, in principle, be observed when greater experimental accuracy will be achieved. Furthermore, the shadow of the TSL solution is generally larger, and the vertical size of the shadow decreases with increasing $a$, in contrast to the Kerr family. 

Considering the optical appearance of the accretion disks around the TSL compact objects, we reach similar conclusions. We have investigated the  geometrically thin and optically thick disk in their vicinity, and showed that its image, as seen by a distant observer, 
does not exhibit observationally relevant qualitative distinctions with  accretion in the  Kerr case.
The largest visible difference in disk images occurs in the near-extreme regime, when the lower edge of the accretion disk is near the ring singularity and we can observe the appearance of new disk images inside the classical Kerr shadow. Moreover, significant deviations  can be detected by measuring the luminosity of the accretion disk. For slow rotation the two types of compact object still look similar with deviations in their observable radiation within $20\, \%$. 
However, for fast rotation, which is considered to be an astrophysically relevant scenario, we can measure about twice larger the observed flux peak values for the Kerr black hole than for its mimicker. 

The KL solution with the scalar-induced oblateness parameter cannot be ruled out by experimental data either, if the value of the scalar charge is less than the transitional value of the existence of the photon region. However, as the charge increases, the difference from the Kerr case increases rapidly. For phantom scalar fields, the deviations are still small and only show up in elongated shadow images as the strain increases with the absolute value of the scalar charge. However, for real scalar fields, we can observe significant effects such as the formation of multiple shadow images.

The optical appearance of thin accretion disks around KL solution  reproduces the qualitative features of shadow images. Disks for the phantom scalar field are more elongated and the effect becomes more pronounced as we increase the absolute value of the scalar charge. The real scalar field may lead to the formation of a complex optical pattern in the central part of the disk image, which resembles multiply connected shadow images. These qualitative effects occur in a similar way with two completely different light source settings. Thus, we can expect them to appear in other physical scenarios as well, e.g. for more diverse models of accretion.
\begin{acknowledgments}
The work was supported by the Russian Foundation for Basic Research on the project 20-52-18012 Bulg-a, and the Scientific and Educational School of Moscow State University “Fundamental and Applied Space Research”. I.B. is also grateful to the Foundation for the Advancement of Theoretical Physics and Mathematics ``BASIS'' for support. The partial support by the Bulgarian National Science Fund Grants KP-06-RUSSIA/13 and KP-06-H38/2 is also gratefully acknowledged.
\end{acknowledgments}

\newpage
\appendix
\section{Shadows, relativistic images and accretion disks}
\FloatBarrier

\begin{figure}[]
\centering
 \subfloat[][$\Sigma^2/M^2=0$]{
  \includegraphics[width=0.3\textwidth]{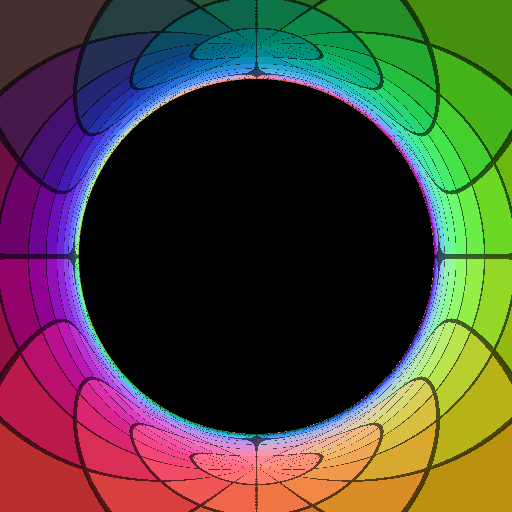} \label{as1}
 }
  \subfloat[][$\Sigma^2/M^2=0.9$]{
  \includegraphics[width=0.3\textwidth]{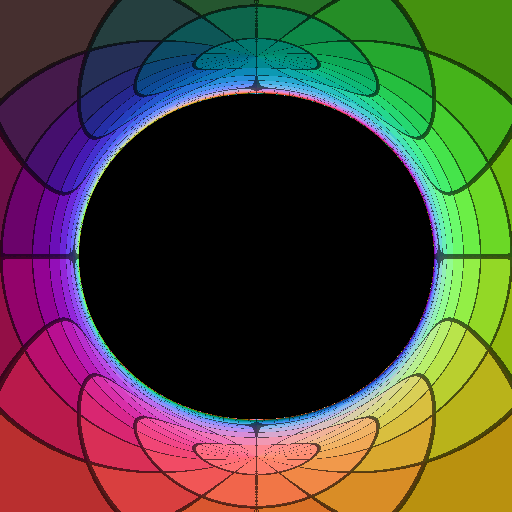} \label{as2}
 }
   \subfloat[][$\Sigma^2/M^2=1.35$]{
  \includegraphics[width=0.3\textwidth]{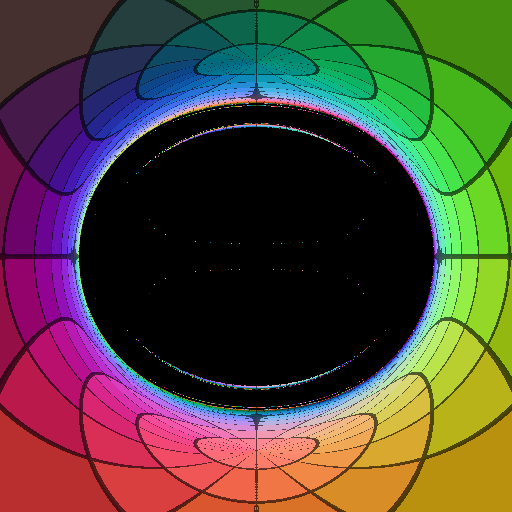} \label{as3}
 }
 
 
  \subfloat[][$\Sigma^2/M^2=1.5$]{
  \includegraphics[width=0.3\textwidth]{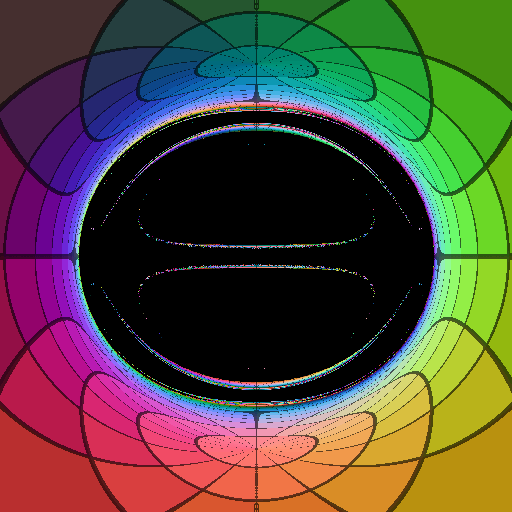} \label{as4}
 }
  \subfloat[][$\Sigma^2/M^2=1.6$]{
  \includegraphics[width=0.3\textwidth]{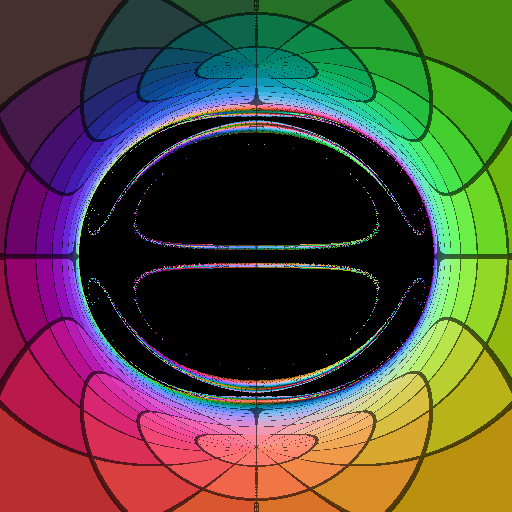} \label{as5}
 }
   \subfloat[][$\Sigma^2/M^2=2$]{
  \includegraphics[width=0.3\textwidth]{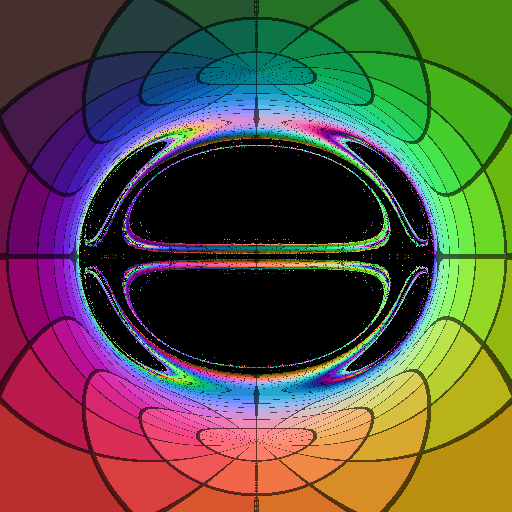} \label{as6}
 }

 
  \subfloat[][$\Sigma^2/M^2=3.5$]{
  \includegraphics[width=0.3\textwidth]{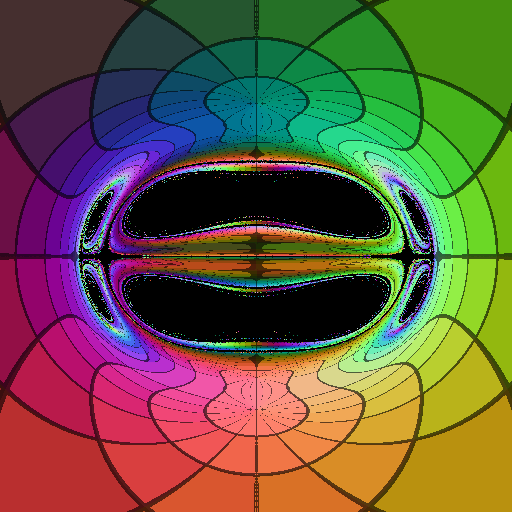} \label{as7}
 }
  \subfloat[][$\Sigma^2/M^2=4$]{
  \includegraphics[width=0.3\textwidth]{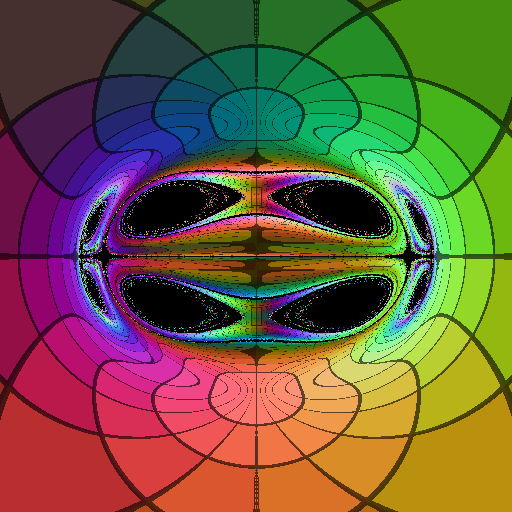} \label{as8}
 }
   \subfloat[][$\Sigma^2/M^2=5$]{
  \includegraphics[width=0.3\textwidth]{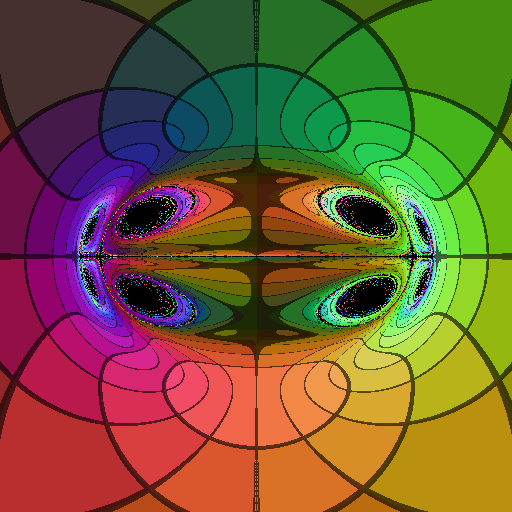} \label{as9}
 }
\caption{Shadows of the KL solution (\ref{eq:solution_I}) in the non-rotating limit for an equatorial observer $r_{O}/M=10\,000$, $\theta_O=\pi/2$ for different values of the scalar charge $\Sigma$.}
\label{RT1}
\end{figure}

\begin{figure}[]
\centering
 \subfloat[][$\Sigma^2/M^2=0$]{
  \includegraphics[width=0.3\textwidth]{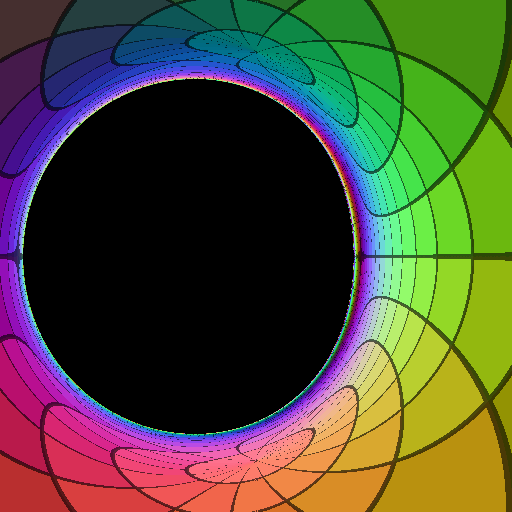} \label{as10}
 }
   \subfloat[][$\Sigma^2/M^2=0.2$]{
  \includegraphics[width=0.3\textwidth]{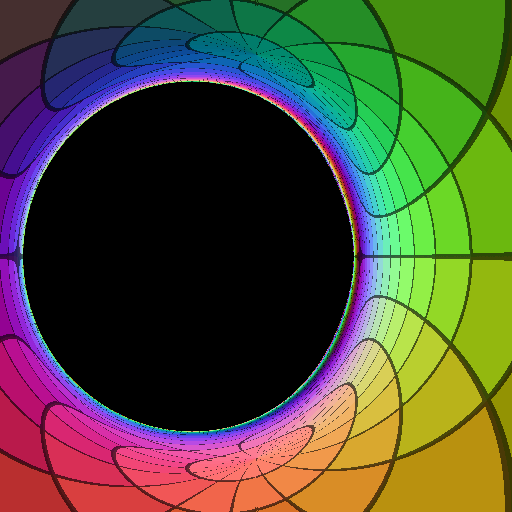} \label{as11}
 }
  \subfloat[][$\Sigma^2/M^2=0.3$]{
  \includegraphics[width=0.3\textwidth]{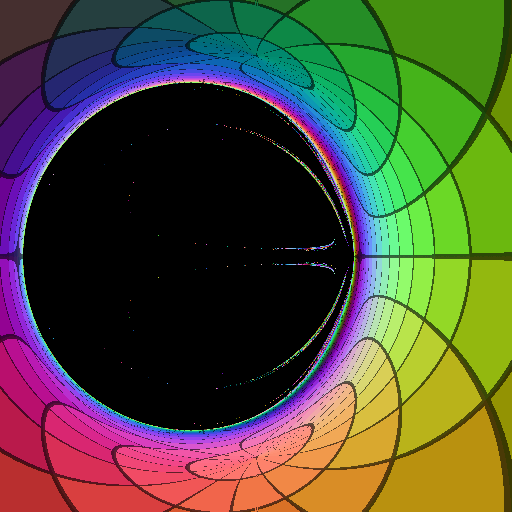} \label{as12}
 }
 
 
\subfloat[][$\Sigma^2/M^2=0.4$]{
  \includegraphics[width=0.3\textwidth]{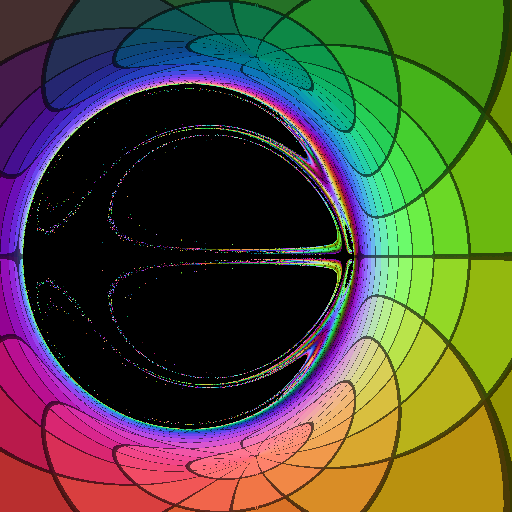} \label{as13}
 } 
 \subfloat[][$\Sigma^2/M^2=0.8$]{
  \includegraphics[width=0.3\textwidth]{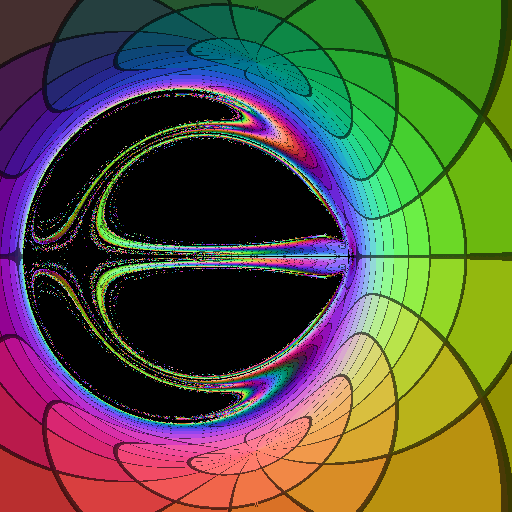} \label{as14}
 } 
  \subfloat[][$\Sigma^2/M^2=2$]{
  \includegraphics[width=0.3\textwidth]{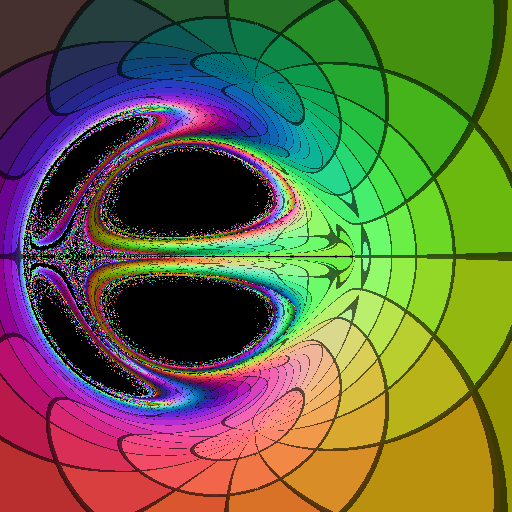} \label{as15}
 } 
\caption{Shadows of the KL solution (\ref{eq:solution_I}) with $a=0.9M$ for an equatorial observer $r_{O}/M=10\,000$, $\theta_O=\pi/2$ for different values of the scalar charge $\Sigma$.}
\label{RT2}
\end{figure}

\begin{figure}[]
\centering
 \subfloat[][$\Sigma^2/M^2=0$, $\theta_O=\pi/2$]{
  \includegraphics[width=0.3\textwidth]{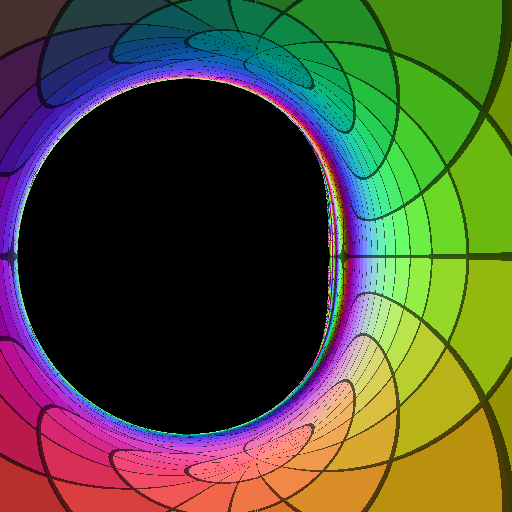} \label{as16}
 }
  \subfloat[][$\Sigma^2/M^2=0.2$, $\theta_O=\pi/2$]{
  \includegraphics[width=0.3\textwidth]{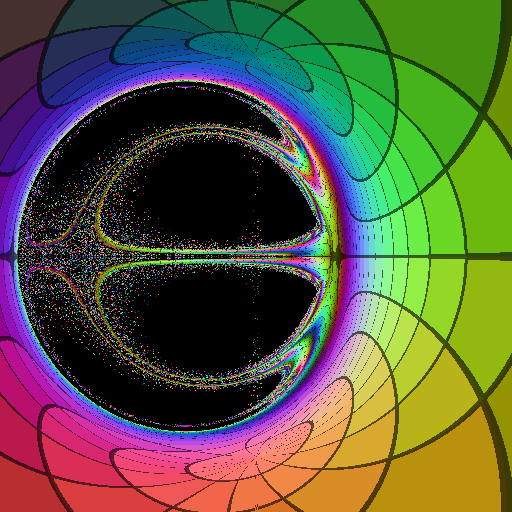} \label{as17}
 }
   \subfloat[][$\Sigma^2/M^2=1$, $\theta_O=\pi/2$]{
  \includegraphics[width=0.3\textwidth]{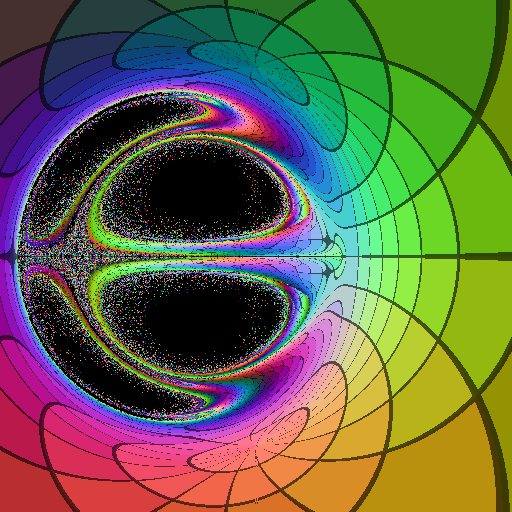} \label{as18}
 }

 \subfloat[][$\Sigma^2/M^2=0$, $\theta_O=4\pi/9$]{
  \includegraphics[width=0.3\textwidth]{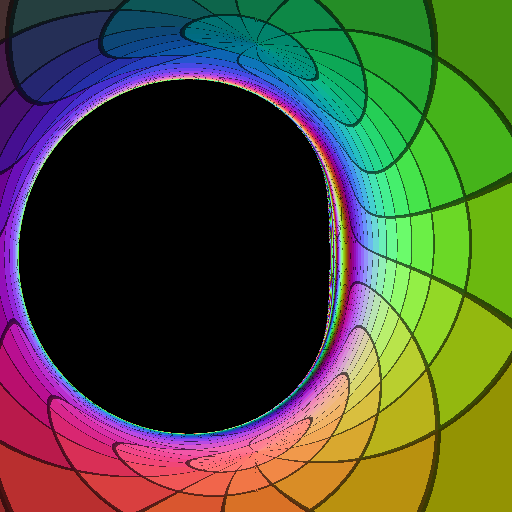} \label{as19i}
 }
  \subfloat[][$\Sigma^2/M^2=0.2$, $\theta_O=4\pi/9$]{
  \includegraphics[width=0.3\textwidth]{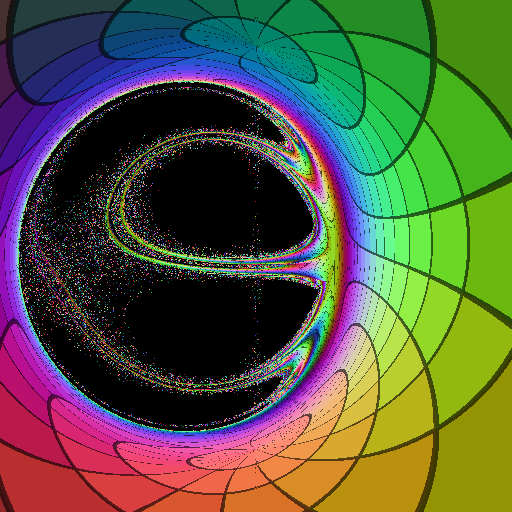} \label{as20i}
 }
   \subfloat[][$\Sigma^2/M^2=1$, $\theta_O=4\pi/9$]{
  \includegraphics[width=0.3\textwidth]{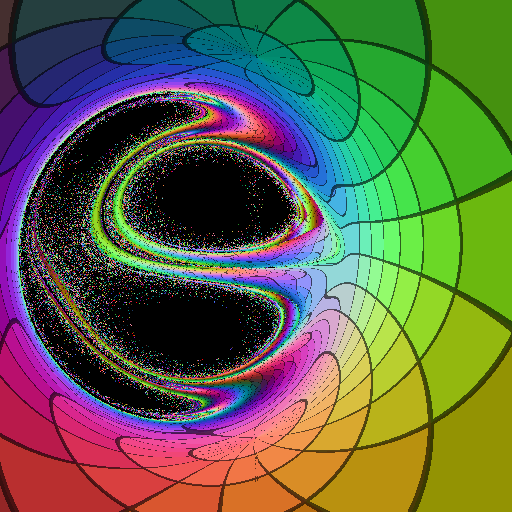} \label{as21i}
 }
\caption{Shadows of the near-extreme KL solution (\ref{eq:solution_I}) with $a=0.998M$ for an observer $r_{O}/M=10\,000$ for different values of the scalar charge $\Sigma$ and the observation angle $\theta_O$.}
\label{RT3}
\end{figure}

\begin{figure}[]
\centering
 \subfloat[][$\Sigma^2/M^2=0$]{
  \includegraphics[width=0.3\textwidth]{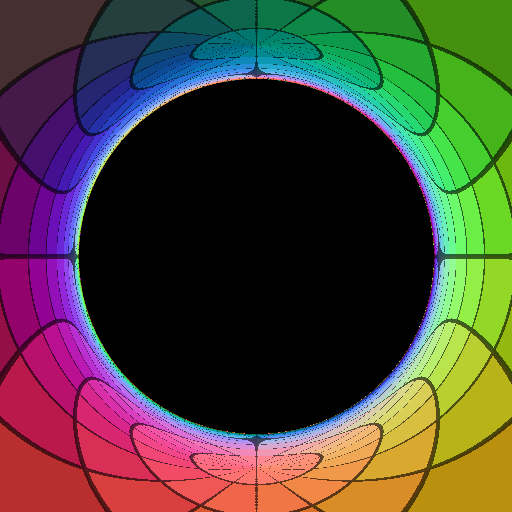} \label{as19}
 }
  \subfloat[][$\Sigma^2/M^2=-1$]{
  \includegraphics[width=0.3\textwidth]{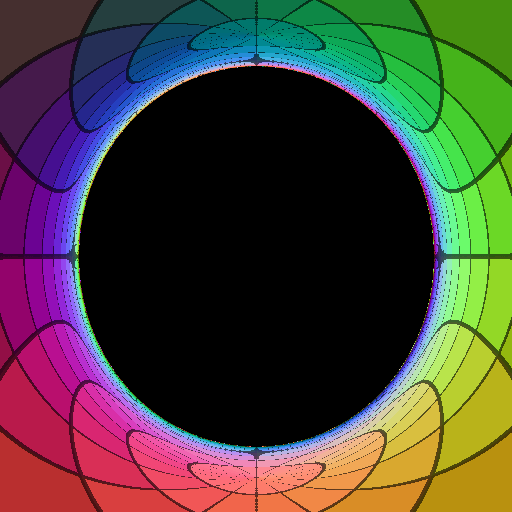} \label{as20}
 }
   \subfloat[][$\Sigma^2/M^2=-1.5$]{
  \includegraphics[width=0.3\textwidth]{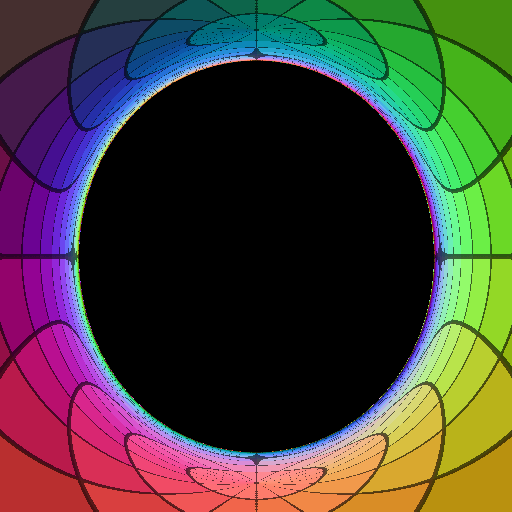} \label{as21}
 }
\caption{Shadows of the static KL solution (\ref{eq:solution_I}) with $a=0$ for an equatorial observer $r_{O}/M=10\,000$, $\theta_O=\pi/2$ for different values of the phantom scalar charge $\Sigma$.}
\label{RT4}
\end{figure}

\begin{figure}[]
\centering
 \subfloat[][$\Sigma^2/M^2=0$]{
  \includegraphics[width=0.3\textwidth]{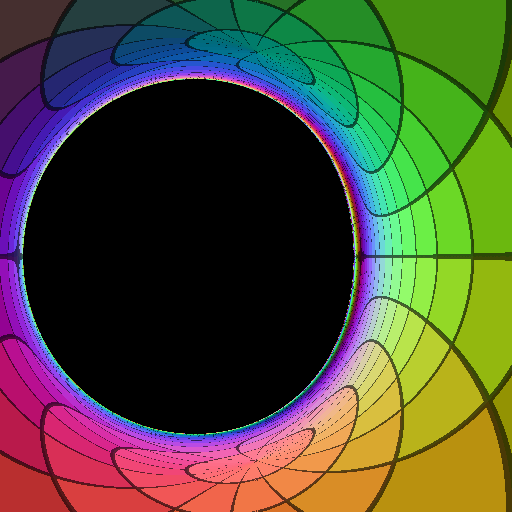} \label{as22}
 }
  \subfloat[][$\Sigma^2/M^2=-0.5$]{
  \includegraphics[width=0.3\textwidth]{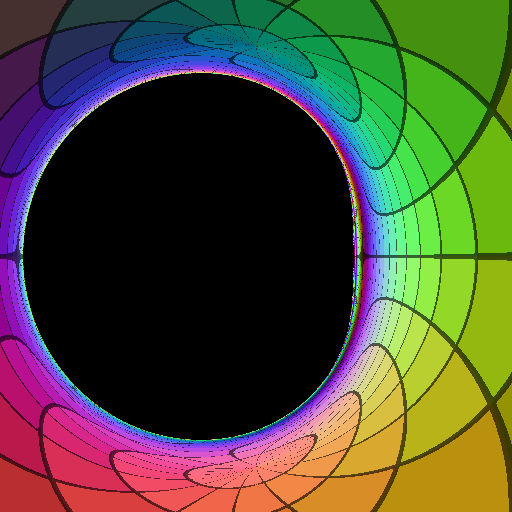} \label{as23}
 }
   \subfloat[][$\Sigma^2/M^2=-2$]{
  \includegraphics[width=0.3\textwidth]{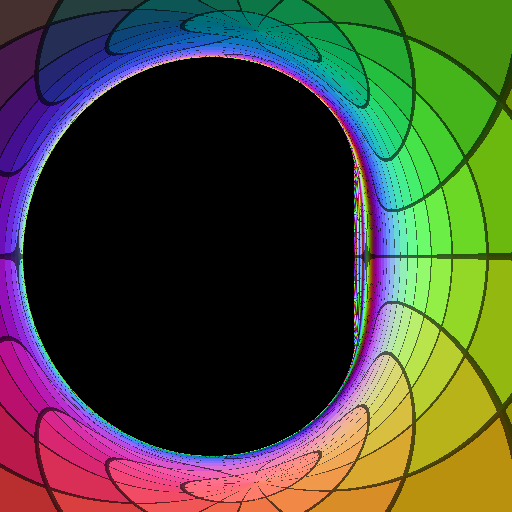} \label{as24}
 }
\caption{Shadows of the extreme KL solution (\ref{eq:solution_I}) with $a=0.9M$ for an equatorial observer $r_O/M=10\,000$, $\theta_O=\pi/2$ for different values of the phantom scalar charge $\Sigma$.}
\label{RT5}
\end{figure}

\begin{figure}[]
\centering
 \subfloat[][Kerr $a=0$]{
  \includegraphics[width=0.3\textwidth]{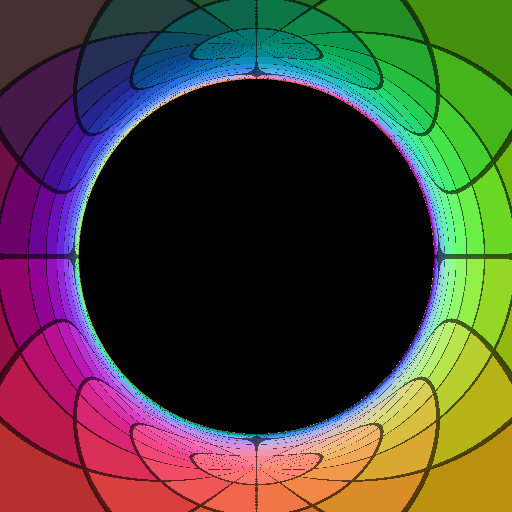} \label{kr0}
 }
  \subfloat[][Kerr $a=0.5$]{
  \includegraphics[width=0.3\textwidth]{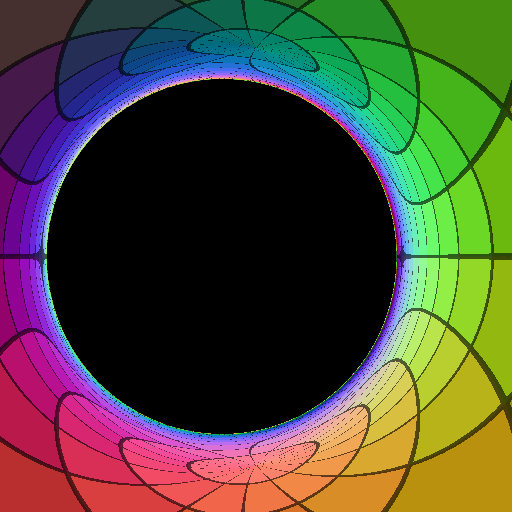} \label{kr5}
 } 
   \subfloat[][Kerr $a=0.998$]{
  \includegraphics[width=0.3\textwidth]{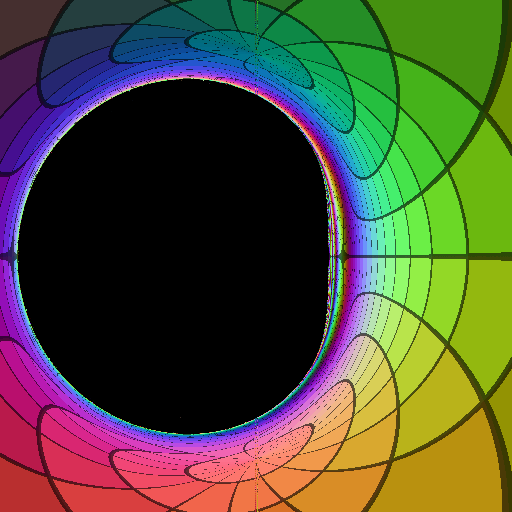} \label{kr10}
 }
 \\
 \subfloat[][TSL $a=0$]{
  \includegraphics[width=0.3\textwidth]{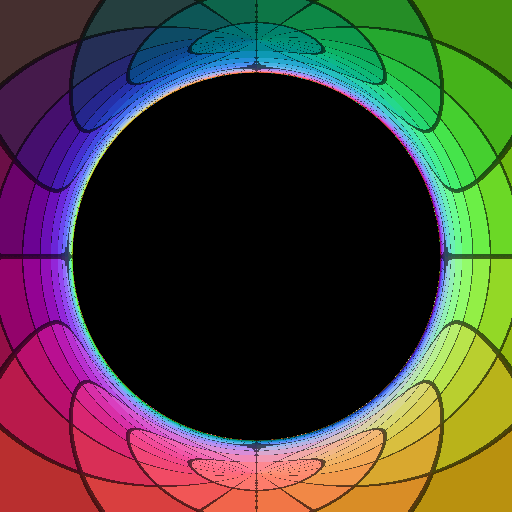} \label{ts0}
  }
 \subfloat[][TSL $a=0.5$]{
  \includegraphics[width=0.3\textwidth]{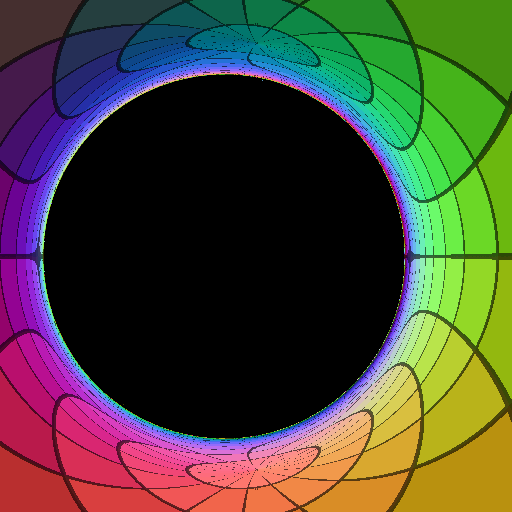} \label{ts5}
 }
  \subfloat[][TSL $a=0.998$]{
  \includegraphics[width=0.3\textwidth]{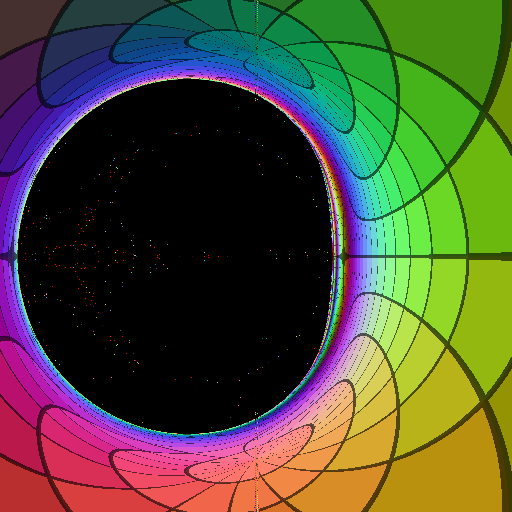} \label{ts10}
 }
\caption{Shadows of (d-f) the TSL solution (\ref{eq:metric}) and (a-c) Kerr metric with $M=1$ and $a=0,\,0.5,\,0.998$ for an equatorial observer $r_O/M=10\,000$, $\theta_O=\pi/2$.}
\label{TS1}
\end{figure}

\begin{figure}[]
\setlength{\tabcolsep}{ 0 pt }{\footnotesize\tt
		\begin{tabular}{ cc}
           \includegraphics[width=1\textwidth]{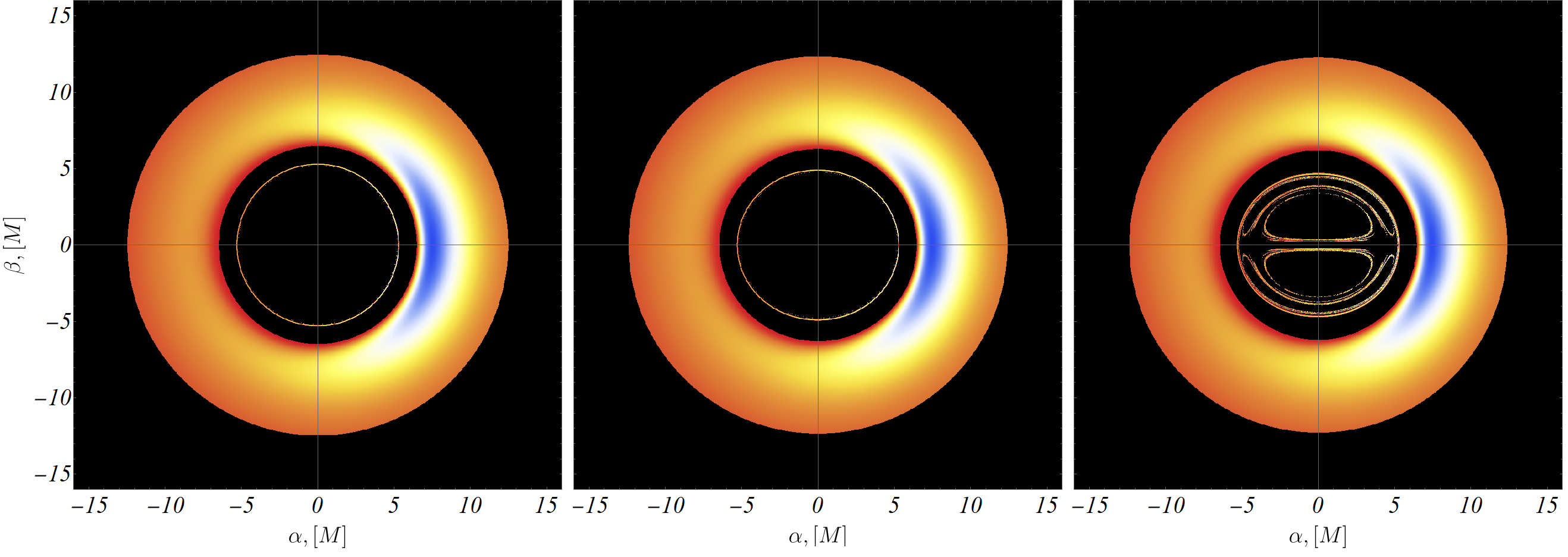} \\[0mm]
              \hspace{-0.5cm}  $(\mathrm{a})  \;\,  \Sigma^2/M^2=0 \;\, (\mathrm{Schwarzschild})$
              \hspace{1.2cm}  $(\mathrm{b})  \;\,  \Sigma^2/M^2=0.9$ 
              \hspace{2.3cm}  $(\mathrm{c})  \;\,  \Sigma^2/M^2=1.35$ \\[2mm]
           \includegraphics[width=1\textwidth]{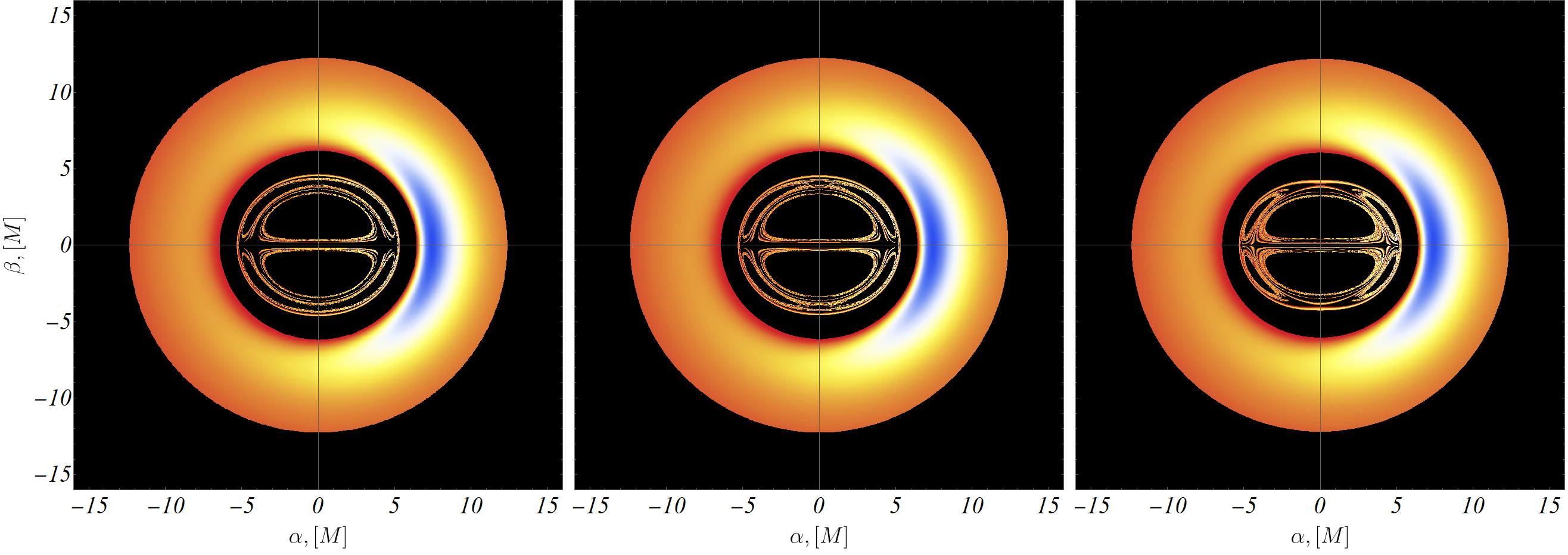} \\[0mm]
               \hspace{0.5cm}  $(\mathrm{d})  \;\,  \Sigma^2/M^2=1.5$ 
               \hspace{2.4cm}  $(\mathrm{e})  \;\,  \Sigma^2/M^2=1.6$ 
               \hspace{2.4cm}  $(\mathrm{f})  \;\,  \Sigma^2/M^2=2.0$ \\[2mm]
           \includegraphics[width=1\textwidth]{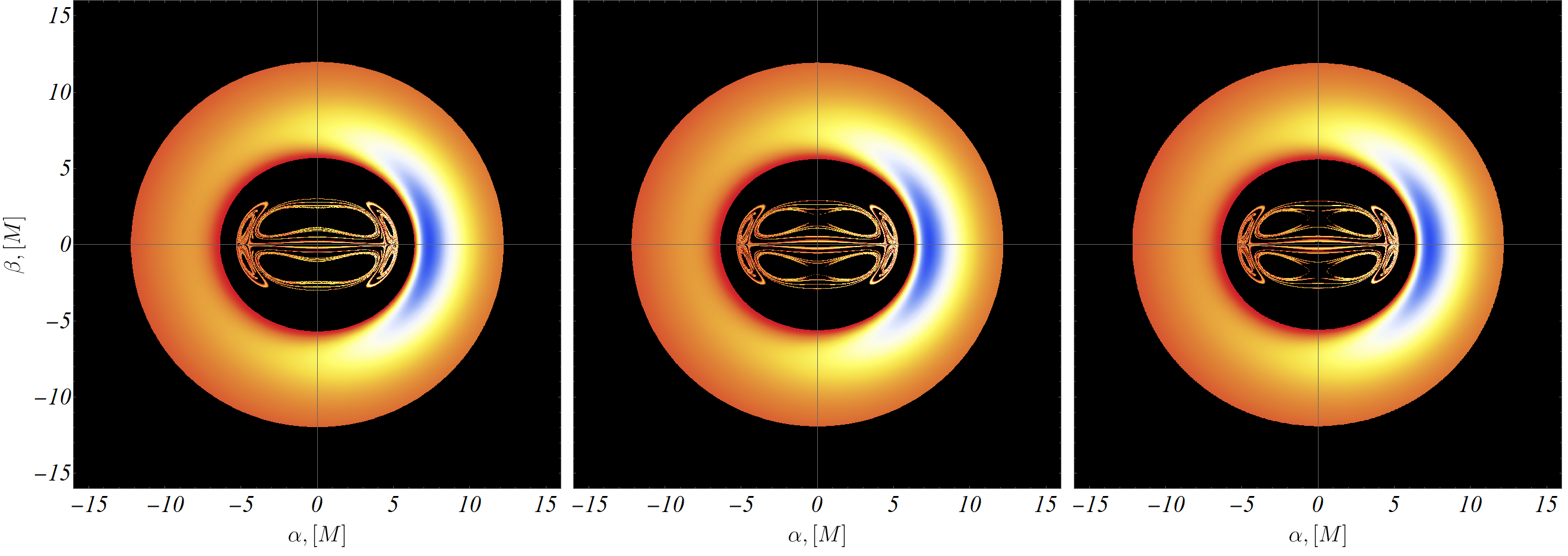} \\[0mm]
              \hspace{0.5cm}  $(\mathrm{g})  \;\,  \Sigma^2/M^2=3.5$ 
              \hspace{2.4cm}  $(\mathrm{h})  \;\,  \Sigma^2/M^2=3.8$ 
              \hspace{2.4cm}  $(\mathrm{i})  \;\,  \Sigma^2/M^2=3.9$ \\[0mm]
		\end{tabular}}
\caption{\label{fig:CD_KL_I_a=0_90deg} Continuous distribution of the apparent radiation flux for the KL black hole/naked singularity for various values of $\Sigma^2/M^2$ in the static case $a=0$. The accretion disk is placed in the range $r_{ISCO}<r<30M$, where $r_{ISCO}=6M$ coincide with Schwarschild for any $\Sigma$. The observer is located at $\theta_{O}=\pi/2$, $r_{O}/M=10\,000$. The flux distribution function is normalized by the maximal value of the observable flux $F_O^{\,max}$. The darkest red/blue colors correspond to the minimum/maximum value of the apparent
radiation flux.}
\end{figure}


\begin{figure}[]
\setlength{\tabcolsep}{ 0 pt }{\footnotesize\tt
		\begin{tabular}{ cc}
           \includegraphics[width=1\textwidth]{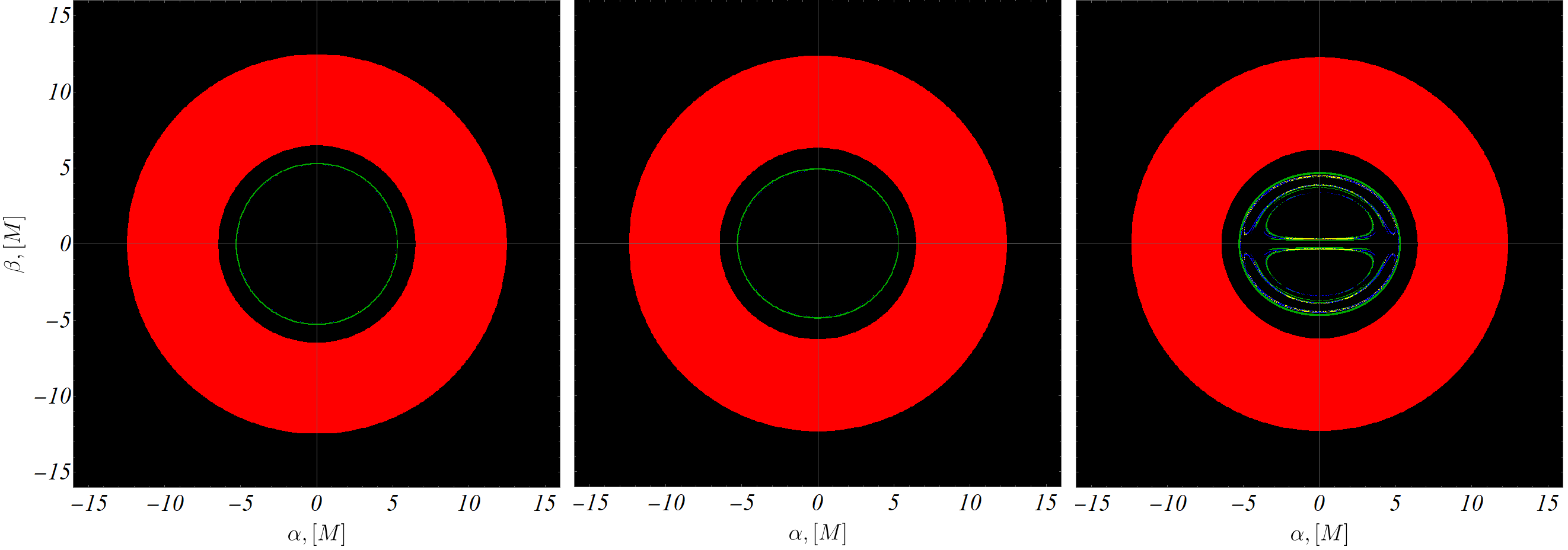} \\[0mm]
              \hspace{-0.5cm}  $(\mathrm{a})  \;\,  \Sigma^2/M^2=0 \;\, (\mathrm{Schwarzschild})$
              \hspace{1.2cm}  $(\mathrm{b})  \;\,  \Sigma^2/M^2=0.9$ 
              \hspace{2.4cm}  $(\mathrm{c})  \;\,  \Sigma^2/M^2=1.35$ \\[2mm]
           \includegraphics[width=1\textwidth]{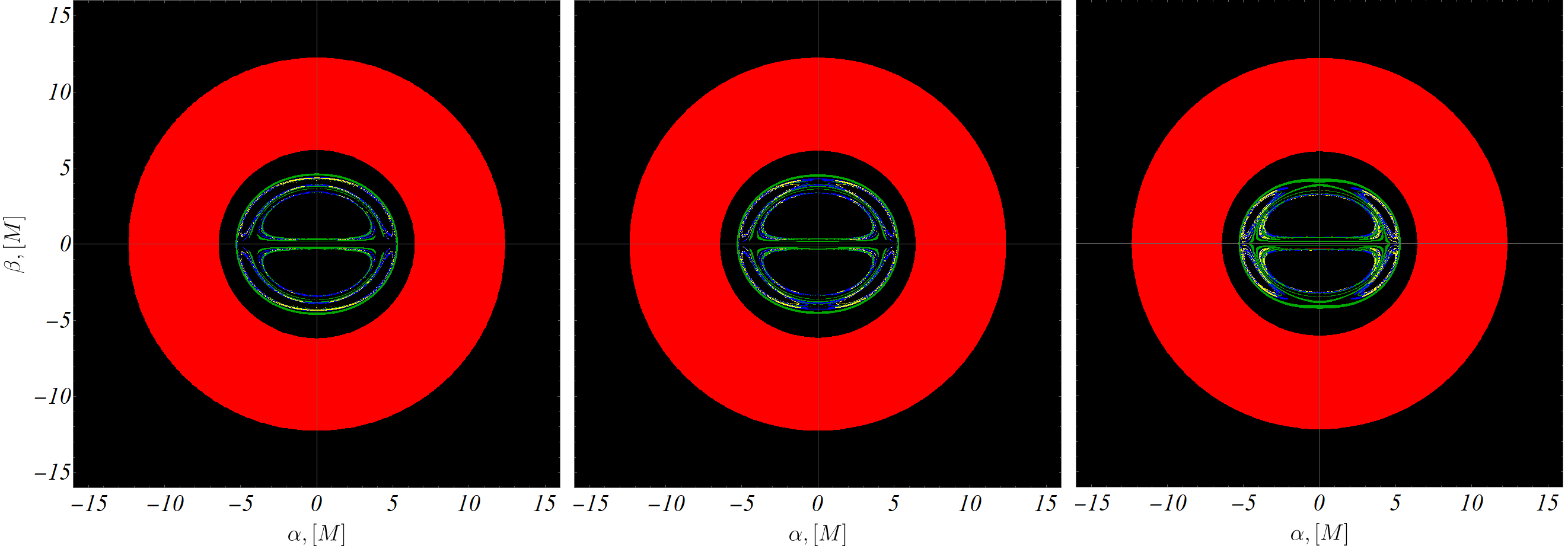} \\[0mm]
                \hspace{0.5cm}  $(\mathrm{d})  \;\,  \Sigma^2/M^2=1.5$ 
               \hspace{2.4cm}  $(\mathrm{e})  \;\,  \Sigma^2/M^2=1.6$ 
               \hspace{2.4cm}  $(\mathrm{f})  \;\,  \Sigma^2/M^2=2.0$ \\[2mm]
           \includegraphics[width=1\textwidth]{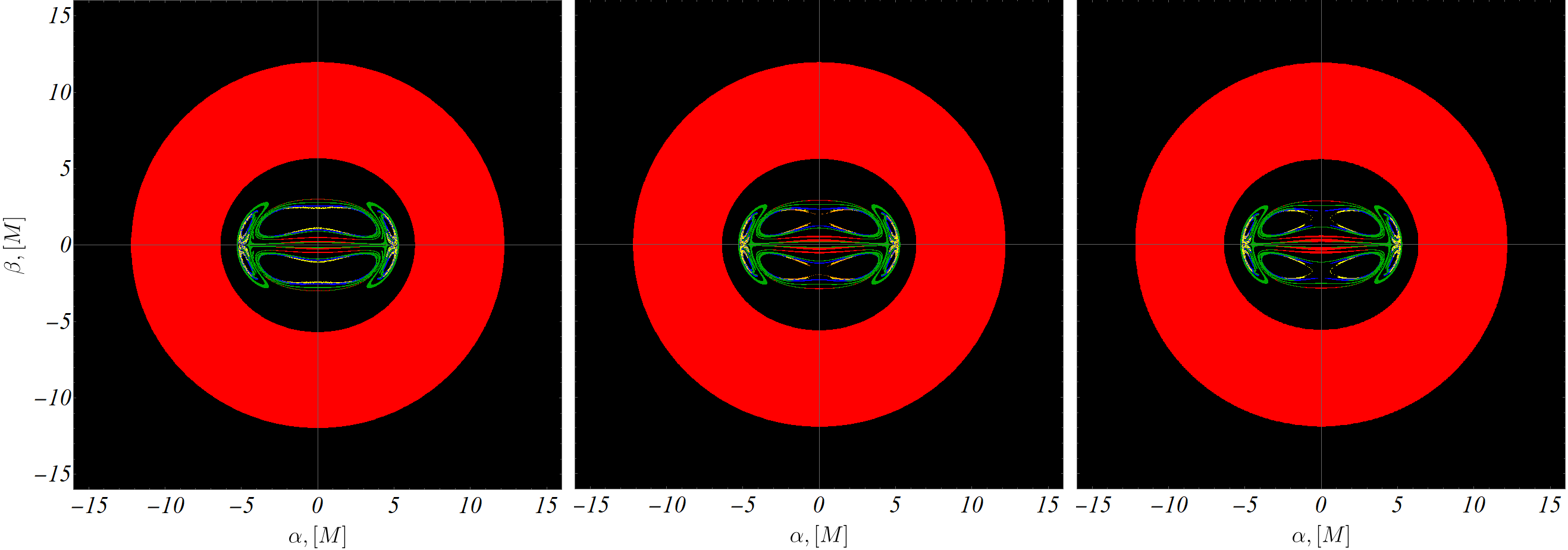} \\[0mm]
              \hspace{0.5cm}  $(\mathrm{g})  \;\,  \Sigma^2/M^2=3.5$ 
              \hspace{2.4cm}  $(\mathrm{h})  \;\,  \Sigma^2/M^2=3.8$ 
              \hspace{2.4cm}  $(\mathrm{i})  \;\,  \Sigma^2/M^2=3.9$ \\[0mm]
		\end{tabular}}
 \caption{\label{fig:KL_I_IsoR_Disk}\small Optical appearance of the thin accretion disk around the KL black hole/naked singularity for various values of the scalar charge $\Sigma^2/M^2$, and an inclination angle of the observer $\theta_{O}=\pi/2$. The direct image of the accretion disk is depicted in red, while the secondary image is in green. Images from third, fourth, fifth and higher-orders are depicted in blue, yellow and orange correspondingly. We use the same conventions as in Fig. \ref{fig:CD_KL_I_a=0_90deg}.}
\end{figure}

\begin{figure}[]
\setlength{\tabcolsep}{ 0 pt }{\footnotesize\tt
		\begin{tabular}{ cc}
           \includegraphics[width=1\textwidth]{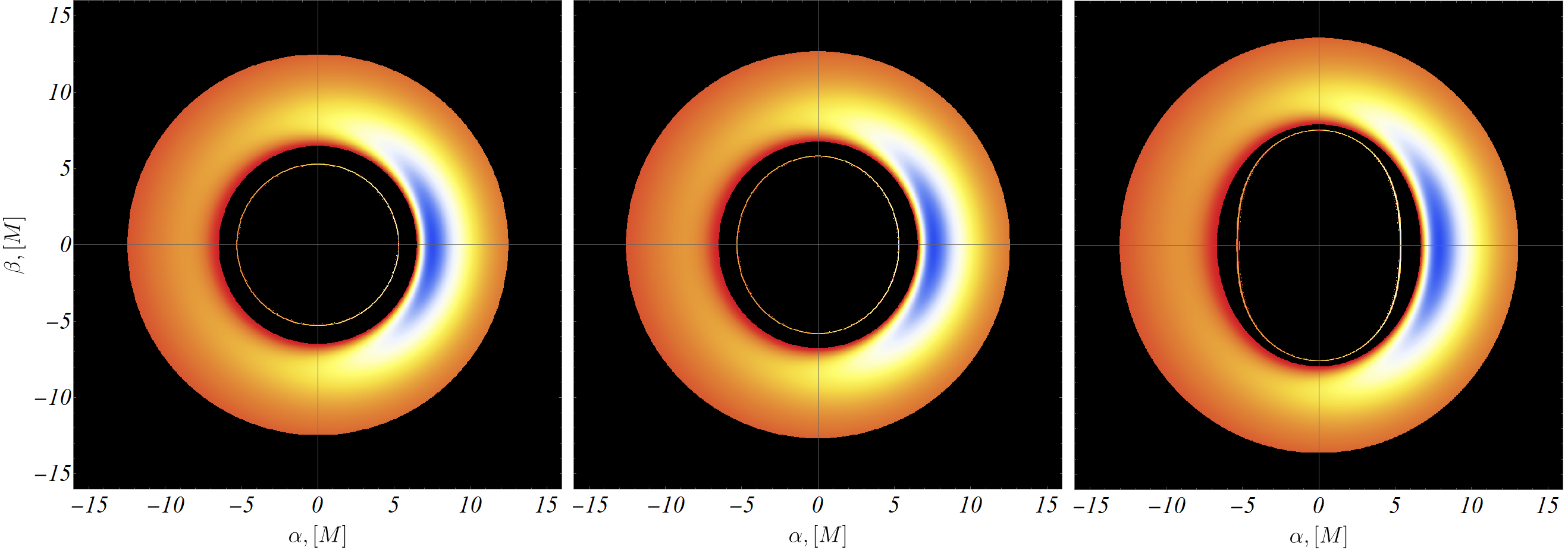} \\[1mm]
              \hspace{-0.6cm}  $(\mathrm{a})  \;\,  \Sigma^2/M^2=0 \;\, (\mathrm{Schwarzschild})$ 
              \hspace{1.0cm}  $(\mathrm{b})  \;\,  \Sigma^2/M^2=-1.5$ 
              \hspace{2.3cm}  $(\mathrm{c})  \;\,  \Sigma^2/M^2=-9$ \\[0mm]
		\end{tabular}}
 \caption{\label{fig:CD_KL_I_Fantom_a=0_90deg}\small Continuous distribution of the apparent radiation flux for the static KL black hole/naked singularity with phantom scalar charge $\Sigma^2/M^2<0$. The inclination angle of the observer is $\theta_{O}=\pi/2$, and their radial position is $r_{O}/M=10\,000$. The flux is normalized by the maximal value of the observable flux distribution for every individual disk. We use the same conventions as in Fig. \ref{fig:CD_KL_I_a=0_90deg}.}
\end{figure}

\begin{figure}[]
\setlength{\tabcolsep}{ 0 pt }{\footnotesize\tt
		\begin{tabular}{ cc}
           \includegraphics[width=1\textwidth]{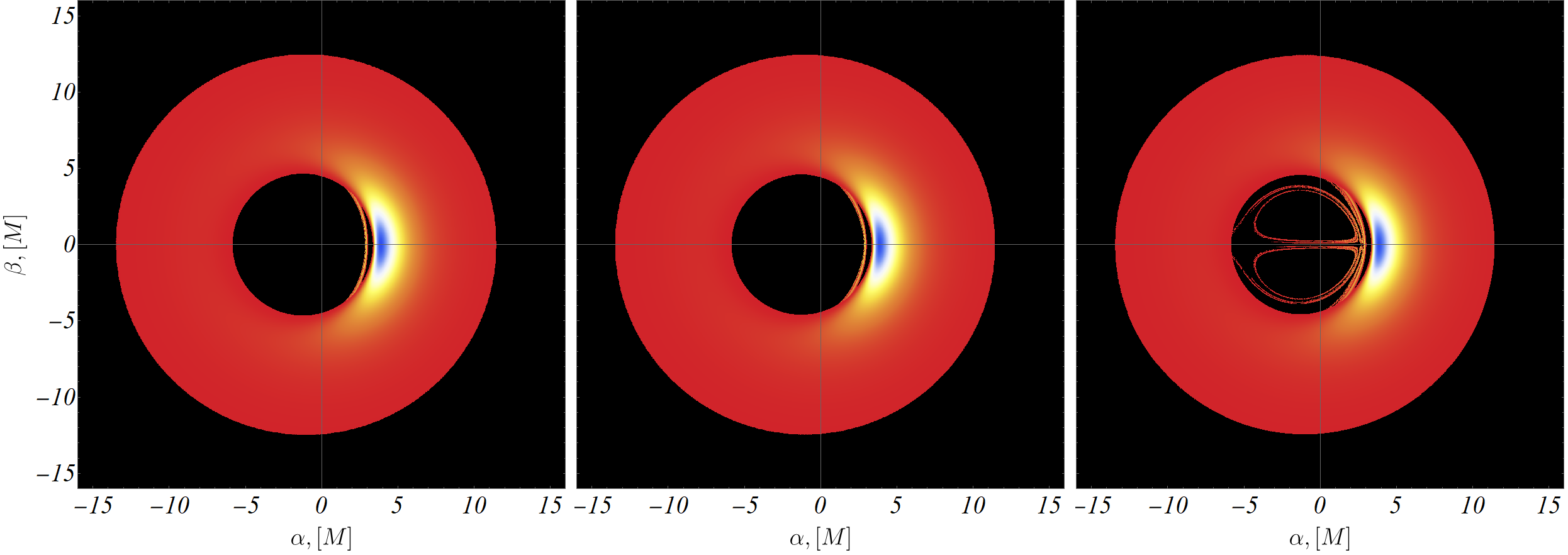} \\[0mm]
              \hspace{0.2cm}  $(\mathrm{a})  \;\,  \Sigma^2/M^2=0 \;\, (\mathrm{Kerr})$ 
              \hspace{1.8cm}  $(\mathrm{b})  \;\,  \Sigma^2/M^2=0.2$ 
              \hspace{2.4cm}  $(\mathrm{c})  \;\,  \Sigma^2/M^2=0.3$ \\[2mm]
           \includegraphics[width=1\textwidth]{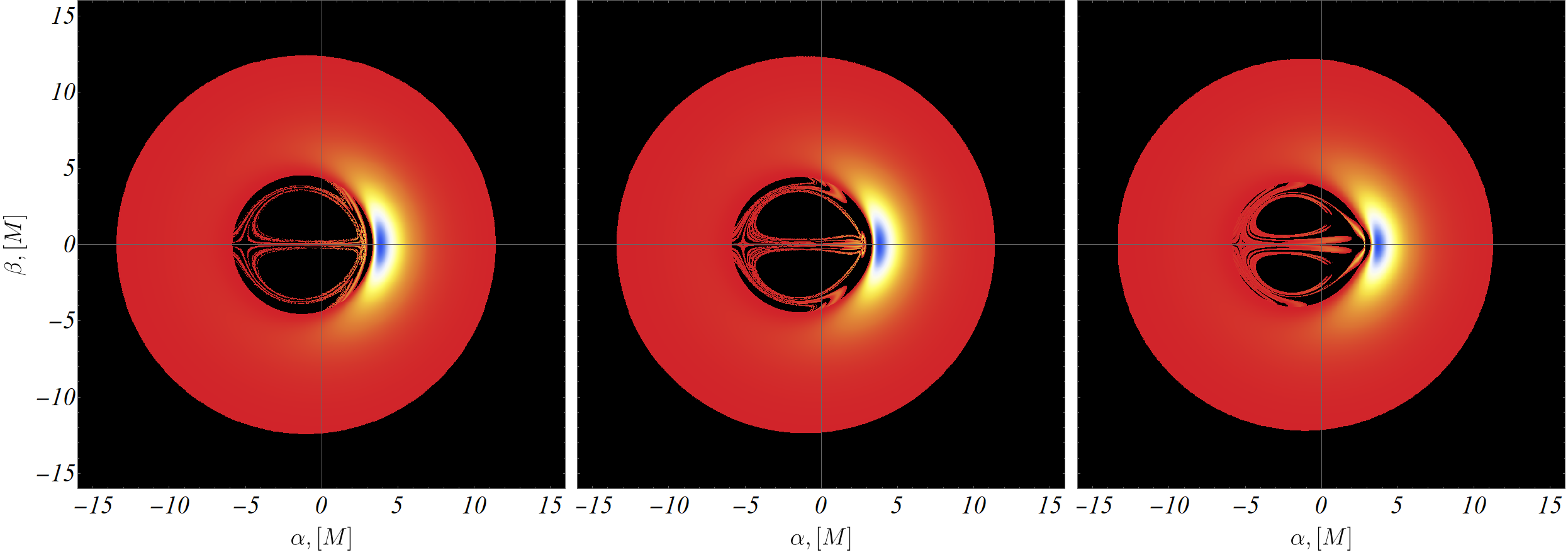} \\[0mm]
               \hspace{0.5cm}  $(\mathrm{d})  \;\,  \Sigma^2/M^2=0.4$ 
               \hspace{2.4cm}  $(\mathrm{e})  \;\,  \Sigma^2/M^2=0.8$ 
               \hspace{2.4cm}  $(\mathrm{f})  \;\,  \Sigma^2/M^2=2$ \\[0mm]
		\end{tabular}}
\caption{\label{fig:CD_KL_I_a=0.9_90deg}\small Continuous distribution of the apparent radiation flux for the KL black hole/naked singularity for a specific angular momentum $a/M=0.9$ and various values of the charge $\Sigma^2/M^2$. The innermost stable circular orbit does not depend on the scalar charge and is located at $r_{ISCO}/M = 2.32$. The inclination angle of the observer is $\theta_{O}=\pi/2$ located at radial position $r_{O}/M=10\,000$.}
\end{figure}


\begin{figure}[]
\setlength{\tabcolsep}{ 0 pt }{\footnotesize\tt
		\begin{tabular}{ cc}
           \includegraphics[width=1\textwidth]{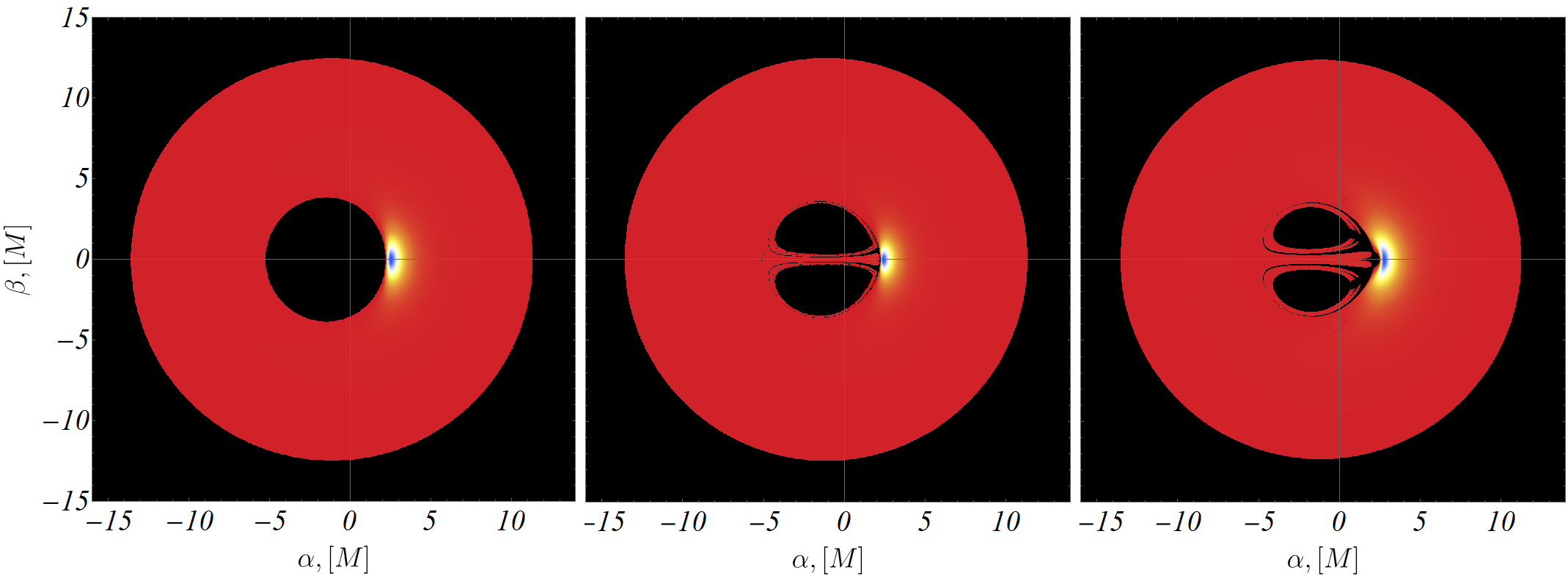} \\[1mm]
              \hspace{0.0cm}  $(\mathrm{a})  \;\,  \Sigma^2/M^2=0 \;\, (\mathrm{Kerr})$ 
              \hspace{1.8cm}  $(\mathrm{b})  \;\,  \Sigma^2/M^2=0.2$ 
              \hspace{2.5cm}  $(\mathrm{c})  \;\,  \Sigma^2/M^2=1$ \\[0mm]
		\end{tabular}}
 \caption{\label{fig:CD_KL_I_а=0.9_90}\small Continuous distribution of the apparent radiation flux for the near-extremal KL black hole/naked singularity $a/M=0.998$ at various values of the scalar charge $\Sigma^2/M^2$. The innermost stable circular orbit does not depend on the scalar charge and is located at $r_{ISCO}/M = 1.24$. The inclination angle of the observer is $\theta_{O}=\pi/2$, and their radial position is $r_{O}/M=10\,000$. The flux is normalized by the maximal value of the observable flux distribution for every individual accretion disk. We use the same conventions as in Fig. \ref{fig:CD_KL_I_a=0.9_90deg}.}
\end{figure}


\begin{figure}[]
\centering
\setlength{\tabcolsep}{ 0 pt }{\footnotesize\tt
		\begin{tabular}{ cc}
           \includegraphics[width=1\textwidth]{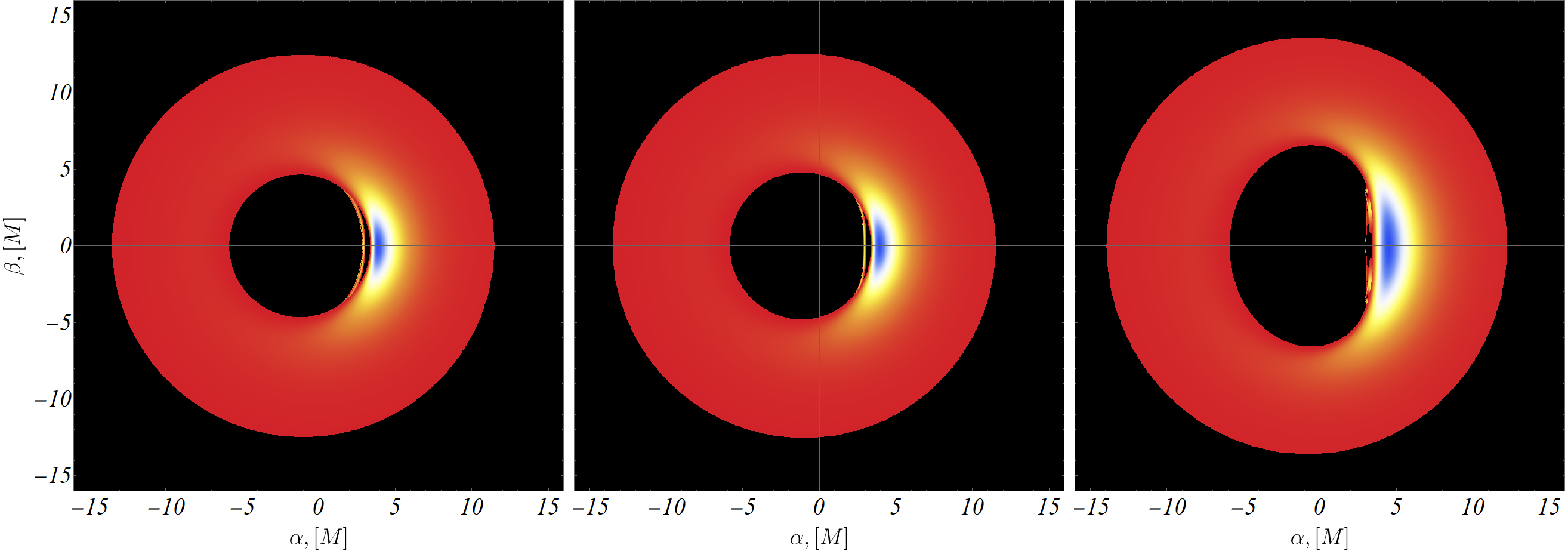}\\[1mm]
              \hspace{0.0cm}  $(\mathrm{a})  \;\,  \Sigma^2/M^2=0 \;\, (\mathrm{Kerr})$ 
              \hspace{1.8cm}  $(\mathrm{b})  \;\,  \Sigma^2/M^2=-0.5$ 
              \hspace{2.3cm}  $(\mathrm{c})  \;\,  \Sigma^2/M^2=-9$ \\[0mm] 
		\end{tabular}}
 \caption{\label{fig:CD_KL_I_Phantom_a=0.9_90deg}\small Continuous distribution of the apparent radiation flux for the KL black hole/naked singularity for a specific angular momentum $a/M=0.9$ and phantom scalar charge $\Sigma^2/M^2\leq 0$. The inclination angle of the observer is $\theta_{O}=\pi/2$, and their radial position is $r_{O}/M=10\,000$. The flux distribution is normalized by the maximal value of the observable flux for each individual disk.}
\end{figure}

\begin{figure}[]
\setlength{\tabcolsep}{ 0 pt }{\footnotesize\tt
		\begin{tabular}{ cc}
           \includegraphics[width=1\textwidth]{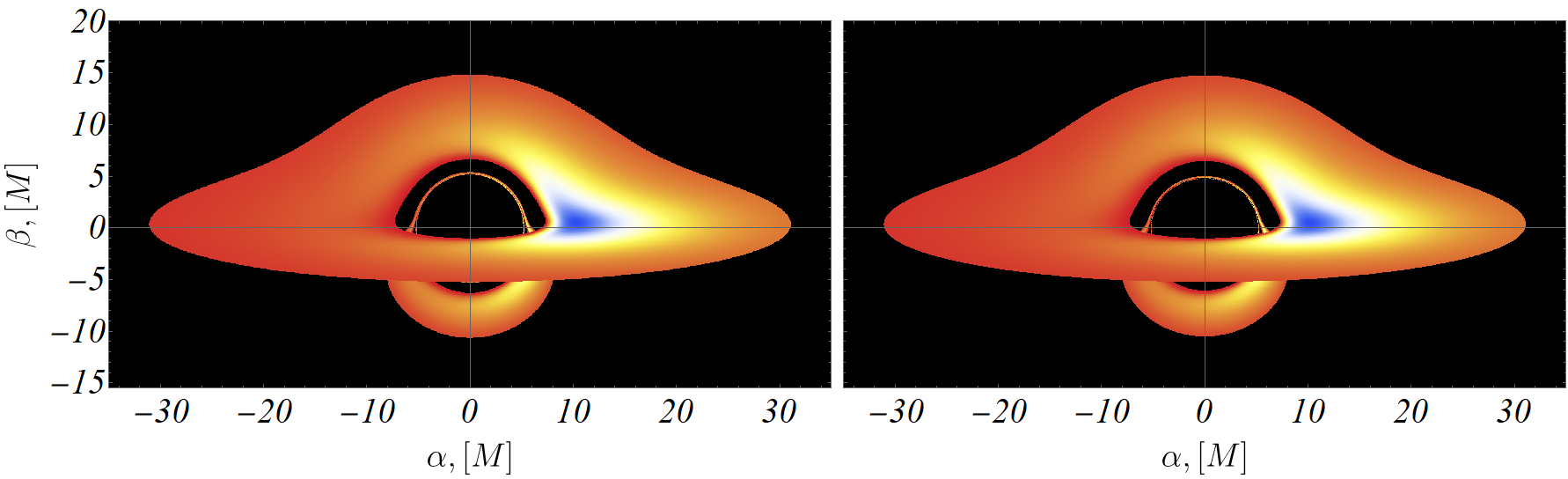}  \\[1mm]
              \hspace{-0.0cm}  $(\mathrm{a})  \;\,  \Sigma^2/M^2=0 \;\, (\mathrm{Schwarzschild})$
              \hspace{3.6cm}  $(\mathrm{b})  \;\,  \Sigma^2/M^2=0.9$ \\[3mm]
           \includegraphics[width=1\textwidth]{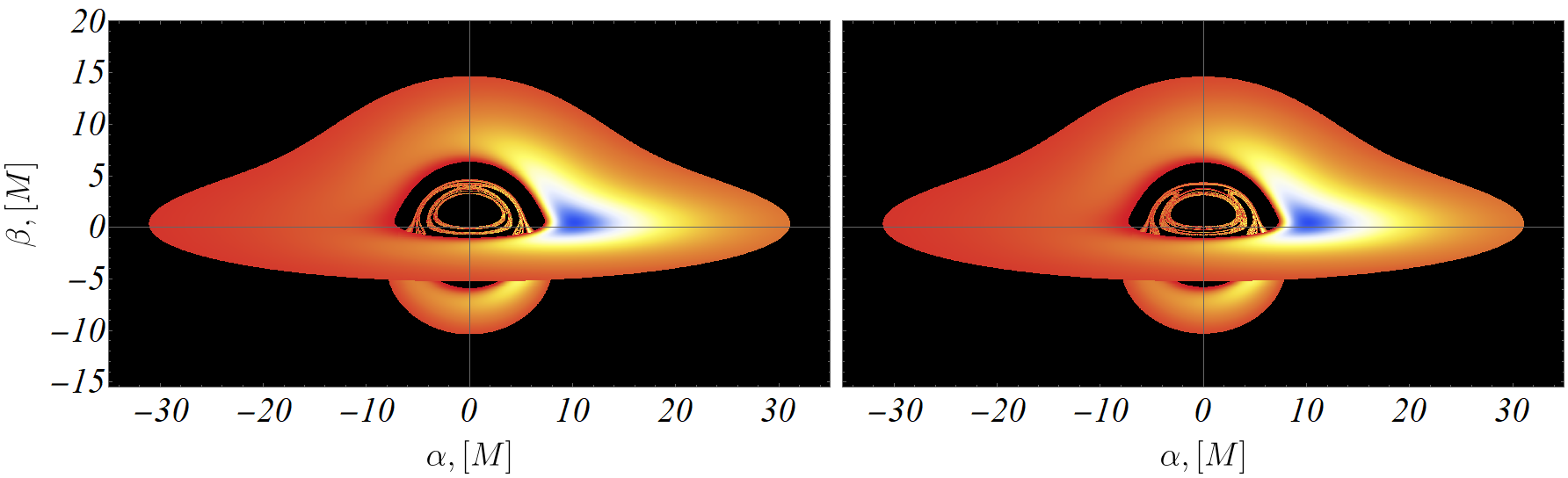} \\[1mm]
              \hspace{0.9cm}  $(\mathrm{c})  \;\,  \Sigma^2/M^2=1.6$ 
              \hspace{4.9cm}  $(\mathrm{d})  \;\,  \Sigma^2/M^2=2.0$ \\[3mm]
           \includegraphics[width=1\textwidth]{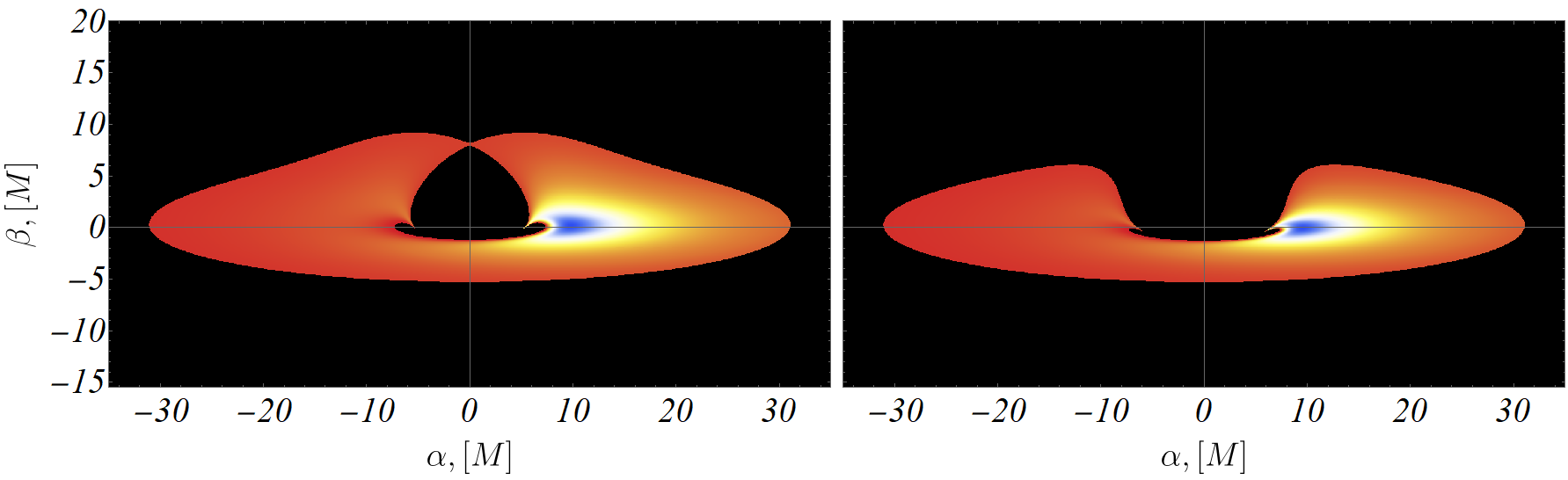}
           \\[1mm]
              \hspace{0.6cm}  $(\mathrm{e})  \;\,  \Sigma^2/M^2=31.8$ 
              \hspace{5.0cm}  $(\mathrm{f})  \;\,  \Sigma^2/M^2=50$ \\[3mm]
		\end{tabular}}
 \caption{\small Continuous distribution of the apparent radiation flux for the KL black hole/naked singularity for various values of the scalar charge $\Sigma^2/M^2$. The inclination angle of the observer is $\theta_{O}=4\pi/9$, and their radial position is $r_{O}/M=10\,000$. The flux distribution is normalized by the maximal value of the observable flux for every individual disk. We use the same conventions as in Fig. \ref{fig:CD_KL_I_a=0_90deg}\label{fig:KL_I_a=0_80deg}.}
\end{figure}


\begin{figure}[]
\setlength{\tabcolsep}{ 0 pt }{\footnotesize\tt
		\begin{tabular}{ cc}
           \includegraphics[width=1\textwidth]{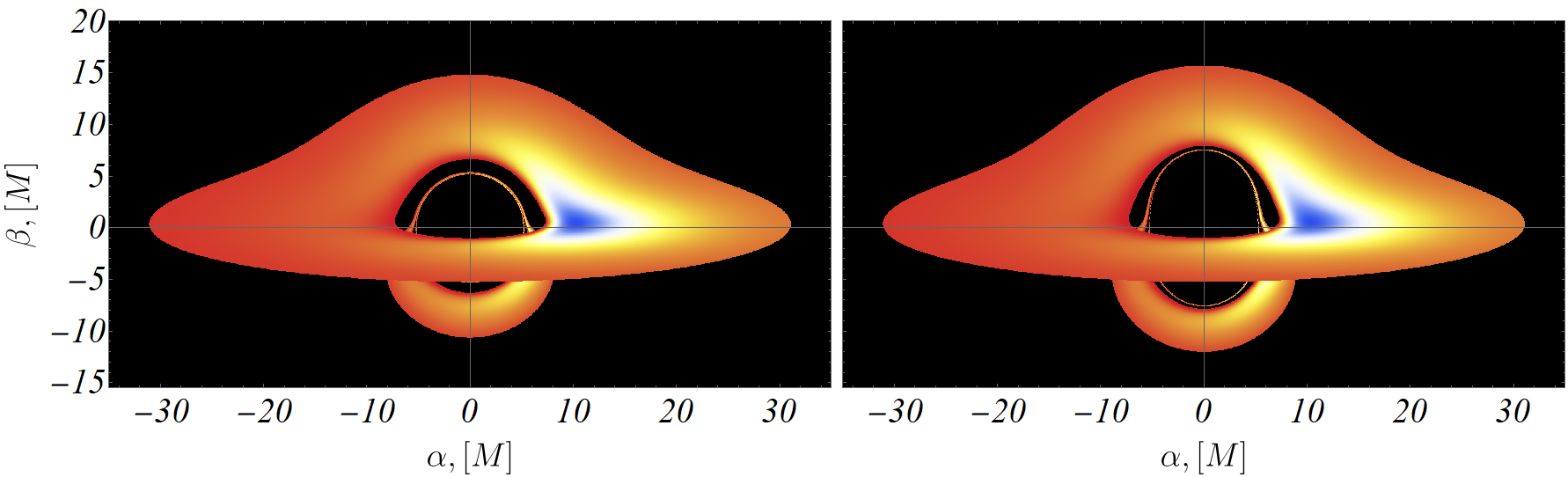} \\[1mm]
              \hspace{-0.3cm}  $(\mathrm{a})  \;\,  \Sigma^2/M^2=0 \;\, (\mathrm{Schwarzschild})$ 
              \hspace{3.8cm}  $(\mathrm{b})  \;\,  \Sigma^2/M^2=-9$ \\[0mm]
		\end{tabular}}
 \caption{\label{fig:KL_I_Stat_Phantom_80deg_a}\small Continuous distribution of the apparent radiation flux for the static KL black hole/naked singularity with phantom scalar charge $\Sigma^2/M^2$. The inclination angle of the observer is $\theta_{O}=4\pi/9$, and their radial position is $r_{O}/M=10\,000$. The flux distribution is normalized by the maximal value of the observable flux for every individual disk. We use the same conventions as in Fig. \ref{fig:CD_KL_I_a=0_90deg}.}
\end{figure}

\begin{figure}[]
\setlength{\tabcolsep}{ 0 pt }{\footnotesize\tt
		\begin{tabular}{ cc}
           \includegraphics[width=1\textwidth]{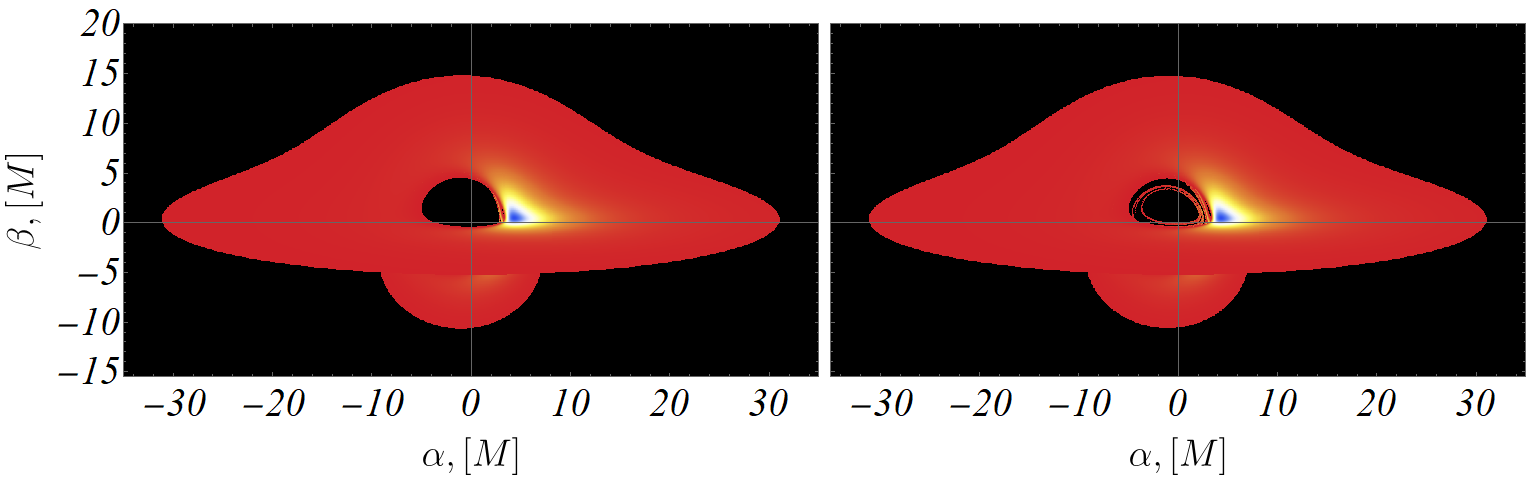}\\[1mm]
              \hspace{0.7cm}  $(\mathrm{a})  \;\,  \Sigma^2/M^2=0 \;\, (\mathrm{Kerr})$  
              \hspace{4.3cm}  $(\mathrm{b})  \;\,  \Sigma^2/M^2=0.3$ \\[3mm]
              \includegraphics[width=1\textwidth]{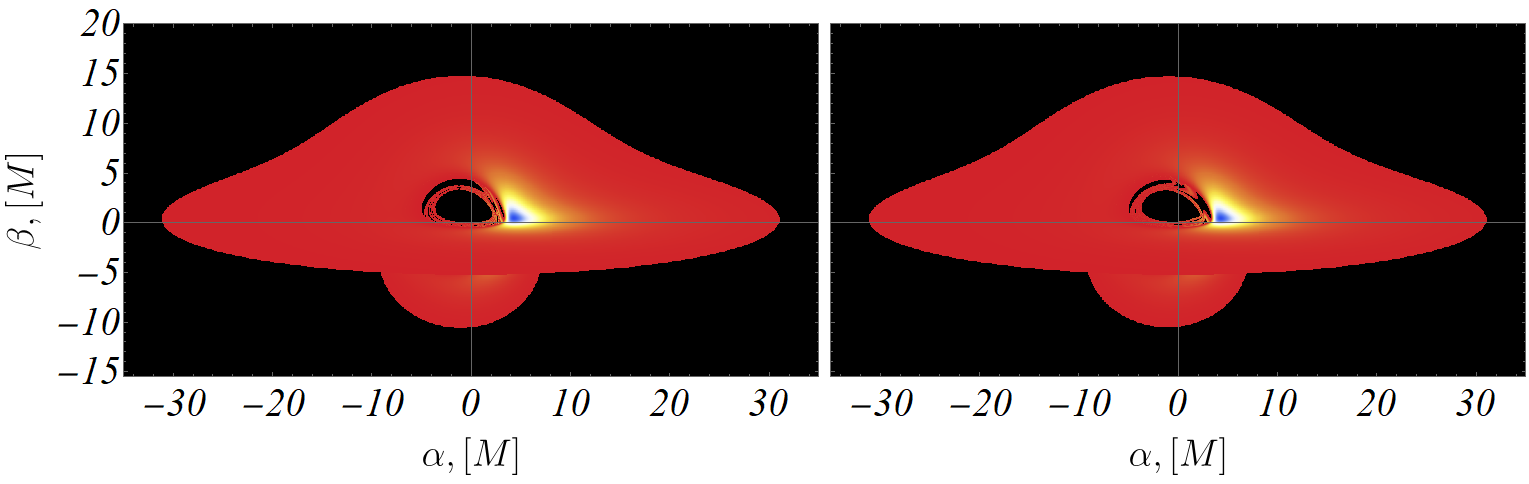}\\[1mm]
              \hspace{1.0cm}  $(\mathrm{c})  \;\,  \Sigma^2/M^2=0.4$ 
              \hspace{4.85cm}  $(\mathrm{d})  \;\,  \Sigma^2/M^2=0.8$ \\[0mm]
		\end{tabular}}
 \caption{\label{fig:CD_KL_I_a=0.9_80deg}\small Continuous distribution of the apparent radiation flux for the KL black hole/naked singularity for a specific angular momentum $a/M=0.9$ and various values of the scalar charge $\Sigma^2/M^2$. The inclination angle of the observer is $\theta_{O}=4\pi/9$, and their radial position is $r_{O}/M=10\,000$. }
\end{figure}

\begin{figure}[]
\setlength{\tabcolsep}{ 0 pt }{\footnotesize\tt
		\begin{tabular}{ cc}
           \includegraphics[width=0.63\textwidth]{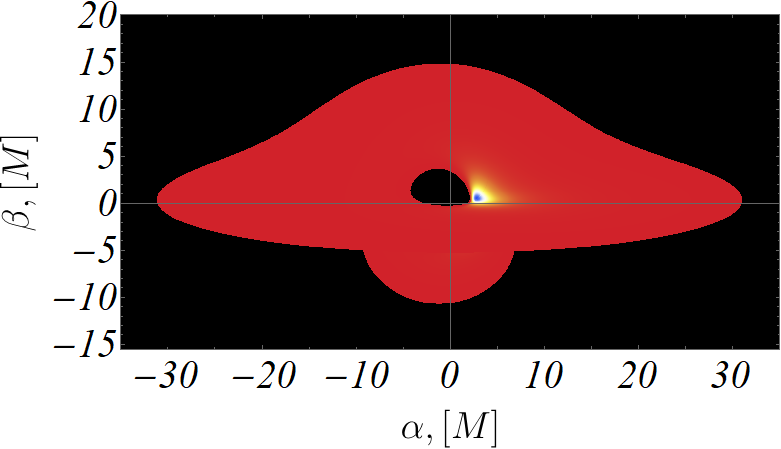}  \\[0mm]
                \hspace{1.2cm}  $(\mathrm{a})  \;\,  \Sigma^2/M^2=0 \;\, (\mathrm{Kerr})$       \\[1mm]
           \includegraphics[width=0.63\textwidth]{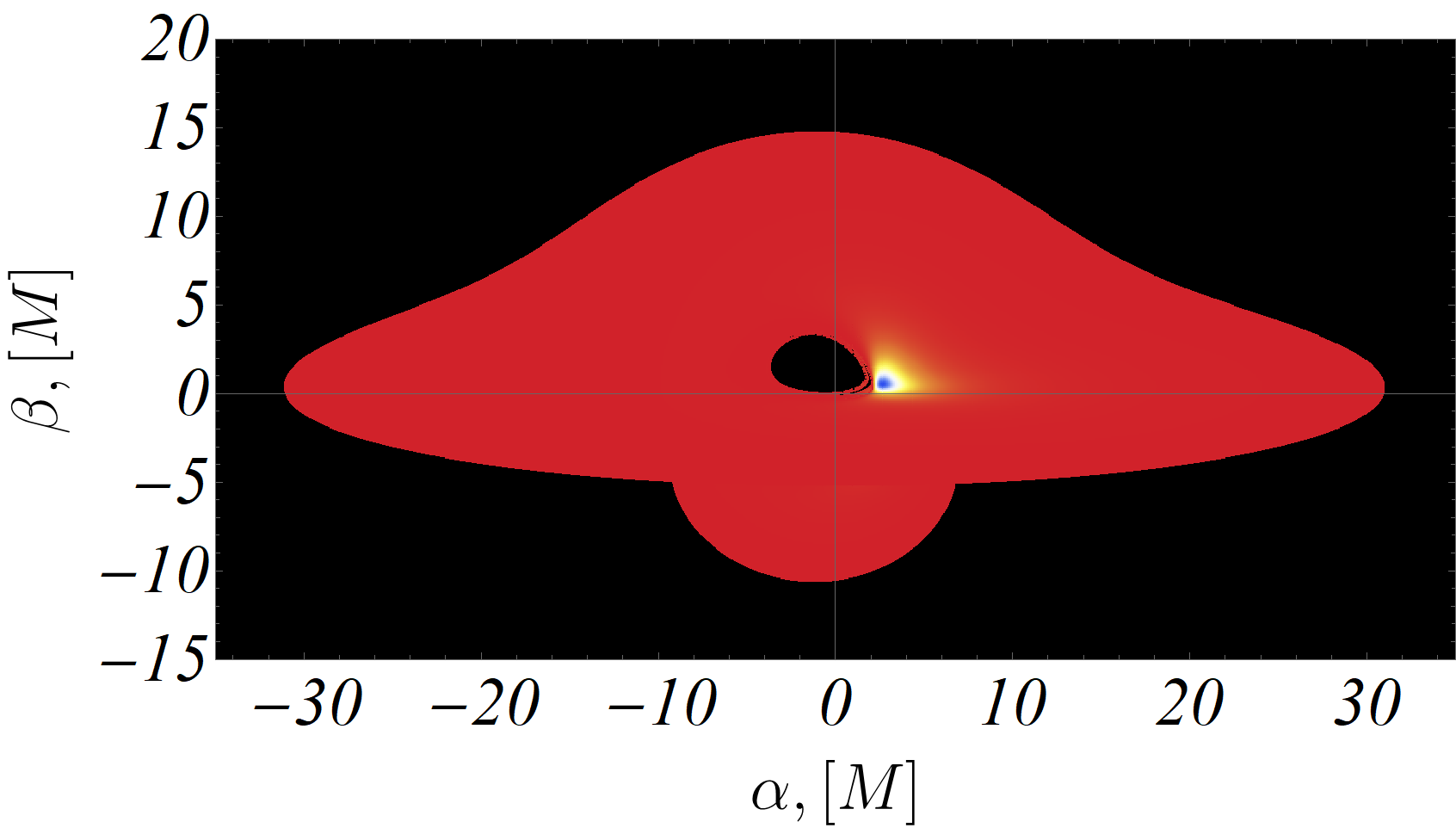}  \\[0mm]
                \hspace{1.1cm}  $(\mathrm{c})  \;\,  \Sigma^2/M^2=0.2$     \\[0mm]
           \includegraphics[width=0.63\textwidth]{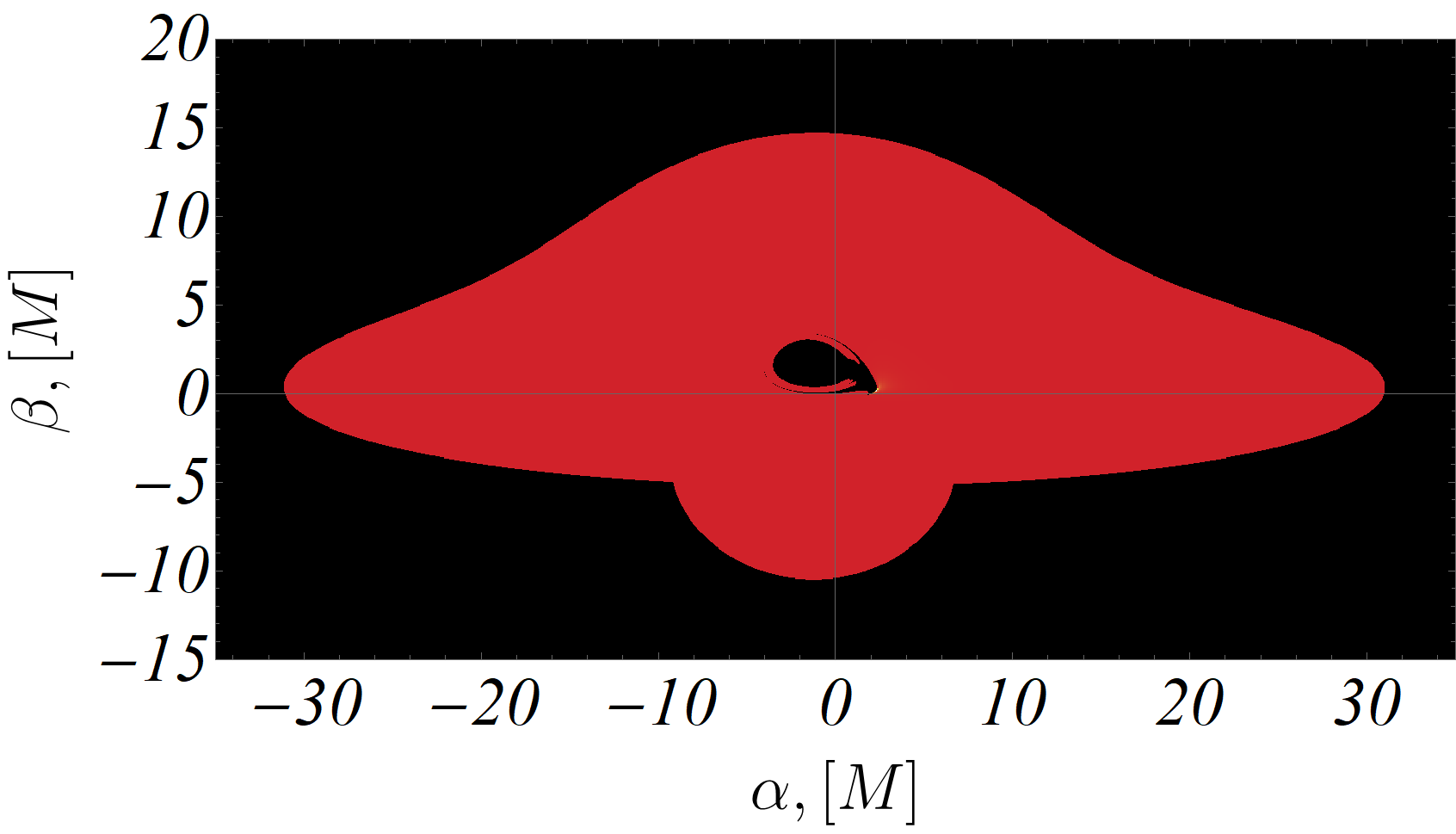}  \\[0mm]
                \hspace{1.2cm}  $(\mathrm{e})  \;\,  \Sigma^2/M^2=1$    \\[0mm]
		\end{tabular}}
		\vspace{-0.2cm}
 \caption{\small Continuous distribution of the apparent radiation flux for the near-extremal KL black hole/naked singularity $a/M=0.998$ for various values of the scalar charge $\Sigma^2/M^2$. The innermost stable circular orbit does not depend on the scalar charge and is located at $r_{ISCO}/M = 1.24$. The inclination angle of the observer is $\theta_{O}=4\pi/9$, and their radial position is $r_{O}/M=10\,000$. The flux is normalized by the maximal value of the observable flux distribution for every individual accretion disk. \label{fig:KL_I_Extremal_a=0_80deg}}
\end{figure}


\begin{figure}[]
\setlength{\tabcolsep}{ 0 pt }{\footnotesize\tt
		\begin{tabular}{ cc}
           \includegraphics[width=1\textwidth]{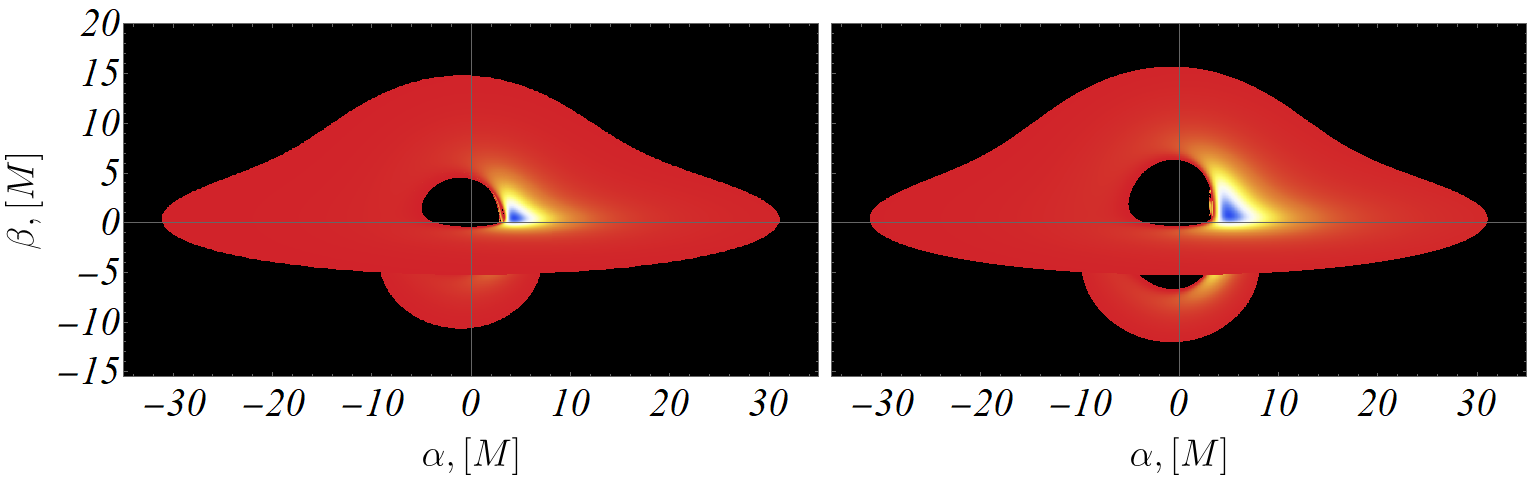} \\[1mm]
              \hspace{0.6cm}  $(\mathrm{a})  \;\,  \Sigma^2/M^2=0 \;\, (\mathrm{Kerr})$ 
              \hspace{4.4cm}  $(\mathrm{b})  \;\,  \Sigma^2/M^2=-9$ \\[0mm]
		\end{tabular}}
 \caption{\label{fig:KL_I_Rot_Phantom_80deg}\small Continuous distribution of the apparent radiation flux for the KL black hole/naked singularity for a specific angular momentum $a/M=0.9$, and phantom scalar charge $\Sigma^2/M^2$. The inclination angle of the observer is $\theta_{O}=4\pi/9$, and their radial position is $r_{O}/M=10\,000$.}
\end{figure}

\begin{figure}[]
\setlength{\tabcolsep}{ 0 pt }{\footnotesize\tt
		\begin{tabular}{ cc}
           \includegraphics[width=1\textwidth]{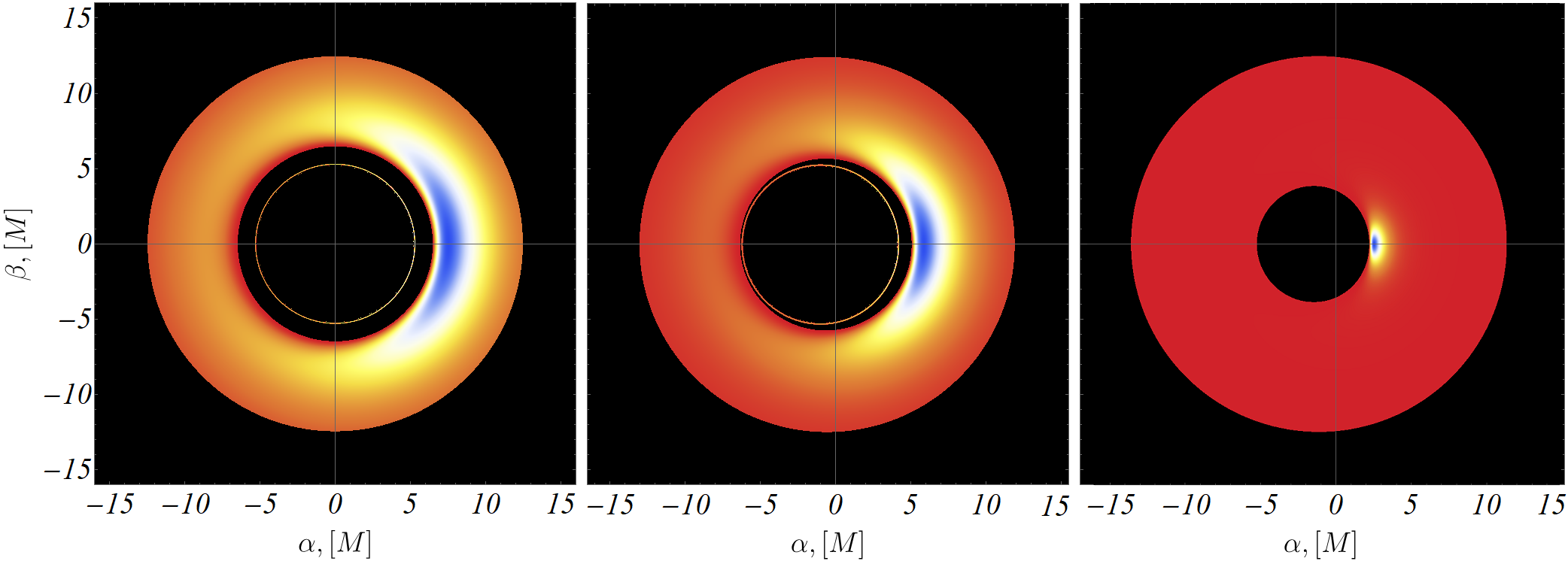} \\[1mm]
              \hspace{0.6cm}  $(\mathrm{a})  \;\,  a=0     \;\, (\mathrm{Schwarzschild})$ 
              \hspace{1.3cm}  $(\mathrm{b})  \;\,  a/M=0.5 \;\, (\mathrm{Kerr})$
              \hspace{1.5cm}  $(\mathrm{c})  \;\,  a/M=0.998   \;\, (\mathrm{Kerr})$ \\[3mm]
           \includegraphics[width=1\textwidth]{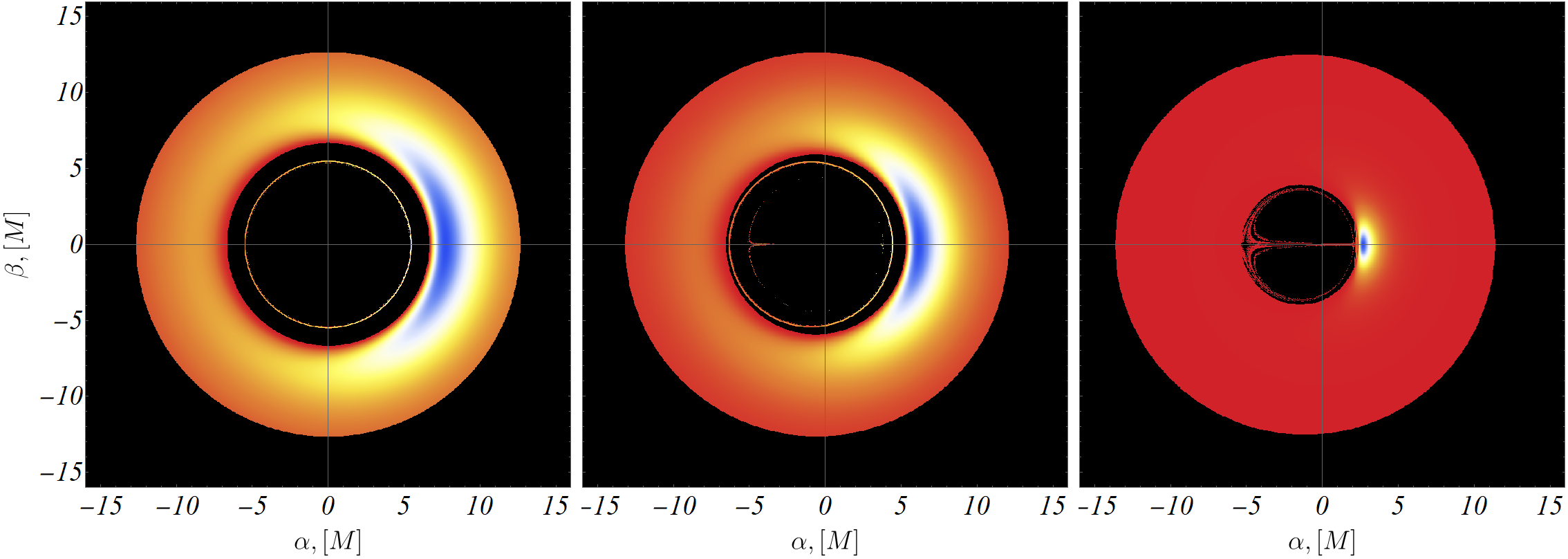} \\[1mm]
              \hspace{1.0cm}  $(\mathrm{d})  \;\,  a/M=0   \;\, (\mathrm{TSL})$
              \hspace{1.8cm}  $(\mathrm{e})  \;\,  a/M=0.5 \;\, (\mathrm{TSL})$
              \hspace{1.5cm}  $(\mathrm{f})  \;\,  a/M=0.998   \;\, (\mathrm{TSL})$ \\[3mm]
		\end{tabular}}
 \caption{\label{fig:ColorDisk_KL_II_90deg}\small Continuous distribution of the apparent radiation flux for Kerr black hole (first row) and TSL naked singularity/naked singularity (second row) for various values of the specific angular momentum $a$. The inclination angle of the observer is $\theta_{O}=\pi/2$ and their radial position is $r_{O}/M=10\,000$.}
\end{figure}


\begin{figure}[]
\setlength{\tabcolsep}{ 0 pt }{\footnotesize\tt
		\begin{tabular}{ cc}
           \includegraphics[width=1\textwidth]{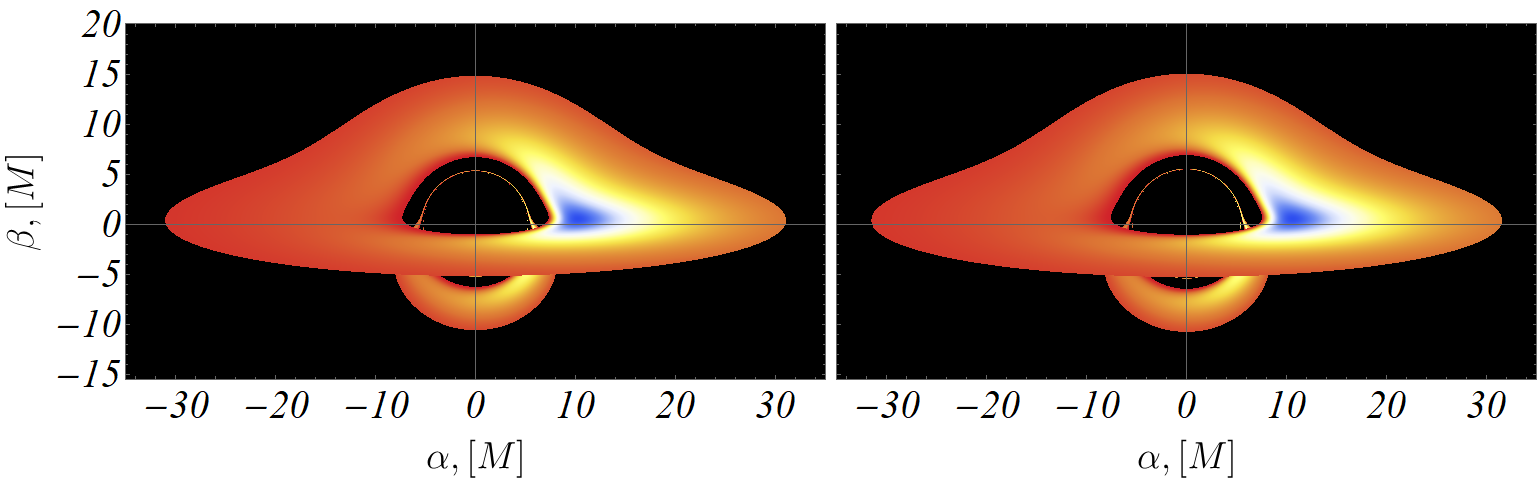} \\[1mm]
              $(\mathrm{a})  \;\,  a/M=0$: $\mathrm{Schwarzschild \; (left)}$, $\mathrm{TSL \; (right)}$ \\[3mm]
           \includegraphics[width=1\textwidth]{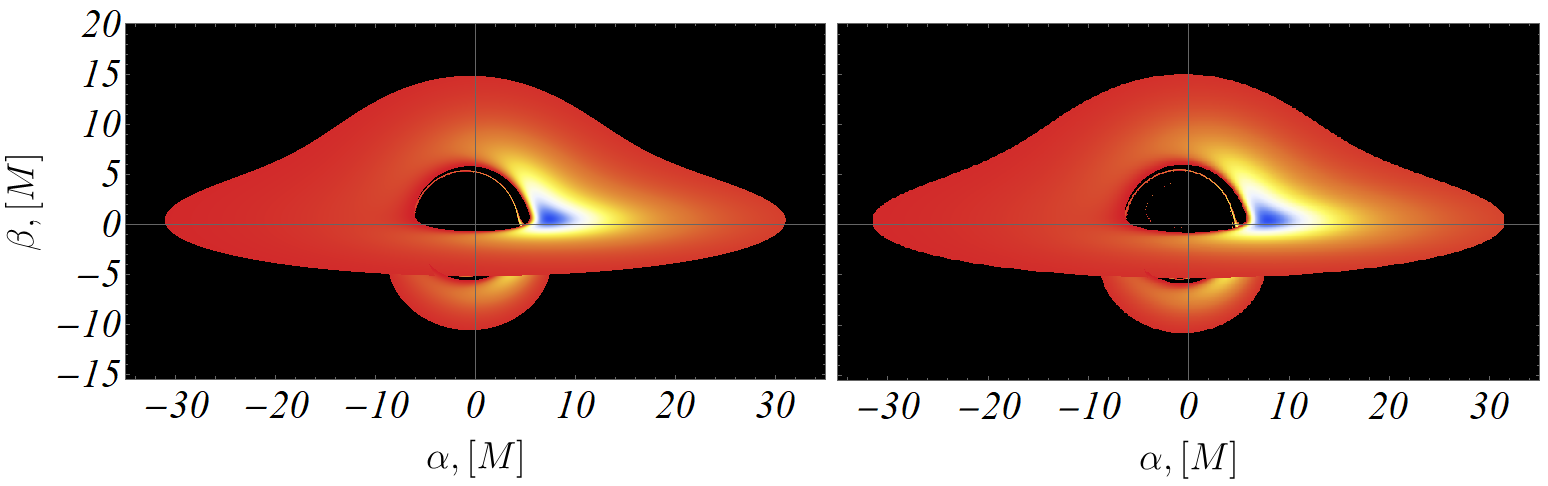} \\[1mm]
              $(\mathrm{b})  \;\,  a/M=0.5$: $\mathrm{Kerr \; (left)}$, $\mathrm{TSL \; (right)}$ \\[1mm]
           \includegraphics[width=1\textwidth]{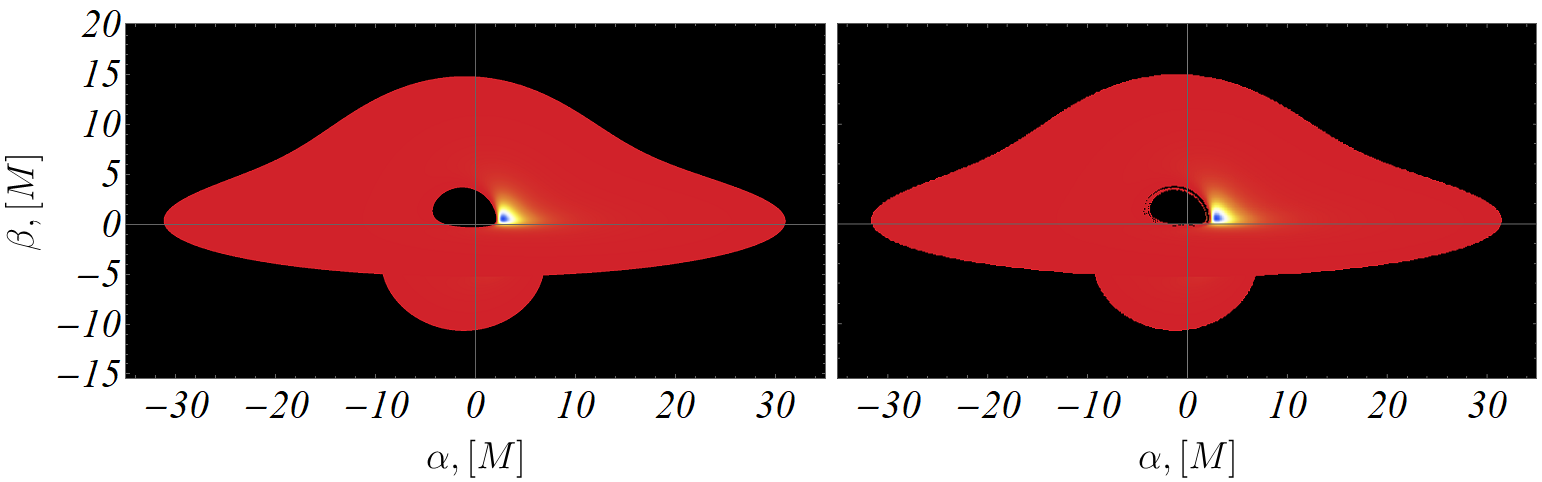}\\[1mm]
              $(\mathrm{c})  \;\,  a/M=0.998$: $\mathrm{Kerr \; (left)}$, $\mathrm{TSL \; (right)}$ \\[1mm]
		\end{tabular}}
 \caption{\label{fig:ColorDisk_KL_II_80deg}\small Continuous distribution of the apparent radiation flux for Kerr black hole (left column) and TSL naked singularity (right column) for various values of the specific angular momentum $a$. The inclination angle of the observer is $\theta_{O}=4\pi/9$, and their radial position is $r_{O}/M=10\,000$.}
\end{figure}

\FloatBarrier

\end{document}